\documentclass[aps,prd,amsmath,twocolumn,amssymb,showpacs]{revtex4}
\usepackage{amssymb}
\usepackage{txfonts}
\usepackage{mathbbold}
\usepackage{amsfonts}
\usepackage{mathrsfs}
\usepackage{epsfig,bm,dcolumn}
\usepackage{graphicx}
\usepackage{color}
\usepackage{amsmath}

\begin{document}
\title{BCS-BEC crossover in relativistic Fermi systems}
\author{Lianyi He$^{1}$, Shijun Mao$^{2}$ and Pengfei Zhuang$^{3}$}
\affiliation{1 Frankfurt Institute for Advanced Studies, 60438 Frankfurt am Main, Germany\\
2 School of Science, Xi¡¯an Jiaotong University, Xi¡¯an 710049, China\\
3 Physics Department, Tsinghua University and Collaborative Innovation Center of Quantum Matter, Beijing 100084, China}
\date{\today}

\begin{abstract}
We review the BCS-BEC crossover in relativistic Fermi systems,
including the QCD matter at finite density. In the first part we
study the BCS-BEC crossover in a relativistic four-fermion
interaction model and show how the relativistic effect affects the
BCS-BEC crossover. In the second part, we investigate both two-color
QCD at finite baryon density and pion superfluid at finite isospin
density, by using an effective Nambu--Jona-Lasinio model. We will
show how the model describes the weakly interacting diquark and pion
condensates at low density and the BEC-BCS crossover at high
density.\\

\end{abstract}
 \maketitle

\section {Introduction}
\label{s1}
It is generally believed that, by tuning the attractive strength in a Fermi system,  one can realize a smooth crossover from the Bardeen--Cooper--Schrieffer (BCS) superfluidity at weak attraction to Bose--Einstein condensation (BEC) of tightly bound difermion molecules at strong attraction~\cite{Eagles,Leggett,BCSBEC1,BCSBEC2,BCSBEC3,BCSBEC4,BCSBEC5,BCSBEC6,BCSBEC7}. The typical system is a dilute atomic Fermi gas
in three dimensions, where the effective range $r_0$ of the short-range attractive interaction is much smaller than the inter-particle distance.
Therefore, the system can be characterized by a dimensionless parameter $1/(k_{\rm f}a_s)$, where $a_s$ is the $s$-wave scattering length of
the short-range interaction and $k_{\rm f}$ is the Fermi momentum in the absence of interaction. The BCS-BEC crossover occurs when the parameter $1/(k_{\rm f}a_s)$ is tuned from negative to positive values, and the BCS and BEC limits correspond to the cases $1/(k_{\rm f}a_s)\rightarrow-\infty$ and $1/(k_{\rm f}a_s)\rightarrow+\infty$, respectively.

This BCS-BEC crossover phenomenon has been successfully demonstrated
in ultracold fermionic atoms, where the $s$-wave scattering length
and hence the parameter $1/(k_{\rm f}a_s)$ are tuned by means of the
Feshbach resonance~\cite{BCSBECexp1,BCSBECexp2,BCSBECexp3}. At the
resonant point or the so-called unitary point with
$a_s\rightarrow\infty$, the only length scale of the system is the
inter-particle distance ($\sim k_{\rm f}^{-1}$). Therefore, the
properties of the system at the unitary point $1/(k_{\rm f}a_s)=0$
become universal, i.e., independent of the details of the
interactions. All physical quantities, scaled by their counterparts
for the non-interacting Fermi gases, become universal constants.
Determining these universal constants has been one of the most
intriguing topics in the research of cold Fermi
gases~\cite{Unitary,Unitary1,Unitary11,Unitary2,Unitary21}.

The BCS-BEC crossover has become an interesting and important issue
for the studies of dense and strongly interacting matter, i.e.,
nuclear or quark matter which may exist in the core of compact
stars~\cite{BCSBECNM,BCSBECNM1,BCSBECNM2,BCSBECNM3,pairsize,kitazawa,kitazawa1,kitazawa2,kitazawa3,Abuki,RBCSBEC,RBCSBEC1,RBCSBEC2,RBCSBEC3,RBCSBEC4,RBCSBEC5,RBCSBEC6,RBCSBEC7,RBCSBEC8,RBCSBEC9,RBCSBEC10,BCSBECQCD,BCSBECQCD1,BCSBECQCD2,BCSBECQCD3,BCSBECQCD4,BCSBECQCD5,BCSBECQCD6,BCSBECQCD7,BCSBECQCD8,BCSBECQCD9}.
By analogy with the usual superfluid, the BCS-BEC crossover in dense
quark matter can be theoretically
described~\cite{pion2,pion21,pion22, pion1,pion11} by the quark
chemical potential which is positive in BCS and negative in BEC, the
size of the Cooper pair which is large in BCS and small in BEC, and
the scaled pair condensate which is small in BCS and large in BEC.
However, unlike the fermion-fermion scattering in cold atom systems,
quarks are unobservable degrees of freedom. The quark-quark
scattering can not be measured or used to experimentally identify
the BCS-BEC crossover, and its function to characterize the BCS-BEC
crossover at quark level is replaced by the difermion molecules
scattering~\cite{maozh,detmold}. In the BCS quark
superconductor/superfluid, the large and overlapped pairs lead to a
large pair-pair cross section, and the small and individual pairs in
the BEC superconductor/superfluid interact weakly with small cross
section.

A good knowledge of Quantum Chromodynamics (QCD) at finite temperature and density~\cite{kapusta} is significant for us to understand a wide range of physical phenomena in thermal and dense nuclear matter and quark matter. To understand the evolution of the early universe in the first few seconds after the big bang, we need the nature of the QCD phase transitions at temperature $T\sim170$ MeV and nearly vanishing density. On the other hand, to understand the physics of compact stars we need the knowledge of the equation of state and dynamics of QCD matter at high density and low temperature.

Color superconductivity in dense quark matter is due to the
attractive interaction in certain diquark
channels~\cite{CSCearly,CSCbegin,CSCbegin1,CSCreview,CSCreview1,CSCreview2,CSCreview3,CSCreview4,CSCreview5,CSCreview6}.
Taking into account only the screened (color) electric interaction
which is weakened at the Debye mass scale $g\mu$ ($g$ is the QCD
coupling constant and $\mu$ the quark chemical potential), the early
studies~\cite{CSCearly} predicted a rather small pairing gap
$\Delta\sim 1$ MeV at moderate density with
$\mu\sim\Lambda_{\text{QCD}}$ ($\sim 300$ MeV as the QCD energy
scale). The breakthrough in the field was made
in~\cite{CSCbegin,CSCbegin1} where it was observed that the pairing
gap is about 2 orders of magnitude larger than the previous
prediction, by using the instanton-induced interactions and the
phenomenological four-fermion interactions. On the other hand, it
was first pointed out by Son that, at asymptotic high density the
unscreened magnetic interaction becomes dominant~\cite{CSCgap}. This
leads to a non-BCS gap $\Delta\sim \mu
g^{-5}\exp{\left(-c/g\right)}$~\cite{pQCD,pQCD1,pQCD2,pQCD3,pQCD4,pQCD5,pQCD6}
with $c=3\pi^2/\sqrt{2}$, which matches the large magnitude of
$\Delta$ at moderate density. The same phenomena happen in pion
superfluid at moderate isospin density~\cite{ISO}.

Such gaps at moderate density are so large that they may fall
outside of applicability range of the usual BCS-like mean field
theory. It was estimated that the size of the Cooper pairs or the
superconducting coherence length $\xi_c$ becomes comparable to the
averaged inter-quark distance $d$~\cite{pairsize} at moderate
density with $\mu\sim \Lambda_{\text{QCD}}$. This feature is quite
different from the standard BCS superconductivity in metals with
$\xi_c\gg d$. Qualitatively, we can examine the ratio of the
superconducting transition temperature $T_c$ to the Fermi energy
$E_{\text f}$, $\kappa=T_c/E_{\text f}$~\cite{TC}. There are
$\kappa\sim 10^{-5}$ for ordinary BCS superconductors and
$\kappa\sim 10^{-2}$ for high temperature superconductors. For quark
matter at moderate density, taking $E_{\text f}\simeq 400$ MeV and
$T_c\simeq 50$ MeV~\cite{TCQ}, we find that $\kappa$ is even higher,
$\kappa\sim 10^{-1}$, which is close to that for the resonant
superfluidity in strongly interacting atomic Fermi gases~\cite{TC}.
This indicates that the color superconductor and pion superfluid at
moderate density are likely in the strongly coupled region or the
BCS-BEC crossover region. It is known that the pairing fluctuation
effects play important role in the BCS-BEC crossover~\cite{BCSBEC5}.
The effects of the pairing fluctuations on the quark spectral
properties, including possible pseudogap formation in heated quark
matter (above the color superconducting transition temperature), was
elucidated by Kitazawa, Koide, Kunihiro, and
Nemoto~\cite{kitazawa,kitazawa1,kitazawa2,kitazawa3}.

Since the dense quark matter is relativistic, it is interesting to
study how the relativistic effect affects the BCS-BEC crossover.
This is recently
investigated~\cite{Abuki,RBCSBEC,RBCSBEC1,RBCSBEC2,RBCSBEC3,RBCSBEC4,RBCSBEC5,RBCSBEC6,RBCSBEC7,RBCSBEC8,RBCSBEC9,RBCSBEC10}
in the Nozieres-Schmitt-Rink (NSR) theory above the critical
temperature, the boson-fermion model, and the BCS-Leggett mean field
theory at zero temperature. It is shown that, not only the BCS
superfluidity and the nonrelativistic BEC for heavy molecules but
also the nonrelativistic and relativistic BEC for nearly massless
molecules can be smoothly connected. In the first part of this
article, we will review the studies on the BCS-BEC crossover in a
relativistic four-fermion interaction model. By using the
generalized mean field theory or the so-called pseudogap theory at
finite temperature, we are able to predict the size of the pseudogap
energy in color superconducting quark matter at moderate baryon
density.

Lattice simulation of QCD at finite temperature and vanishing
density has been successfully performed. However, at large baryon
density the lattice simulation has not yet been successfully done
due to the sign problem~\cite{Lreview,Lreview1}: the fermion
determinant is not positively definite in the presence of a nonzero
baryon chemical potential $\mu_{\text B}$. To study the nature of
QCD matter at finite density, we first study some special theories
which possess positively definite fermion determinant and can be
simulated on the lattice. One case is the so-called QCD-like
theories at finite $\mu_{\rm
B}$~\cite{QC2D,QC2D1,QC2D2,QC2D3,QC2D4,QC2D5,QC2D6, QL03, ratti,
2CNJL04} where quarks are in a real or pseudo-real representation of
the gauge group, including two-color QCD with quarks in the
fundamental representation and QCD with quarks in the adjoint
representation. While these theories are not real QCD, they can be
simulated on the
lattice~\cite{LQC2D,LQC2D1,LQC2D2,LQC2D3,LQC2D4,LBECBCS} and may
give us some information of real QCD at finite baryon density.
Notice that the lightest baryon state in two-color QCD is just the
scalar diquarks. Another interesting case is real QCD at finite
isospin chemical potential $\mu_{\text I}$~\cite{ISO}, where the
chemical potentials for light $u$ and $d$ quarks have opposite signs
and hence the fermion determinant is positively definite. For both
two-color QCD at finite $\mu_{\rm B}$ and QCD at finite $\mu_{\rm
I}$, chiral perturbation theory and other effective models predict a
continuous quantum phase transition from the vacuum to the matter
phase when $\mu_{\rm B}$ or $\mu_{\text I}$ is equal to the pion
mass $m_\pi$~\cite{QC2D,QC2D1,QC2D2,QC2D3,QC2D4,QC2D5,QC2D6, QL03,
ratti,
2CNJL04,ISO,ISOother01,ISOother011,ISOother012,ISOother013,ISOother014,ISOother015,ISOother016,ISOother017,ISOother018,ISOother019,ISOother0110,ISOother0111,ISOother0112,ISOother0113,ISOother0114,ISOother0115,boser,ISOother02,ISOother021},
in contrast to real QCD at finite $\mu_{\rm B}$ where the phase
transition takes place when $\mu_{\text B}$ is approximately equal
to the nucleon mass $m_{\rm N}$. This transition has also been
verified by lattice
simulations~\cite{LQC2D,LQC2D1,LQC2D2,LQC2D3,LQC2D4,LBECBCS,Liso,Liso1,Liso2,Liso3}.
The resulting matter near the quantum phase transition is expected
to be a dilute Bose condensate with weakly repulsive
interactions~\cite{Bose01}.

The Bose-Einstein condensation (BEC) phenomenon is believed to
widely exist in dense and strongly interacting matter. For instance,
pions or Kaons can condense in neutron stars if the electron
chemical potential exceeds the effective mass of pions or
Kaons~\cite{PiC,PiC1,PiC2,PiC3,PiC4}. However, the condensation of
pions and Kaons in neutron stars is rather complicated due to the
meson-nucleon interactions in dense nuclear medium. On the other
hand, at asymptotically high density, perturbative QCD calculations
show that the ground state of dense QCD is a weakly coupled BCS
superfluid with condensation of overlapping Cooper
pairs~\cite{pQCD,pQCD1,pQCD2,pQCD3,pQCD4,pQCD5,pQCD6,ISO}. For
two-color QCD at finite baryon density or QCD at finite isospin
density, the BCS superfluid state at high density and the Bose
condensate of diquarks or pions have the same symmetry breaking
pattern and thus are smoothly connected with one another~\cite{ISO}.
The BCS and BEC state are both characterized by the nonzero
expectation value $\langle qq\rangle\neq0$ for two-color QCD at
finite baryon density or $\langle\bar{u}i\gamma_5d\rangle\neq0$ for
QCD at finite isospin density. This phenomenon is just the BCS-BEC
crossover discussed by Eagles~\cite{Eagles} and
Leggett~\cite{Leggett} in condensed matter physics.

While lattice simulations of two-color QCD at finite baryon chemical
potential~\cite{LQC2D,LQC2D1,LQC2D2,LQC2D3,LQC2D4,LBECBCS} and QCD
at finite isospin chemical potential~\cite{Liso,Liso1,Liso2,Liso3}
can be performed, it is still interesting to employ some effective
models to describe the crossover from the Bose condensate at low
density to the BCS superfluidity at high density. The chiral
perturbation theories as well as the linear sigma
models~\cite{QC2D,QC2D1,QC2D2,QC2D3,QC2D4,QC2D5,QC2D6, QL03, ratti,
2CNJL04,ISO,ISOother01,ISOother011,ISOother012,ISOother013,ISOother014,ISOother015,ISOother016,ISOother017,ISOother018,ISOother019,ISOother0110,ISOother0111,ISOother0112,ISOother0113,ISOother0114,ISOother0115,boser,ISOother02,ISOother021},
which describe only the physics of Bose condensate, does not meet
our purpose. The Nambu--Jona-Lasinio (NJL) model~\cite{NJL} with
quarks as elementary blocks, which describes well the chiral
symmetry breaking and low energy phenomenology of the QCD vacuum, is
generally believed to work at low and moderate temperatures and
densities~\cite{NJLreview,NJLreview1,NJLreview2,NJLreview3}. In the
second part of this article, we present our study of the BEC-BCS
crossover in two-color QCD at finite baryon density in the frame of
NJL model. Pion superfluid and the BEC-BCS crossover at finite
isospin density are discussed in this model too.

The article is organized as follows. In Section \ref{s2} we present our study on the BCS-BEC crossover in relativistic Fermi systems by using a four-fermion interaction model. Both zero temperature and finite temperature cases are studied.  In Section \ref{s3} the two-color QCD at finite baryon density and pion superfluid at finite isospin density are studied in the NJL model. We summarize in Section \ref{s4}. Throughout the article, we use $K\equiv(i\omega_n,{\bf k})$ and $Q\equiv(i\nu_m,{\bf q})$ to denote the four momenta for fermions and bosons, respectively, where $\omega_n=(2n+1)\pi T$ and
$\nu_m=2m\pi T$ ($m, n$ integer) are the Matsubara frequencies. We will use
the following notation for the frequency summation and momentum integration,
\begin{equation}
\sum_P= T\sum_l\sum_{\bf p},\ \ \ \ \sum_{\bf p}=\int\frac{d^3{\bf p}}{(2\pi)^3},\ \ \ P=K,Q,\ \ l=n,m.
\end{equation}

\section {BCS-BEC crossover with relativistic fermions}
\label{s2}
The motivation of studying BCS-BEC crossover with relativistic fermions is mostly due to the study of dense and
hot quark matter which may exist in compact stars and can be created in heavy ion collisions. However, we shall
point out that a relativistic theory is also necessary for non-relativistic systems when the attractive
interaction strength becomes super strong. To this end, let us first review the nonrelativistic theory of
BCS-BEC crossover in dilute Fermi gases with $s$-wave interaction.

The Leggett mean field theory~\cite{Leggett} is successful to describe the BCS-BEC crossover at zero temperature
in dilute Fermi gases with short-range $s$-wave interaction. The BCS-BEC crossover can be realized in a
dilute two-component Fermi gas with fixed total density $n=k_{\rm f}^3/(3\pi^2)$ ($k_{\rm f}$ is the Fermi momentum) by
tuning the $s$-wave scattering length $a_s$ from negative to positive. Theoretically, the BCS-BEC crossover
can be seen if we self-consistently solve the gap and number equations for the pairing gap $\Delta_0$ and the
fermion chemical potential $\mu_{\rm n}$~\cite{BCSBEC3},
\begin{eqnarray}
-\frac{m}{4\pi a_s}&=&\sum_{\bf k}\left({\frac{1}{2\sqrt{\xi_{\bf k}^2+\Delta_0^2}}-\frac{m}{{\bf k}^2}}\right),\nonumber\\
\frac{k_{\rm f}^3}{3\pi^2}&=&\sum_{\bf k}\left(1-\frac{\xi_{\bf k}}{\sqrt{\xi_{\bf k}^2+\Delta_0^2}}\right)
\end{eqnarray}
with $\xi_{\bf k}={\bf k}^2/(2m)-\mu_{\rm n}$ and $a_s$ being the s-wave scattering length. The fermion mass $m$
plays a trivial role here, since the BCS-BEC crossover depends only on a dimensionless parameter
$\eta=1/(k_{\rm f}a_s)$. This is the so-called universality for such a nonrelativistic syetem. The BCS-BEC crossover
can be characterized by the behavior of the chemical potential $\mu_{\rm n}$: it coincides with the Fermi energy
$\epsilon_{\rm f}=k_{\rm f}^2/(2m)$ in the BCS limit $\eta\rightarrow -\infty$, but becomes negative in the BEC region.
In the BEC limit $\eta\rightarrow+\infty$, we have $\mu_{\rm n}\rightarrow-E_b/2$ with $E_b=1/(ma_s^2)$
being the molecular binding energy. Therefore, in the nonrelativistic theory the chemical potential will tend to
be negatively infinity in the strong coupling BEC limit.

Then a problem arises if we look into the physics of the BEC limit from a relativistic point of view. In the
relativistic description, the fermion dispersion (without pairing) becomes $\xi_{\bf k}^\pm=\sqrt{{\bf k}^2+m^2}\pm\mu$,
where $\mp$ correspond to fermion and anti-fermion degrees of freedom, and $\mu$ is the chemical potential in
relativistic theory~\cite{kapusta}. In the non-relativistic limit $|{\bf k}|\ll m$, if $|\mu-m|\ll m$, we can
neglect the anti-fermion degrees of freedom and recover the nonrelativistic dispersion
$\xi_{\bf k}^-\simeq{\bf k}^2/(2m)-(\mu-m)$. Therefore, the quantity $\mu-m$ plays the role of the chemical
potential $\mu_{\rm n}$ in nonrelativistic theory. While $\mu_{\rm n}$ can be arbitrarily negative in the
nonrelativistic theory, $\mu$ is under some physical constraint in the relativistic theory. Since the molecule
binding energy $E_b$ can not exceed two times the constituent mass $m$, even at super strong coupling the absolute
value of the nonrelativistic chemical potential $\mu_{\rm n}=\mu-m\simeq-E_b/2$ can not exceed the fermion mass
$m$, and the relativistic chemical potential $\mu$ should be always positive. If the system is dilute, i.e., the
Fermi momentum satisfies $k_{\rm f}\ll m$, we expect that the non-relativistic theory works well when the attraction is
not very strong (the binding energy $E_b\ll 2m$). However, if the attraction is strong enough to ensure $E_b\sim 2m$,
relativistic effects will appear. From $E_b\sim1/(ma_s^2)$, we can roughly estimate that the nonrelativistic
theory becomes unphysical for $a_s\sim1/m$, corresponding to super strong attraction. This can be understood
if we consider $1/m$ as the Compton wavelength $\lambda_c$ of a particle with mass $m$.

What will then happen in the strong attraction limit in an
attractive Fermi gas?  If the attraction is so strong that
$E_b\rightarrow2m$ and $\mu\rightarrow0$, the excitation spectra
$\xi_{\bf k}^-$ and $\xi_{\bf k}^+$ for fermions and anti-fermions
become nearly degenerate, and non-relativistic limit cannot be
reached even though $k_{\rm f}\ll m$. This means that the
anti-fermion pairs can be excited by strong attraction and the
condensed bosons and anti-bosons become both nearly massless.
Therefore, without any model dependent calculation, we observe an
important relativistic effect on the BCS-BEC crossover: there exists
a relativistic BEC (RBEC)
state~\cite{Abuki,RBCSBEC,RBCSBEC1,RBCSBEC2,RBCSBEC3,RBCSBEC4,RBCSBEC5,RBCSBEC6,RBCSBEC7,RBCSBEC8,RBCSBEC9,RBCSBEC10}
which is smoothly connected to the nonrelativistic BEC (NBEC) state.
The RBEC is not a specific phenomenon for relativistic Fermi
systems, it should appear in any Fermi system if the attraction can
be strong enough, even though the initial non-interacting gas
satisfies the non-relativistic condition $k_{\rm f}\ll m$.

We now study the BCS-BEC crossover in a relativistic four-fermion interaction model, which we expect to recover the
nonrelativistic theory in a proper limit. The Lagrangian density of the model is given by
\begin{equation}
{\cal L}=\bar{\psi}(i\gamma^\mu\partial_\mu-m)\psi
+{\cal L}_{\rm I}(\bar{\psi},\psi),
\end{equation}
where $\psi,\bar{\psi}$ denote the Dirac fields with mass $m$ and
${\cal L}_{\rm I}$ describes the attractive interaction among the
fermions. For the sake of simplicity, we consider only the dominant
interaction in the scalar $J^P=0^+$ channel
~\cite{RBCS,RBCS1,RBCS2}, which in the nonrelativistic limit
recovers the $s$-wave interaction in the nonrelativistic theory. The
interaction Lagrangian ${\cal L}_{\rm I}$ can be modeled by a
contact
interaction~\cite{Abuki,RBCSBEC,RBCSBEC1,RBCSBEC2,RBCSBEC3,RBCSBEC4,RBCSBEC5,RBCSBEC6,RBCSBEC7,RBCSBEC8,RBCSBEC9,RBCSBEC10}
\begin{equation}
{\cal L}_{\rm I}=\frac{g}{4}\left(\bar{\psi} i\gamma_5C\bar{\psi}^{\text
T}\right)\left(\psi^{\text T}C i\gamma_5\psi\right) ,
\label{li}
\end{equation}
where $g>0$ is the coupling constant and $C=i\gamma_0\gamma_2$ is the charge conjugation matrix. Generally, by increasing
the attractive coupling $g$, the crossover from condensation of spin-zero Cooper pairs at weak coupling to the
Bose-Einstein condensation of bound bosons at strong coupling can be realized.

We start our calculation from the partition function ${\cal Z}$ in the imaginary time formalism,
\begin{equation}
{\cal Z}=\int [d\bar{\psi}] [d\psi] e^{\int dx({\cal L}+\mu\bar{\psi}\gamma_0\psi)}
\end{equation}
with $\int dx\equiv\int_0^{1/T} d\tau\int d^3{\bf r}$, $x=(\tau, {\bf r})$
and $\mu$ being the chemical potential conjugating to the charge density operator $\psi^\dagger\psi=\bar{\psi}\gamma_0\psi$.
Similar to the method in the study of superconductivity, we introduce the Nambu-Gor'kov spinors~\cite{nam}
\begin{equation}
\Psi = \left(\begin{array}{cc} \psi\\
C\bar{\psi}^{\text{T}}\end{array}\right),\ \ \ \bar{\Psi} =
\left(\begin{array}{cc} \bar{\psi} &
\psi^{\text{T}}C\end{array}\right)
\end{equation}
and the auxiliary pair field $\Delta(x)=(g/2)\psi^{\text
T}(x)Ci\gamma_5\psi(x)$. After performing the Hubbard-Stratonovich
transformation~\cite{hs,hs1}, we rewrite the partition function as
\begin{equation}
{\cal Z}=\int [d\bar{\Psi}] [d\Psi] [d\Delta] [d\Delta^*] e^{-{\cal A}_{\rm eff}}
\end{equation}
with
\begin{equation}
{\cal A}_{\rm eff}=\int dx\frac{|\Delta(x)|^2}{g}-\frac{1}{2}\int dx \int dx^\prime
\bar{\Psi}(x){\bf G}^{-1}(x,x^\prime)\Psi(x^\prime)
\end{equation}
and the inverse fermion propagator ${\bf G}^{-1}$
\begin{eqnarray}
{\bf G}^{-1}=\left(\begin{array}{cc} i\gamma^\mu\partial_\mu-m+\mu\gamma_0&i\gamma_5\Delta(x)\\
i\gamma_5\Delta^*(x)& i\gamma^\mu\partial_\mu-m-\mu\gamma_0\end{array}\right)\delta(x-x^\prime).
\end{eqnarray}
Integrating out the fermion fields, we obtain
\begin{equation}
{\cal Z}=\int [d\Delta] [d\Delta^*] e^{-{\cal S}_{\rm eff}\left[\Delta,\Delta^*\right]}
\end{equation}
with the bosonized effective action
\begin{eqnarray}
\label{eff} {\cal S}_{\rm eff}=\int dx\frac{|\Delta(x)|^2}{g}-\frac{1}{2}\text{Tr}\ln\left[{\bf G}^{-1}(x,x^\prime)\right].
\end{eqnarray}

\subsection {Zero temperature analysis}
While in general case the pairing fluctuations are
important~\cite{BCSBEC1,BCSBEC2,BCSBEC5}, the Leggett mean field
theory is already good to describe qualitatively the BCS-BEC
crossover at zero temperature. This can be naively seen from the
fact that the dominant contribution of fluctuations to the
thermodynamic potential at low temperature is from the massless
Goldstone modes~\cite{gold,gold1} and is therefore proportional to
$T^4$~\cite{BCSBEC3}. At zero temperature it vanishes. However,
since the quantum fluctuations are not taken into account, the mean
field theory cannot predict quantitatively the universal constants
in the unitary limit and the boson-boson scattering length in the
BEC limit. Since our goal in this paper is to study the BCS-BEC
crossover with relativistic fermions on a qualitative level, we
shall take the mean field approximation for the study of the zero
temperature case.

In the mean field approximation, we consider the uniform and static saddle point $\Delta(x)=\Delta_0$ which serves as the order parameter of the
fermionic superfluidity. Due to the U$(1)$ symmetry of the Lagrangian, the phase of the order parameter can be chosen arbitrarily and we therefore
set $\Delta_0$ to be real without loss of generality. The thermodynamic potential density $\Omega=T{\cal S}_{\rm eff}(\Delta_0)/ V$ at the saddle point can be evaluated as\cite{kapusta}
\begin{eqnarray}
\Omega ={\Delta_0^2\over g}-\sum_{\bf k}\left(E_{\bf k}^-+E_{\bf k}^+-\xi_{\bf k}^--\xi_{\bf
k}^+\right),
\end{eqnarray}
where the relativistic BCS-like excitation spectra read $E_{\bf k}^\pm=\sqrt{(\xi_{\bf k}^\pm)^2+\Delta_0^2}$ and
$\xi_{\bf k}^\pm=\epsilon_{\bf k}\pm\mu$ with $\epsilon_{\bf k}=\sqrt{{\bf k}^2+m^2}$, and the superscripts - and + correspond to the contributions from fermions and anti-fermions, respectively. Minimizing $\Omega$ with respect to $\Delta_0$, we obtain the gap equation to determine the physical $\Delta_0$,
\begin{equation}
\label{gap0} \frac{1}{g}={1\over 2}\sum_{\bf k}\left(\frac{1}{E_{\bf k}^-}+\frac{1}{E_{\bf k}^+}\right).
\end{equation}
From the thermodynamic relation we also obtain the number equation for the fermion density $n=k_f^3/(3\pi^2)$~\cite{bcs},
\begin{equation}
\label{number0} n=\sum_{\bf k}\left[\left(1-\frac{\xi_{\bf k}^-}{E_{\bf k}^-}\right)
-\left(1-\frac{\xi_{\bf k}^+}{E_{\bf k}^+}\right)\right].
\end{equation}

The relativistic model with contact four-fermion interaction is non-renormalizable and a proper regularization is needed. In this study we introduce a momentum cutoff $\Lambda$ to regularize the divergent momentum integrals. The cutoff $\Lambda$ then serves as a model parameter. For a more realistic model such as fermions interact via exchange of bosons (Yukawa coupling), the cutoff does not appear in principle but the calculation becomes much more complicated. To compare with the nonrelativistic theory, we also replace the coupling constant $g$ by a ``renormalized" coupling $U$, which is given by
\begin{equation}
-\frac{1}{U}=\frac{1}{g}-\frac{1}{2}\sum_{\bf k}\left(\frac{1}{\epsilon_{\bf k}-m}+\frac{1}{\epsilon_{\bf k}+m}\right).
\end{equation}
This corresponds to subtracting the right hand side of the gap equation (\ref{gap0}) at $\Delta_0=0$ and $\mu=m$. Such a subtraction is consistent with the formula derived from the relativistic two-body scattering matrix~\cite{Abuki}. The effective $s$-wave scattering length $a_s$ can be defined as $U=4\pi a_s/m$. Actually, by defining the new coupling constant $U$, we find that $a_s$ recovers the definition of the $s$-wave scattering length in the nonrelativistic limit. While this is a natural extension of the coupling constant renormalization in nonrelativistic theory, we keep in mind that the ultraviolet divergence cannot be completely removed, and the momentum cutoff $\Lambda$ still exists in the relativistic theory. For our convenience, we define the relativistic Fermi energy $E_{\rm f}$ as $E_{\rm f}=\sqrt{k_{\rm f}^2+m^2}$, which recovers the Fermi kinetic energy
$E_{\rm f}-m=\epsilon_{\rm f}\simeq k_{\rm f}^2/(2m)$ in nonrelativistic limit.

In the nonrelativistic theory for BCS-BEC crossover in dilute Fermi gases, there are only two characteristic lengths, i.e., $k_{\rm f}^{-1}$ and $a_s$.  The BCS-BEC crossover then shows the universality: after a proper scaling, all physical quantities depend only on the dimensionless coupling
$\eta=1/(k_{\rm f} a_s)$. Especially, in the unitary limit $a_s\rightarrow\infty$, all scaled physical quantities become universal constants. However, unlike in the nonrelativistic theory where the fermion mass $m$ plays a trivial role, in the relativistic theory a new length scale, namely the Compton wavelength $\lambda_c=m^{-1}$ appears. As a consequence, the BCS-BEC crossover should depend on not only the dimensionless coupling
$\eta$, but also the relativistic parameter $\zeta=k_{\rm f}/m=k_{\rm f}\lambda_c$. Since the cutoff $\Lambda$ is needed, the result also depends on $\Lambda/m$ or $\Lambda/k_{\rm f}$. By scaling all energies by $\epsilon_f$ and momenta by $k_f$, the gap and number equations (\ref{gap0}) and (\ref{number0}) become dimensionless,
\begin{eqnarray}
\label{scale}
-\frac{\pi}{2}\eta &=& \int_0^zx^2dx\left[\left(\frac{1}{E_x^-}-\frac{1}{\epsilon_x-2\zeta^{-2}}\right)+\left(\frac{1}{E_x^+}-\frac{1}{\epsilon_x+2\zeta^{-2}}\right)\right],\nonumber\\
\frac{2}{3}&=&\int_0^zx^2dx\left[\left(1-\frac{\xi_x^-}{E_x^-}\right)-\left(1-\frac{\xi_x^+}{E_x^+}\right)\right],
\end{eqnarray}
with $E_x^\pm=\sqrt{\left(\xi_x^\pm\right)^2+\left(\Delta_0/\epsilon_{\rm f}\right)^2}$, $\xi_x^\pm=\epsilon_x\pm\mu/\epsilon_{\rm f}$,
$\epsilon_x=2\zeta^{-1}\sqrt{x^2+\zeta^{-2}}$, and $z=\zeta^{-1}\Lambda/m=\Lambda/k_{\rm f}$. It becomes now clear that the BCS-BEC crossover in such a relativistic system is characterized by three dimensionless parameters, $\eta, \zeta$, and $\Lambda/m$.

We now study in what condition we can recover the nonrelativistic theory in the limit $\zeta\ll1$. Expanding $\epsilon_x$ in powers of $\zeta$, $\epsilon_x=x^2+2\zeta^{-2}+O(\zeta^2)$, we can recover the nonrelativistic version of $\xi_x$. However, we cannot simply neglect the terms corresponding to anti-fermions, namely the second terms on the right hand sides of Eq(\ref{scale}). Such terms can be neglected only when
$|\mu-m|\ll m$ and $\Delta_0\ll m$. When the coupling is very strong, we expect that $\mu\rightarrow 0$, these conditions cannot be met and the contribution from anti-fermions becomes significant. Therefore, we find that the so-called nonrelativistic condition $\zeta\ll 1$ for free Fermi
gas is not sufficient to recover the nonrelativistic limit of the BCS-BEC crossover. Actually, another important condition $a_s\ll 1/m$ should be imposed to guarantee the molecule binding energy $E_b\ll 2m$. With the dimensionless coupling $\eta$, this condition becomes
\begin{eqnarray}
\eta=1/(k_{\rm f}a_s)=(m/k_{\rm f})(1/m a_s)\ll m/k_{\rm f} =\zeta^{-1}.
\end{eqnarray}
Therefore, the complete condition for the nonrelativistic limit of the BCS-BEC crossover can be expressed as
\begin{eqnarray}
\zeta\ll 1 \ \ \ \ {\rm and} \ \ \ \ \ \eta\ll \zeta^{-1}.
\end{eqnarray}

\begin{figure}[!htb]
\begin{center}
\includegraphics[width=7.5cm]{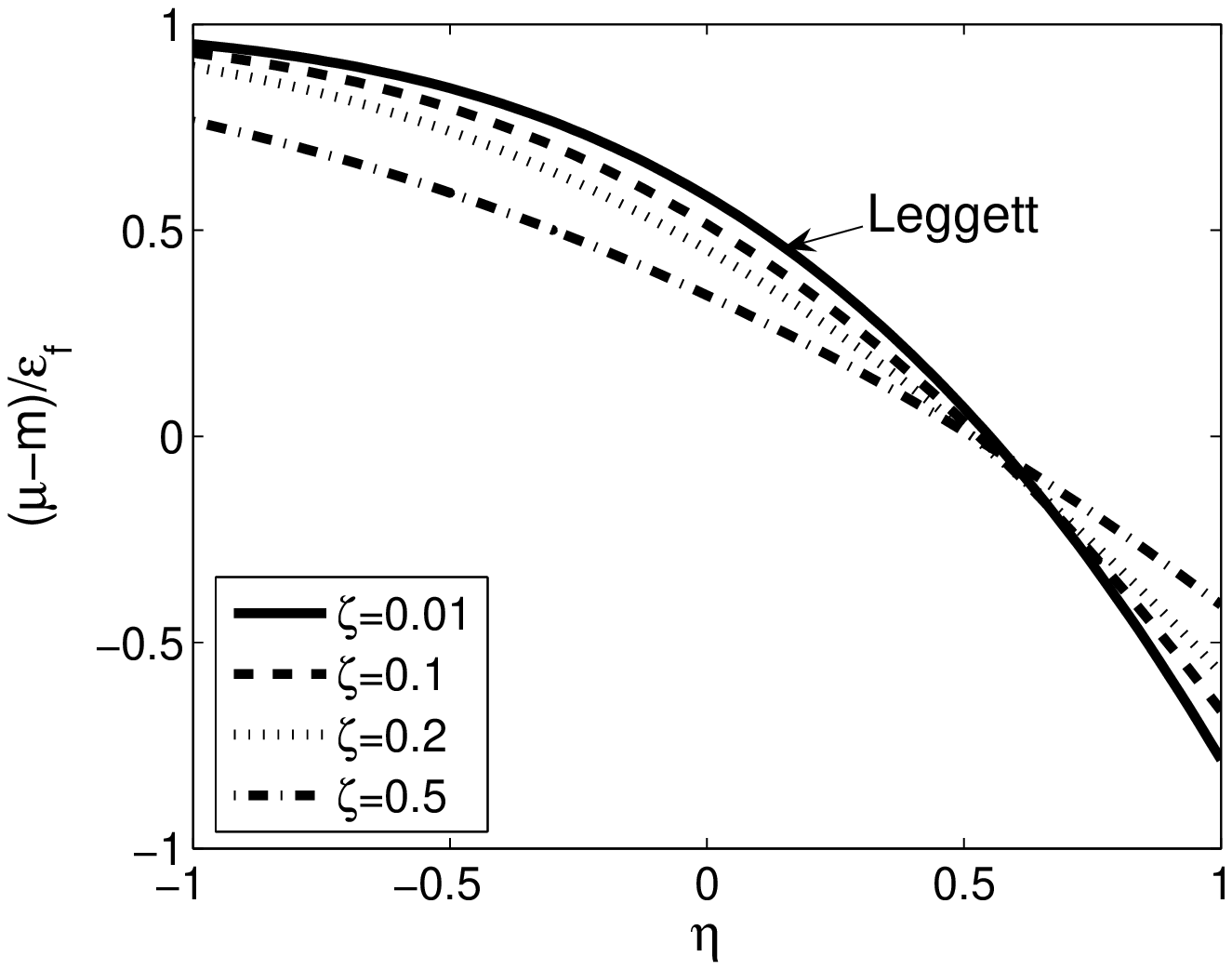}
\includegraphics[width=7.5cm]{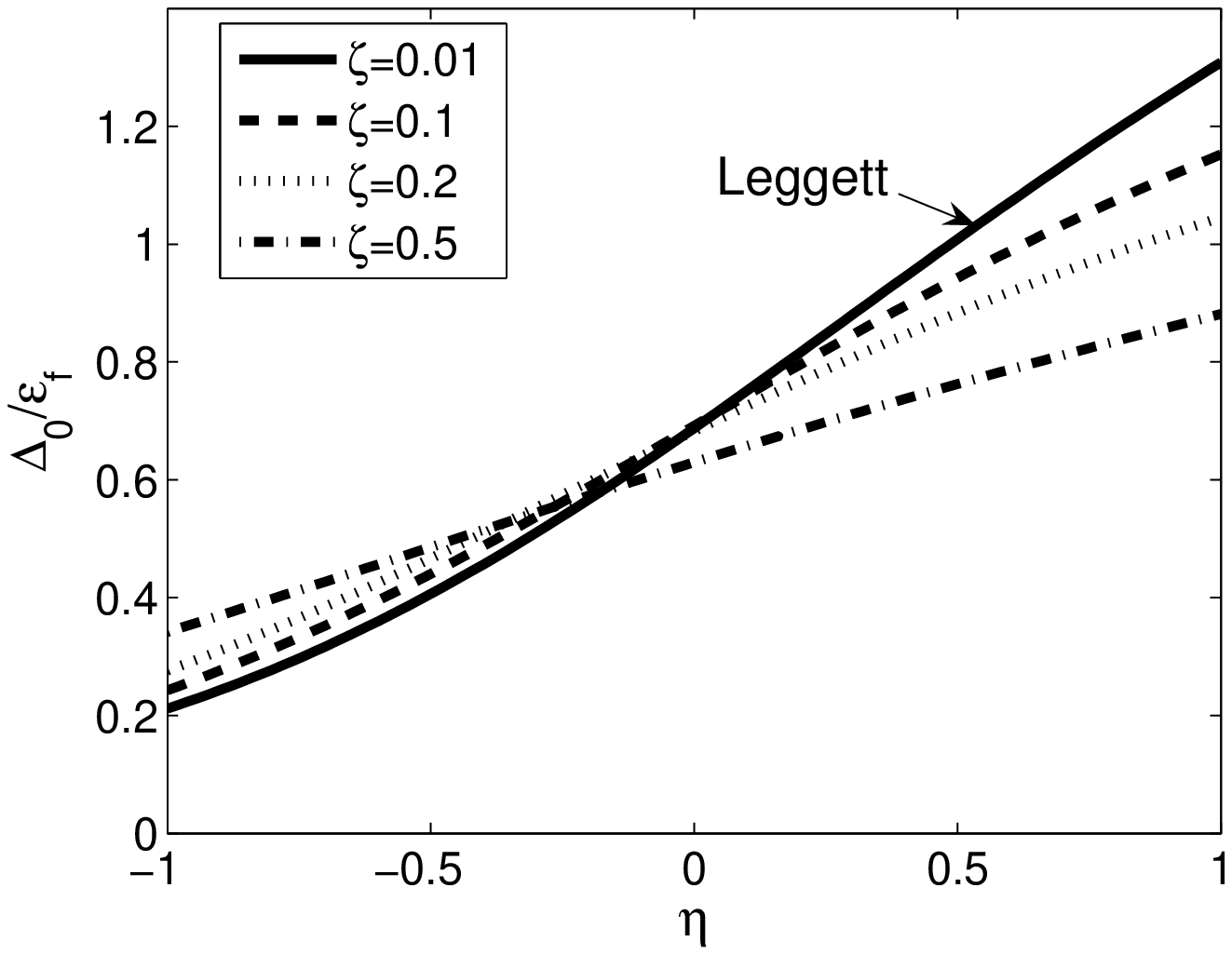}
\caption{The condensate $\Delta_0$ and nonrelativistic chemical
potential $\mu-m$, scaled by the nonrelativistic Fermi energy
$\epsilon_{\rm f}$, as functions of $\eta$ in the region $-1<\eta<1$ for
several values of $\zeta$. In the calculations we set
$\Lambda/m=10$. \label{fig1-1}}
\end{center}
\end{figure}

\begin{figure}[!htb]
\begin{center}
\includegraphics[width=7.5cm]{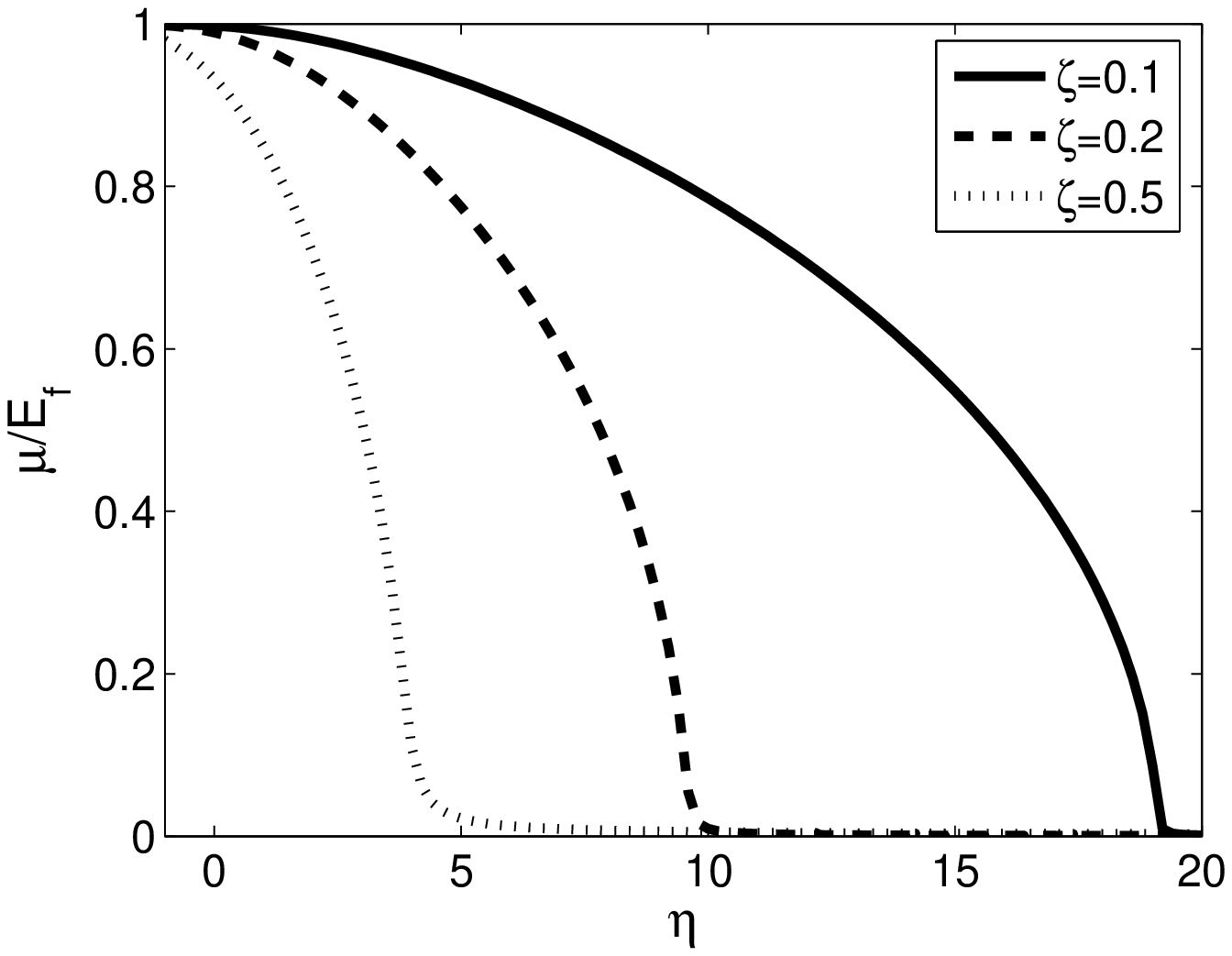}
\includegraphics[width=7.5cm]{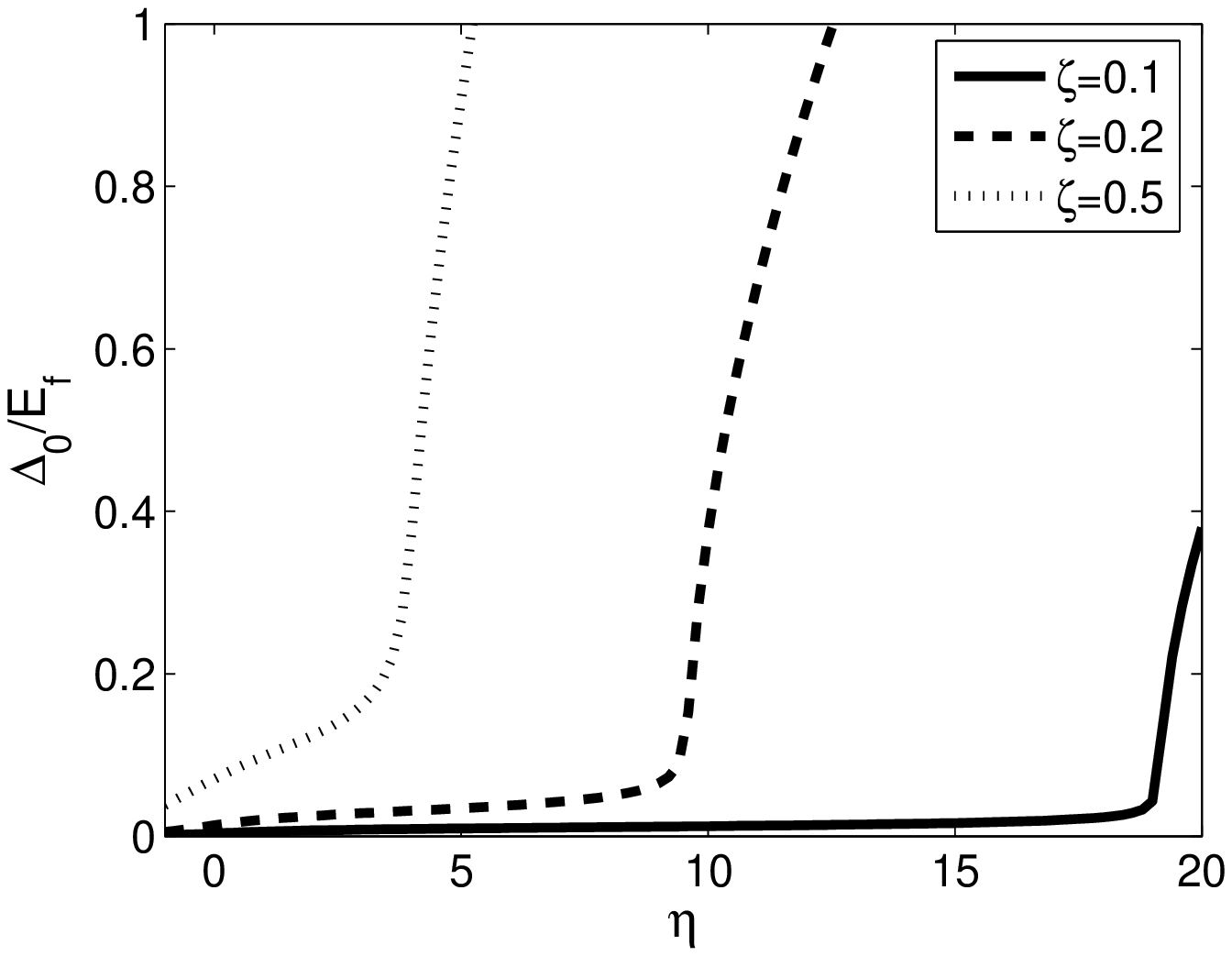}
\caption{The condensate $\Delta_0$ and chemical potential $\mu$,
scaled by the relativistic Fermi energy $E_{\rm f}$, as functions of
$\eta$ in a wide region $-1<\eta<20$ for several values of
$\zeta$. In the calculations we set $\Lambda/m=10$. \label{fig1-2}}
\end{center}
\end{figure}

To confirm the above conclusion we solve the gap and number equations numerically. In Fig.\ref{fig1-1} we show the condensate $\Delta_0$ and the nonrelativistic chemical potential $\mu-m$ as functions of $\eta$ in the region $-1<\eta<1$ for several values of $\zeta<1$. In this region the cutoff $\Lambda$ dependence is weak and can be neglected. For sufficiently small $\zeta$, we really recover the Leggett result of BCS-BEC crossover
in nonrelativistic dilute Fermi gases. With increasing $\zeta$, however, the universality is broken and the deviation becomes more and more
remarkable. On the other hand, when we increase the coupling $\eta$, especially for $\eta \sim\zeta^{-1}$, the difference between our calculation at any fixed $\zeta$ and the Leggett result becomes larger due to relativistic effects. This means that even for the case $\zeta\ll 1$ we cannot recover
the nonrelativistic result when the coupling $\eta$ becomes of order of $\zeta^{-1}$. This can be seen clearly from the $\eta$ dependence of $\Delta_0$ and $\mu$ in a wider $\eta$ region, shown in Fig.\ref{fig1-2}. We find the critical coupling $\eta_c\simeq2\zeta^{-1}$ in our numerical calculations with the cutoff $\Lambda/m=10$,  which is consistent with the above estimation. Beyond this critical coupling, the behavior of the chemical potential $\mu$ changes characteristically and approaches zero, and the condensate $\Delta_0$ becomes of order of the relativistic
Fermi energy $E_{\rm f}$ or the fermion mass $m$ rapidly. In the region $\eta\sim\eta_c$, the relativistic effect already becomes important, even
though the initial noninteracting Fermi gas satisfies the nonrelativistic condition $\zeta\ll1$. On the other hand, the smaller the parameter $\zeta$, the stronger the coupling to exhibit the relativistic effects. For the experiments of ultra cold atomic Fermi gases, it seems that it is very hard to show this effect, since the system is very dilute and we cannot reach such strong attraction.


We can derive an analytical expression for the critical coupling $\eta_c$ or $U_c$ for the RBEC state. At the critical coupling, we
can take $\mu\simeq 0$ and $\Delta_0\ll m$ approximately, and the gap equation becomes
\begin{equation}
\frac{1}{U_c}\simeq\frac{m^2}{2\pi^2}\int_0^\Lambda \frac{dk}{\sqrt{k^2+m^2}}.
\end{equation}
Completing the integral we obtain
\begin{eqnarray}\label{couplingc}
U_c^{-1}&=&\frac{2}{\pi}U_0^{-1}f(\Lambda/m),\nonumber\\
\eta_c&=&\frac{2}{\pi}\zeta^{-1}f(\Lambda/m)
\end{eqnarray}
with $f(x)=\ln(x+\sqrt{x^2+1})$ and $U_0=4\pi/m^2=4\pi\lambda_c^2$. While in the nonrelativistic region with $\eta\ll\zeta^{-1}$ the result
is almost cutoff independent, in the RBEC region $\eta\sim \zeta^{-1}$ the solution becomes sensitive to the cutoff $\Lambda$. For a realistic
model such as Yukawa coupling model, the solution at super strong coupling should be sensitive to the model parameters.

To see what happens in the region with $\eta\sim \zeta^{-1}$, we first discuss the fermion and anti-fermion momentum distributions
$n_-({\bf k})$ and $n_+({\bf k})$. From the number equation we have
\begin{equation}
n_\pm({\bf k})=\frac{1}{2}\left(1-\frac{\xi_{\bf k}^\pm}{E_{\bf k}^\pm}\right).
\end{equation}
In the nonrelativistic BCS and BEC regions with $\eta\ll\zeta^{-1}$, we have $\Delta_0\ll E_{\rm f}$, and the anti-fermion degree of freedom can be
safely neglected, i.e., $n_+({\bf k})\simeq 0$. In the weak coupling BCS limit we have $\Delta_0\ll \epsilon_{\rm f}$. Therefore $n_-({\bf k})$ deviates slightly from the standard Fermi distribution at the Fermi surface, especially we have $n_-({\bf 0})\simeq 1$. In the deep NBEC region, we have $\Delta_0\sim\eta\epsilon_{\rm f}$ and $|\mu-m|\sim\eta^2\epsilon_{\rm f}$. From $|\mu-m|\gg\Delta_0$, we find that $n_-({\bf 0})\ll 1$ and $n_-({\bf k})$ become very smooth in the momentum space. However, in the RBEC region with $\eta\sim \zeta^{-1}$, $\mu$ approaches zero and the anti-fermions become nearly degenerate with the fermions. We have
\begin{equation}
n_-({\bf k})\simeq n_+({\bf k})=\frac{1}{2}\left(1-\frac{\epsilon_{\bf k}}{\sqrt{\epsilon_{\bf k}^2+\Delta_0^2}}\right).
\end{equation}
Therefore, unlike the NBEC case, $n_\pm({\bf 0})$ here can be large as long as $\Delta_0$ is of order of $m$.

In the nonrelativistic BCS and BEC regions, the total net density $n=n_--n_+$ is approximately $n\simeq n_-=\sum_{\bf k}n_-({\bf k})$, and the contribution from the anti-fermions can be neglected, i.e., $n_+=\sum_{\bf k}n_+({\bf k})\simeq 0$. However, when we approach the RBEC region, the contributions from fermions and anti-fermions are almost equally important. Near the onset of the RBEC region with $\Delta_0<m$ we can estimate
\begin{equation}
n_-\simeq n_+ \simeq \frac{\Delta_0^2}{8\pi^2}\int_0^\Lambda dk\frac{k^2}{k^2+m^2}=\frac{\Delta_0^2\Lambda}{8\pi^2}\left(1-\frac{m}{\Lambda}\arctan{\frac{\Lambda}{m}}\right).
\end{equation}
For $m\ll\Lambda$, the second term in the bracket can be omitted, and we get $n_-\simeq n_+\simeq\Delta_0^2\Lambda/(8\pi^2)$. Therefore, in the RBEC region, the system in fact becomes very dense even though the net density $n$ is dilute: the densities of fermions and anti-fermions are both much larger than $n$ but their difference produces a small net density $n$.

In the nonrelativistic theory of BCS-BEC crossover in dilute Fermi gases, the attraction strength and number density are reflected in the theory in a
compact way through the combined dimensionless quantity $\eta=1/(k_{\rm f} a_s)$, and therefore changing the density of the system cannot induce a BCS-BEC crossover. However, if the universality is broken, there would exist an extra density dependence which may induce a BCS-BEC crossover. In nonrelativistic Fermi systems, the breaking of the universality can be induced by a finite-range interaction~\cite{density}. In the relativistic theory, the universality is naturally broken by the $\zeta=k_{\rm f}/m$ dependence which leads to the extra density effect.

\begin{figure}[!htb]
\begin{center}
\includegraphics[width=7.5cm]{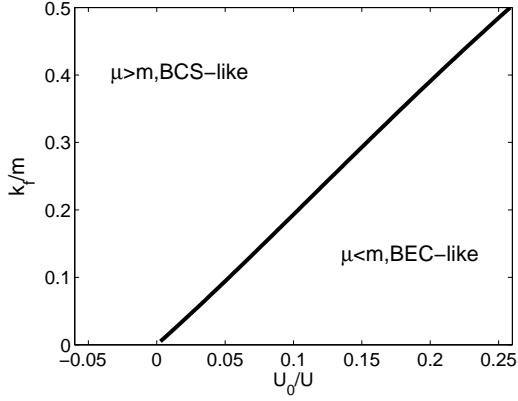}
\caption{The phase diagram in the $U_0/U-k_{\rm f}/m$ plane. The line which
separates the BCS-like region and the BEC-like region is defined as
$\mu=m$. \label{fig1-3}}
\end{center}
\end{figure}

To study this phenomenon, we calculate the ``phase diagram" in the $U_0/U-k_{\rm f}/m$ plane where $U_0/U$ reflects the pure coupling constant effect and $k_{\rm f}/m$ reflects the pure density effect. The reason why we do not present the phase diagram in the $\eta-\zeta$ plane is that both
$\eta=1/(k_{\rm f} a_s)$ and $\zeta=k_{\rm f}/m$ include the density effect. To identify the BCS-like and BEC-like phases, we take a look at the lower branch of the excitation spectra, i.e., $E_{\bf k}^-=\sqrt{(\epsilon_{\bf k}-\mu)^2+\Delta_0^2}$. This excitation spectrum is qualitatively different for $\mu>m$ and $\mu<m$. Actually, the minimum of the dispersion is located at nonzero momentum $|{\bf k}|=\sqrt{\mu^2-m^2}$ for $\mu>m$ (BCS-like) and located at zero momentum $|{\bf k}|=0$ for $\mu<m$ (BEC-like). Therefore, the BCS state and the BEC state can be separated by the line determined by the condition $\mu=m$. The phase diagram is shown in Fig.\ref{fig1-3}. The BEC state is below the line $\mu=m$ and the BCS-like state is above the line. We see clearly two ways to realize the BCS-BEC crossover, by changing the attraction strength at some fixed density and changing the density at  some fixed attraction strength. Note that we only plot the line which separates the BCS-like region and the BEC-like region. Above and close to the line $\mu=m$ there should exist a crossover region, like the phase diagram given in~\cite{density}.

The density induced BCS-BEC crossover can be realized in dense QCD
or QCD-like theories, such as QCD at finite isospin
density~\cite{ISO,ISOother01,ISOother011,ISOother012,ISOother013,ISOother014,ISOother015,ISOother016,ISOother017,ISOother018,ISOother019,ISOother0110,ISOother0111,ISOother0112,ISOother0113,ISOother0114,ISOother0115,boser,ISOother02,ISOother021}
and two-color QCD at finite baryon
density~\cite{QC2D,QC2D1,QC2D2,QC2D3,QC2D4,QC2D5,QC2D6, QL03, ratti,
2CNJL04}. The new feature in QCD and QCD-like theories is that the
effective quark mass $m$ decreases with increasing density due to
the effect of chiral symmetry restoration at finite density, which
would lower the BCS-BEC crossover line in the phase diagram.

To study the evolution of the collective modes in the BCS-BEC crossover, we investigate the fluctuations around the saddle point $\Delta(x)=\Delta_0$. To this end, we write $\Delta(x)=\Delta_0+\phi(x)$ and expand the effective action ${\cal S}_{\rm eff}$ to the
quadratic terms in $\phi$ (Gaussian fluctuations). We obtain
\begin{equation}
{\cal S}_{\rm Gauss}[\phi]={\cal S}_{\rm eff}[\Delta_0]+\frac{1}{2}\sum_Q\Phi^\dagger(Q)
{\bf M}(Q)\Phi(Q),
\end{equation}
where $\Phi$ is defined as $\Phi^\dagger(Q)=(\phi^*(Q),\phi(-Q))$. The matrix ${\bf M}(Q)$ then determines the spectra of the collective
bosonic excitations.

The inverse propagator ${\bf M}$ of the collective modes is a $2\times2$ matrix. The elements are given by
\begin{eqnarray}
{\bf M}_{11}(Q)&=& \frac{1}{g}+\frac{1}{2}\sum_K\text{Tr}
\left[i\gamma_5{\cal G}_{11}(K+Q)i\gamma_5{\cal G}_{22}(K)\right],\nonumber\\
{\bf M}_{12}(Q)&=& \frac{1}{2}\sum_K\text{Tr}
\left[i\gamma_5{\cal G}_{12}(K+Q)i\gamma_5{\cal G}_{12}(K)\right],\nonumber\\
{\bf M}_{21}(Q)&=&{\bf M}_{12}(-Q),\nonumber\\
{\bf M}_{22}(Q)&=&{\bf M}_{11}(-Q),
\end{eqnarray}
where ${\cal G}_{ij} (i,j=1,2)$ are the elements of the fermion propagator ${\cal G}={\bf G}[\Delta_0]$ in the Nambu-Gor'kov space. The explicit form of the fermion propagator is given by
\begin{eqnarray}
{\cal G}_{11}&=& {i\omega_n+\xi_{\bf k}^-\over(i\omega_n)^2-(E_{\bf k}^-)^2}\Lambda_+\gamma_0
+{i\omega_n-\xi_{\bf k}^+\over (i\omega_n)^2-(E_{\bf k}^+)^2}\Lambda_-\gamma_0,\nonumber\\
{\cal G}_{12}&=& {i\Delta_0\over (i\omega_n)^2-(E_{\bf k}^-)^2}\Lambda_+\gamma_5+
{i\Delta_0\over (i\omega_n)^2-(E_{\bf k}^+)^2}\Lambda_-\gamma_5,\nonumber\\
{\cal G}_{22}&=&{\cal G}_{11}(\mu\rightarrow-\mu),\nonumber\\
{\cal G}_{21}&=&{\cal G}_{12}(\mu\rightarrow-\mu),
\end{eqnarray}
where the energy projectors are defined as
\begin{equation}
\Lambda_{\pm}({\bf k}) = {1\over 2}\left[1\pm{\gamma_0\left(\mbox{\boldmath{$\gamma$}}\cdot{\bf k}+m\right)\over
\epsilon_{\bf k}}\right].
\end{equation}
At zero temperature, ${\bf M}_{11}$ and ${\bf M}_{12}$ can be evaluated as
\begin{widetext}
\begin{eqnarray}
{\bf M}_{11}(Q)&=&\frac{1}{g}+\sum_{\bf k}\Bigg[\left(\frac{(\upsilon_{\bf k}^-)^2(\upsilon_{{\bf k}+{\bf q}}^-)^2}{i\nu_m-E_{\bf k}^--E_{{\bf k}+{\bf q}}^-}
-\frac{(u_{\bf k}^-)^2(u_{{\bf k}+{\bf q}}^-)^2}{i\nu_m+E_{\bf k}^-+E_{{\bf k}+{\bf q}}^-}\right){\cal T}_+
+\left(\frac{(u_{\bf k}^+)^2(u_{{\bf k}+{\bf q}}^+)^2}{i\nu_m-E_{\bf k}^+-E_{{\bf k}+{\bf q}}^+}
-\frac{(\upsilon_{\bf k}^+)^2(\upsilon_{{\bf k}+{\bf q}}^+)^2}{i\nu_m+E_{\bf k}^++E_{{\bf k}+{\bf q}}^+}\right){\cal T}_+\nonumber\\
&&\ \ \ \ \ \ \ \ \ \ \ \  +\left(\frac{(\upsilon_{\bf k}^-)^2(u_{{\bf k}+{\bf q}}^+)^2}{i\nu_m-E_{\bf k}^--E_{{\bf k}+{\bf q}}^+}
-\frac{(u_{\bf k}^-)^2(\upsilon_{{\bf k}+{\bf q}}^+)^2}{i\nu_m+E_{\bf k}^-+E_{{\bf k}+{\bf q}}^+}\right){\cal T}_-
+\left(\frac{(u_{\bf k}^+)^2(\upsilon_{{\bf k}+{\bf q}}^-)^2}{i\nu_m-E_{\bf k}^+-E_{{\bf k} +{\bf q}}^-}
-\frac{(\upsilon_{\bf k}^+)^2(u_{{\bf k}+{\bf q}}^-)^2}{i\nu_m+E_{\bf k}^++E_{{\bf k}+{\bf q}}^-}\right){\cal T}_-\Bigg],\nonumber\\
{\bf M}_{12}(Q)&=&\sum_{\bf k}\Bigg[\left(\frac{u_{\bf k}^-\upsilon_{\bf k}^-u_{{\bf k}+{\bf q}}^-\upsilon_{{\bf k}+{\bf q}}^-}
{i\nu_m+E_{\bf k}^-+E_{{\bf k}+{\bf q}}^-}-\frac{u_{\bf k}^-\upsilon_{\bf k}^-u_{{\bf k}+{\bf q}}^-\upsilon_{{\bf k}+{\bf q}}^-}
{i\nu_m-E_{\bf k}^--E_{{\bf k}+{\bf q}}^-}\right){\cal T}_+
+\left(\frac{u_{\bf k}^+\upsilon_{\bf k}^+u_{{\bf k}+{\bf q}}^+\upsilon_{{\bf k}+{\bf q}}^+}{i\nu_m+E_{\bf k}^++E_{{\bf k}+{\bf q}}^+}
-\frac{u_{\bf k}^+\upsilon_{\bf k}^+u_{{\bf k}+{\bf q}}^+\upsilon_{{\bf k}+{\bf q}}^+}{i\nu_m-E_{\bf k}^+-E_{{\bf k}+{\bf q}}^+}\right){\cal T}_+
\nonumber\\
&&\ \ \ \ \   +\left(\frac{u_{\bf k}^-\upsilon_{\bf k}^-u_{{\bf k}+{\bf q}}^+\upsilon_{{\bf k}+{\bf q}}^+}
{i\nu_m+E_{\bf k}^-+E_{{\bf k}+{\bf q}}^+}-\frac{u_{\bf k}^-\upsilon_{\bf k}^-u_{{\bf k}+{\bf q}}^+\upsilon_{{\bf k}+{\bf q}}^+}
{i\nu_m-E_{\bf k}^--E_{{\bf k}+{\bf q}}^+}\right){\cal T}_-
+\left(\frac{u_{\bf k}^+\upsilon_{\bf k}^+u_{{\bf k}+{\bf q}}^-\upsilon_{{\bf k}+{\bf q}}^-}{i\nu_m+E_{\bf k}^++E_{{\bf k}+{\bf q}}^-}
-\frac{u_{\bf k}^+\upsilon_{\bf k}^+u_{{\bf k}+{\bf q}}^-\upsilon_{{\bf k}+{\bf q}}^-}{i\nu_m-E_{\bf k}^+-E_{{\bf k}+{\bf q}}^-}\right)
{\cal T}_-\Bigg],
\end{eqnarray}
\end{widetext}
where $(u_{\bf k}^{\pm})^2=(1+\xi_{\bf k}^\pm/E_{\bf k}^\pm)/2$ and $(\upsilon_{\bf k}^{\pm})^2=(1-\xi_{\bf k}^\pm/E_{\bf k}^\pm)/2$ are the BCS distributions and ${\cal T}_\pm=1/2\pm({\bf k}\cdot{\bf q}+\epsilon_{\bf k}^2)/(2\epsilon_{\bf k}\epsilon_{{\bf k}+{\bf q}})$. At ${\bf q}=0$, we have ${\cal T}_+=1$ and ${\cal T}_-=0$.

Taking the analytical continuation $i\nu_m\rightarrow\omega+i0^+$, the excitation spectrum $\omega=\omega({\bf q})$ of the collective mode is
obtained by solving the equation
\begin{equation}
\det{{\bf M}[{\bf q},\omega({\bf q})]}=0.
\end{equation}
To make the result more physical, we decompose the complex fluctuation field $\phi(x)$ into its amplitude mode $\lambda(x)$ and phase mode $\theta(x)$, i.e., $\phi(x)=(\lambda(x)+i\theta(x))/\sqrt{2}$. Then in terms of the phase and amplitude fields the Gaussian part of the effective action takes the form
\begin{equation}
\left(\begin{array}{cc} \lambda^*&\theta^*\end{array}\right)\left(\begin{array}{cc} {\bf M}_{11}^++{\bf M}_{12}&i{\bf M}_{11}^-\\
-i{\bf M}_{11}^- & {\bf M}_{11}^+-{\bf M}_{12}\end{array}\right)\left(\begin{array}{c} \lambda\\ \theta\end{array}\right)
\end{equation}
with ${\bf M}_{11}^\pm({\bf q},\omega)=({\bf M}_{11}({\bf q},\omega)\pm{\bf M}_{11}({\bf q},-\omega))/2$. Note that ${\bf M}_{11}^+({\bf q},\omega)$ and ${\bf M}_{11}^-({\bf q},\omega)$ are even and odd functions of $\omega$, respectively. Considering ${\bf M}_{11}^-({\bf q},0)=0$, the amplitude and phase modes decouple exactly at $\omega=0$. Furthermore, using the gap equation for $\Delta_0$ we find ${\bf M}_{11}^+(0,0)={\bf M}_{12}(0,0)$, which ensures the gapless phase mode at ${\bf q}=0$, i.e., the Goldstone mode corresponding to the spontaneous breaking of the global $U(1)$ symmetry of the Lagrangian density (\ref{li}).

We now determine the excitation spectrum of the Goldstone mode at low momentum and frequency, i.e., $\omega,|{\bf q}|\ll
\text{min}_{\bf k}\{E_{\bf k}^\pm\}$. In this region, the dispersion takes the linear form $\omega({\bf q})=c|{\bf q}|$. The behavior of the Goldstone mode velocity $c$ in the BCS-BEC crossover is most interesting since it determines the low temperature behavior of the thermodynamic quantities. To calculate the velocity $c$, we make a small ${\bf q}$ and $\omega$ expansion of the effective action, that is,
\begin{eqnarray}
{\bf M}_{11}^++{\bf M}_{12} &=& A+C|{\bf q}|^2-D\omega^2+\cdots,\nonumber\\
{\bf M}_{11}^+-{\bf M}_{12} &=& H|{\bf q}|^2-R\omega^2+\cdots,\nonumber\\
{\bf M}_{11}^- &=& B\omega+\cdots.
\end{eqnarray}
The Goldstone mode velocity $c$ can be shown to be
\begin{equation}
c=\sqrt{H\over B^2/A+R}.
\end{equation}
The corresponding eigenvector of ${\bf M}$ is $(\lambda,\theta)=(-ic|{\bf q}|B/A,1)$, which is a pure phase mode at ${\bf q}=0$ but has an admixture of the amplitude mode controlled by $B$ at finite ${\bf q}$. The explicit forms of $A,\ B,\ R$, and $H$ can be calculated as
\begin{eqnarray}
A &=& 4\Delta_0^2 R,\nonumber\\
B &=& {1\over 4}\sum_{\bf k}\left({\xi_{\bf k}^-\over E_{\bf k}^{-3}}-{\xi_{\bf k}^+\over E_{\bf k}^{+3}}\right),\nonumber\\
R &=& {1\over 8}\sum_{\bf k}\left({1\over E_{\bf k}^{-3}}+{1\over E_{\bf k}^{+3}}\right),\nonumber\\
H &=&{1\over 16}\sum_{\bf k}\Bigg[\frac{1}{E_{\bf k}^{-3}}\left(\frac{\xi_{\bf k}^-}{\epsilon_{\bf k}}+\left(\frac{\Delta_0^2}{E_{\bf k}^{-2}}
-\frac{\xi_{\bf k}^-}{3\epsilon_{\bf k}}\right)\frac{{\bf k}^2}{\epsilon_{\bf k}^2}\right)\nonumber\\
&&+\frac{1}{E_{\bf k}^{+3}}\left(\frac{\xi_{\bf k}^+}{\epsilon_{\bf k}}
+\left(\frac{\Delta_0^2}{E_{\bf k}^{+2}}-\frac{\xi_{\bf k}^+}{3\epsilon_{\bf k}}\right)
\frac{{\bf k}^2}{\epsilon_{\bf k}^2}\right)\nonumber\\
&&+2\left(\frac{1}{E_{\bf k}^-}+\frac{1}{E_{\bf k}^+}-2\frac{E_{\bf k}^-E_{\bf k}^+-\xi_{\bf k}^-\xi_{\bf k}^++\Delta_0^2}
{E_{\bf k}^-E_{\bf k}^+(E_{\bf k}^-+E_{\bf k}^+)}\right)\nonumber\\
&&\times\frac{1}{\epsilon_{\bf k}^2}\left(1-\frac{{\bf k}^2}{3\epsilon_{\bf k}^2}\right)\Bigg].
\end{eqnarray}

In the nonrelativistic limit with $\zeta\ll 1$ and $\eta\ll\zeta^{-1}$, we have $|\mu-m|,\Delta_0\ll m$, all the terms that include anti-fermion energy can be neglected. The fermion dispersion relation $\xi_{\bf k}^-$ and $E_{\bf k}^-$ can be well approximated as $\xi_{\bf k}={\bf k}^2/(2m)-(\mu-m)$ and $E_{\bf k}=\sqrt{\xi_{\bf k}^2+\Delta_0^2}$, and we can take ${\cal T}_+\simeq 1$. In this limit the functions ${\bf M}_{11}$ and ${\bf M}_{12}$ are the same as those obtained in nonrelativistic theory~\cite{BCSBEC3}. In this case, we have
\begin{eqnarray}
B &=& {1\over 2}\sum_{\bf k}{\xi_{\bf k}\over E_{\bf k}^3},\nonumber\\
R &=& {1\over 8}\sum_{\bf k}{1\over E_{\bf k}^3},\nonumber\\
H &=& {1\over 16}\sum_{\bf k}{1\over E_{\bf k}^3}
\left({\xi_{\bf k}\over m}+{\Delta_0^2\over E_{\bf k}^2}{{\bf k}^2\over m^2}\right).
\end{eqnarray}
In the weak coupling BCS limit, all the integrated functions peak near the Fermi surface, we have $B=0$ and  $c=\sqrt{H/R}$. Working out the integrals we recover the well known result $c=\zeta/\sqrt{3}$ for nonrelativistic fermionic superfluidity in the weak coupling limit. In the NBEC region, the Fermi surface does not exist and $B$ becomes nonzero. An explicit calculation shows $c=\zeta/\sqrt{3\pi\eta}\ll\zeta$~\cite{BCSBEC3}. This result can be rewritten as $c^2=4\pi n_Ba_B/m_B^2$, where $m_B=2m,\ a_B=2a_s$, and $n_B=n/2$ are the corresponding mass, scattering length and density of bosons. This recovers the result for weakly interacting Bose condensate~\cite{Bose01}.

In the RBEC region with $\eta\sim\zeta^{-1}$, the chemical potential $\mu\rightarrow0$. Therefore, the terms include anti-fermion energy become nearly degenerate with the fermion terms and cannot be neglected. Notice that $B$ is an odd function of $\mu$, it vanishes for $\mu\rightarrow0$.
Taking $\mu=0$ we obtain
\begin{eqnarray}
R &=& {1\over 4}\sum_{\bf k}{1\over E_{\bf k}^3},\nonumber\\
H &=& {1\over 8}\sum_{\bf k}{1\over E_{\bf k}^3}\left(3-{{\bf k}^2\over\epsilon_{\bf k}^2}
+{\Delta_0^2\over E_{\bf k}^2}{{\bf k}^2\over\epsilon_{\bf k}^2}\right),
\end{eqnarray}
where $E_{\bf k}=\sqrt{\epsilon_{\bf k}^2+\Delta_0^2}$ is now the degenerate dispersion relation in the limit $\mu\rightarrow 0$. In the RBEC region the Goldstone mode velocity can be well approximated as  $c=\lim_{\Lambda\rightarrow\infty} \sqrt{H/R}$. Therefore, we find $c\rightarrow1$ in
this region. This is consistent with the Goldstone boson velocity for relativistic Bose-Einstein condensation~\cite{boser}.

On the other hand, in the ultra relativistic BCS state with $k_{\rm f}\gg m$ and $\Delta_0\ll\mu\simeq E_{\rm f}$, all the terms that include anti-fermion energy can be neglected again. This case corresponds to color superconductivity in high density quark matter. In this case we have $B\simeq0$ and
\begin{eqnarray}
R &\simeq& {\mu^2\over 16\pi^2}\int_0^\infty dk {1\over \left[(k-\mu)^2+\Delta_0^2\right]^{3/2}},\nonumber\\
H &\simeq& {\mu^2\over 32\pi^2}\int_0^\infty dk{\Delta_0^2\over \left[(k-\mu)^2+\Delta_0^2\right]^{5/2}}.
\end{eqnarray}
For weak coupling, we have $\Delta_0\ll\mu$, and a simple algebra shows $H/R=3$. Therefore we recover the well-known result $c=1/\sqrt{3}$ for
BCS superfluidity in ultra relativistic Fermi gases.

The mixing of amplitude and phase modes undergoes characteristic changes in the BCS-NBEC-RBEC crossover. In the weak coupling BCS region, all the
integrands peak near the Fermi surface and we have $B=0$ due to the particle-hole symmetry. In this region the amplitude and phase modes decouple exactly. In the NBEC region where $\eta\ll\zeta^{-1}$, while the anti-fermion term in $B$ can be neglected, we have $B\neq 0$ since the particle-hole symmetry is lost, which induces strong phase-amplitude mixing. In the RBEC region, while both particle-hole and anti-particle--anti-hole symmetries are lost, they cancel each other and we have again $B=0$. This can be seen from the fact that for $\mu\rightarrow 0$ the first and second terms in $B$ cancel each other. Thus in the RBEC region, the amplitude and phase modes decouple again. The above observation for the phase-amplitude mixing in NBEC and RBEC regions can also be explained in the frame of the bosonic field theory for Bose-Einstein condensation. In the nonrelativistic field theory, the off-diagonal elements of the inverse boson propagator are proportional to $i\omega$~\cite{nao}, which induces a strong phase-amplitude mixing. However, in the relativistic field theory, the off-diagonal elements are proportional to $i\mu\omega$~\cite{kapusta,boser}. Therefore the phase-amplitude mixing is weak for the Bose-Einstein condensation of nearly massless bosons ($\mu\rightarrow0$).

\subsection {Finite temperature analysis}
In this subsection we turn to the finite temperature case. First, we consider the BCS mean field theory at finite temperature. In the mean field approximation, we consider a uniform and static saddle point $\Delta(x)=\Delta_{\text{sc}}$. In this part we denote the superfluid order parameter
by $\Delta_{\text{sc}}$ for convenience. The thermodynamic potential in the mean field approximation can be evaluated as
\begin{eqnarray}
\Omega &=&{\Delta_\text{sc}^2\over g}-\sum_{\bf k}\Bigg\{\left(E_{\bf k}^++E_{\bf k}^--\xi_{\bf k}^+-\xi_{\bf k}^-\right)\nonumber\\
&&-{T}\left[\ln(1+e^{- E_{\bf k}^+/T})+\ln(1+e^{- E_{\bf k}^-/T})\right]\Bigg\},
\end{eqnarray}
where $E_{\bf k}^\pm$ now reads $E_{\bf k}^\pm=\sqrt{(\xi_{\bf k}^\pm)^2+\Delta_{\text{sc}}^2}$. Minimizing $\Omega$ with respect to $\Delta_{\text{sc}}$, we get the gap equation at finite temperature,
\begin{equation}
\label{gapT} \frac{1}{g}=\sum_{\bf k}
\left[\frac{1-2f(E_{\bf k}^-)}{2E_{\bf k}^-}+\frac{1-2f(E_{\bf k}^+)}{2E_{\bf k}^+}\right],
\end{equation}
where $f(x)=1/(e^{ x/T}+1)$ is the Fermi-Dirac distribution function. Meanwhile, the number equation at finite temperature can be expressed as
\begin{equation}
\label{numberT}
n=\sum_{\bf k}\left\{\left[1-\frac{\xi_{\bf k}^-}{E_{\bf k}^-}(1-2f(E_{\bf k}^-))\right]-\left[1-\frac{\xi_{\bf k}^+}{E_{\bf k}^+}(1-2f(E_{\bf k}^+))\right]\right\}.
\end{equation}
We note that the first and second terms in the square bracket on the right hand sides of equations (\ref{gapT}) and (\ref{numberT}) correspond
to fermion and anti-fermion degrees of freedom, respectively.

Generally, we expect that the order parameter $\Delta_{\rm sc}$
vanishes at some critical temperature due to thermal excitation of
fermionic quasiparticles. At weak coupling, the BCS mean field
theory is enough to predict quantitatively the critical temperature.
However, it fails to recover the correct critical temperature for
Bose-Einstein condensation at strong coupling~\cite{add1,add12}.
Therefore, to study the BCS-BEC crossover at finite temperature, the
effects of pairing fluctuations should be considered. There exist
many methods to treat pair fluctuations at finite temperature. In
the NSR theory~\cite{BCSBEC1}, which is also called $G_0 G_0$
theory, the pair fluctuations enter only the number equation, but
the fermion loops which appear in the pair propagator are
constructed by bare Green function $G_0$. As a consequence, such a
theory is, in principle, approximately valid only at $T \ge T_c$.
For the study of BCS-BEC crossover, one needs a theory which is
valid not only above the critical temperature but also in the
symmetry breaking phase. While such a strict theory has not been
reached so far, some T-matrix approaches are developed, see, for
instance \cite{BCSBEC5,add1,add12}. An often used treatment for the
pair fluctuations in these approaches is the asymmetric pair
approximation or the so-called $G_0G$
scheme~\cite{BCSBEC5,G0G,G0G1,G0G2}. The effect of the pair
fluctuations in the $G_0G$ method is treated as a fermion pseudogap
which has been widely discussed in high temperature
superconductivity. In contrast to the $G_0G_0$ scheme, the $G_0G$
scheme keeps the Ward identity~\cite{BCSBEC5}.

To generalize the mean field theory to including the effects of uncondensed pairs, we first reexpress the BCS mean field theory by using the $G_0 G$
formalism~\cite{BCSBEC5}. Such a formalism is convenient for us to go beyond the BCS and include uncondensed pairs at finite temperature.

Let us start from the fermion propagator ${\cal G}$ in the superfluid phase. The inverse fermion propagator reads
\begin{equation}
{\cal G}^{-1}(K)= \left(\begin{array}{cc} G_0^{-1}(K;\mu)&i\gamma_5\Delta_{\text{sc}}\\
i\gamma_5\Delta_{\text{sc}}&G_0^{-1}(K;-\mu)\end{array}\right)
\end{equation}
with the inverse free fermion propagator given by
\begin{equation}
\label{g0}
G_0^{-1}(K;\mu) = (i\omega_n+\mu)\gamma_0-\mbox{\boldmath{$\gamma$}}\cdot{\bf k}-m.
\end{equation}
The fermion propagator can be formally expressed as
\begin{equation}
{\cal G}(K)= \left(\begin{array}{cc} G(K;\mu)& F(K;\mu)\\
F(K;-\mu)&G(K;-\mu)\end{array}\right).
\end{equation}
The normal and anomalous Green's functions can be explicitly expressed as
\begin{eqnarray}
\label{element}
G(K;\mu)= {i\omega_n+\xi_{\bf k}^-\over(i\omega_n)^2-(E_{\bf k}^-)^2}\Lambda_+\gamma_0
+{i\omega_n-\xi_{\bf k}^+\over (i\omega_n)^2-(E_{\bf k}^+)^2}\Lambda_-\gamma_0,\nonumber\\
F(K;\mu)= {i\Delta_\text{sc}\over(i\omega_n)^2-(E_{\bf k}^-)^2}\Lambda_+\gamma_5
+{i\Delta_\text{sc}\over (i\omega_n)^2-(E_{\bf k}^+)^2}\Lambda_-\gamma_5.
\end{eqnarray}

On the other hand, the normal and anomalous Green's functions can also be expressed as
\begin{eqnarray}
G(K;\mu)&=&\left[G_0^{-1}(K;\mu)-\Sigma_{\text{sc}}(K)\right]^{-1},\nonumber\\
F(K;\mu)&=&-G(K;\mu)i\gamma_5\Delta_{\text{sc}}G_0(K;-\mu),
\end{eqnarray}
where the fermion self-energy $\Sigma_\text{sc}(K)$ is given by
\begin{equation}
\Sigma_\text{sc}(K)=i\gamma_5\Delta_\text{sc}G_0(K;-\mu)i\gamma_5\Delta_{\text{sc}}=-\Delta_\text{sc}^2G_0(-K;\mu).
\end{equation}
According to the Green's function relations, the gap equation can be expressed as
\begin{eqnarray}
\label{gap2}
\Delta_\text{sc}&=&\frac{g}{2}\sum_K\text{Tr}\left[i\gamma_5F(K;\mu)\right]\nonumber\\
&=&\frac{g}{2}\Delta_{\text{sc}}\sum_K\text{Tr}\left[G(K;\mu)G_0(-K;\mu)\right].
\end{eqnarray}
The number equation reads
\begin{equation}
n=\sum_K\text{Tr}\left[\gamma_0G(K;\mu)\right].
\end{equation}
Completing the Matsubara frequency summation, we obtain the gap equation (\ref{gapT}) and the number equation (\ref{numberT}).

There are two lessons from the above formalism. First, in the BCS mean field theory, fermion--fermion pairs and anti-fermion--anti-fermion pairs explicitly enter the theory below $T_c$ only through the condensate $\Delta_\text{sc}$. In the $G_0G$ formalism, the fermion self-energy can equivalently be expressed as
\begin{equation}
\Sigma_\text{sc}(K)=\sum_Q t_\text{sc}(Q)G_0(Q-K;\mu)
\end{equation}
associated with a condensed pair propagator given by
\begin{equation}
t_\text{sc}(Q)=-\frac{\Delta_\text{sc}^2}{T}\delta(Q).
\end{equation}

Second, the BCS mean field theory can be associated with a specific pair susceptibility $\chi(Q)$ defined by
\begin{equation}
\chi_\text{BCS}(Q)=\frac{1}{2}\sum_K\text{Tr}\left[G_0(Q-K;\mu)G(K;\mu)\right].
\end{equation}
With this susceptibility, the gap equation can also be expressed as
\begin{equation}
1-g\chi_{\text{BCS}}(Q=0)=0.
\end{equation}
This implies that the uncondensed pair propagator takes the form
\begin{equation}
t(Q)=\frac{-g}{1-g\chi_{\text{BCS}}(Q)},
\end{equation}
and $t^{-1}(Q=0)$ is proportional to the pair chemical potential $\mu_{\text{pair}}$. Therefore, the fact that in the superfluid phase the pair chemical potential vanishes leads to the BEC-like condition
\begin{equation}
t^{-1}(Q=0)=0.
\end{equation}

While the uncondensed pairs do not play any real role in the BCS mean field theory, such a specific choice of the pair susceptibility and the BEC-like condition tell us how to go beyond the BCS mean field theory and include the effects of uncondensed pairs.

\begin{figure}
\centerline{\includegraphics[]{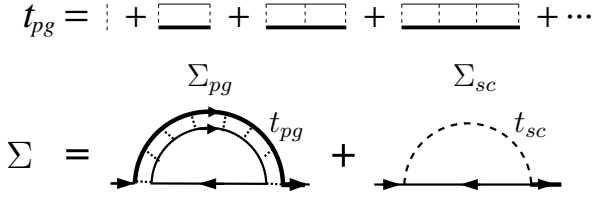}}
\caption{Diagramatic
representation of the propagator $t_\text{pg}$ for the uncondensed
pairs and the fermion self-energy~\cite{BCSBEC5}. The total fermion self-energy
contains contributions from condensed ($\Sigma_\text{sc}$) and
uncondensed ($\Sigma_\text{pg}$) pairs. The dashed, thin solid and
thick solid lines in $t_\text{pg}$ represent, respectively, the
coupling constant $g/2$, bare propagator ${ G}_0$ and full
propagator ${ G}$.
\label{fig1-4}}
\end{figure}

We now go beyond the BCS mean field approximation and include the effects of uncondensed pairs by using the $G_0 G$ formalism. It is clear that, in the BCS mean field approximation, the fermion self-energy $\Sigma_\text{sc}(K)$ includes contribution only from the condensed pairs. At finite temperature, the uncondensed pairs with nonzero momentum can be thermally excited, and the total pair propagator should contain both the condensed (sc) and uncondensed or ``pseudogap"-associated (pg) contributions. Then we write~\cite{BCSBEC5}
\begin{eqnarray}
t(Q)&=&t_\text{pg}(Q)+t_\text{sc}(Q),\nonumber\\
t_\text{pg}(Q)&=&\frac{-g}{1-g\chi(Q)},\ \ \ Q\neq 0,\nonumber\\
t_\text{sc}(Q)&=&-\frac{\Delta_\text{sc}^2}{T}\delta(Q).
\end{eqnarray}
Then the fermion self-energy reads
\begin{equation}
\Sigma(K)=\sum_Q t(Q)G_0(Q-K;\mu)=\Sigma_\text{sc}(K)+\Sigma_\text{pg}(K),
\end{equation}
where the BCS part is
\begin{equation}
\Sigma_\text{sc}(K)=\sum_Q t_\text{sc}(Q)G_0(Q-K;\mu)
\end{equation}
and the pseudogap part reads
\begin{equation}
\Sigma_\text{pg}(K)=\sum_Q t_\text{pg}(Q)G_0(Q-K;\mu).
\end{equation}
The pair susceptibility $\chi(Q)$ is still given by the $G_0G$ form,
\begin{equation}
\label{sus}
\chi(Q)=\frac{1}{2}\sum_K \text{Tr}\left[G_0(Q-K;\mu)G(K;\mu)\right],
\end{equation}
where the full fermion propagator now becomes
\begin{equation}
G(K;\mu)=\left[G_0^{-1}(K;\mu)-\Sigma(K)\right]^{-1}.
\end{equation}
The $G_0 G$ formalism used here can be diagrammatically illustrated in Fig.\ref{fig1-4}. The order parameter $\Delta_\text{sc}$ and the chemical potential $\mu$ are in principle determined by the BEC condition $t_\text{pg}^{-1}(0)=0$ and the number equation
$n=\sum_K\text{Tr}\left[\gamma_0 G(K;\mu)\right]$.

However, since the explicit form of the full propagator $G(K)$ is not known \emph{a priori}, the above equations are no long simple algebra equations and become hard to handle analytically. In the superfluid phase $T\leq T_c$, the BEC condition $t_\text{pg}^{-1}(0)=0$ implies that $t_\text{pg}(Q)$ is strongly peaked around $Q=0$. This allows us to take the approximation
\begin{equation}
\label{appro}
\Sigma(K)\simeq -\Delta^2 G_0(-K;\mu),
\end{equation}
where $\Delta^2$ contains contributions from both the condensed and uncondensed pairs. We have
\begin{equation}
\Delta^2=\Delta_\text{sc}^2+\Delta_\text{pg}^2,
\end{equation}
where the pseudogap energy $\Delta_\text{pg}$ is defined as
\begin{equation}
\label{pg1}
\Delta_\text{pg}^2 = -\sum_{Q\neq 0}t_\text{pg}(Q).
\end{equation}
We should point out that, well above the critical temperature $T_c$ such an approximation is no longer good, since the BEC
condition is not valid in normal phase and $t_\text{pg}(Q)$ becomes no longer peaked around $Q=0$.

Like in the nonrelativistic theory~\cite{BCSBEC5,G0G,G0G1,G0G2}, we
can show that $\Delta_\text{pg}^2$ physically corresponds to the
classical fluctuations of the order parameter field $\Delta(x)$,
i.e.,
\begin{equation}
\Delta_\text{pg}^2\simeq\langle|\Delta|^2\rangle-\langle|\Delta|\rangle^2.
\end{equation}
Therefore, the effects of the quantum fluctuations are not included in such a theory. In fact, at zero temperature $\Delta_{\rm pg}$ vanishes and the theory recovers exactly the BCS mean field theory. Such a theory may be called a generalized mean field theory. But as we will show below that, such a theory is already good to describe the BCS-NBEC-RBEC crossover in relativistic Fermi systems.

Under the approximation (\ref{appro}), the full Green's function $G(K;\mu)$ can be evaluated explicitly as
\begin{eqnarray}
G(K;\mu)= {i\omega_n+\xi_{\bf k}^-\over(i\omega_n)^2-(E_{\bf k}^-)^2}\Lambda_+\gamma_0
+{i\omega_n-\xi_{\bf k}^+\over (i\omega_n)^2-(E_{\bf k}^+)^2}\Lambda_-\gamma_0,
\end{eqnarray}
where the excitation spectra becomes
\begin{eqnarray}
E_{\bf k}^\pm=\sqrt{(\xi_{\bf k}^\pm)^2+\Delta^2}=\sqrt{(\xi_{\bf k}^\pm)^2+\Delta_\text{sc}^2+\Delta_\text{pg}^2}.
\end{eqnarray}
We see clearly that the excitation gap at finite temperature becomes $\Delta$ rather than the superfluid order parameter $\Delta_\text{sc}$. With the explicit form of the full Green's function $G(K;\mu)$, the pairing susceptibility $\chi(Q)$ can be evaluated as
\begin{widetext}
\begin{eqnarray}
\chi(Q)&=&\sum_{\bf k}\Bigg\{\left[\frac{1-f(E_{\bf k}^-)-f(\xi_{{\bf q}-{\bf k}}^-)}{E_{\bf k}^-+\xi_{{\bf q}-{\bf k}}^--q_0}
\frac{E_{\bf k}^-+\xi_{\bf k}^-}{2E_{\bf k}^-}
-\frac{f(E_{\bf k}^-)-f(\xi_{{\bf q}-{\bf k}}^-)} {E_{\bf k}^--\xi_{{\bf q}-{\bf k}}^-+q_0}
\frac{E_{\bf k}^--\xi_{\bf k}^-}{2E_{\bf k}^-}\right]
\left(\frac{1}{2}+\frac{\epsilon_{\bf k}^2-{\bf k}\cdot{\bf q}}{2\epsilon_{\bf k}\epsilon_{{\bf q}-{\bf k}}}\right)
\nonumber\\
&&+\left[\frac{1-f(E_{\bf k}^-)-f(\xi_{{\bf q}-{\bf k}}^+)}{E_{\bf
k}^-+\xi_{{\bf q}-{\bf k}}^++q_0} \frac{E_{\bf k}^--\xi_{\bf
k}^-}{2E_{\bf k}^-}-\frac{f(E_{\bf k}^-)-f(\xi_{{\bf q}-{\bf
k}}^+)} {E_{\bf k}^--\xi_{{\bf q}-{\bf k}}^+-q_0}\frac{E_{\bf
k}^-+\xi_{\bf k}^-}{2E_{\bf k}^-}\right]
\left(\frac{1}{2}-\frac{\epsilon_{\bf k}^2-{\bf k}\cdot{\bf q}}{2\epsilon_{\bf k}\epsilon_{{\bf q}-{\bf k}}}\right)\Bigg\}
\nonumber\\
&&+\left(E_{\bf k}^\pm,\xi_{\bf k}^\pm,q_0\rightarrow E_{\bf
k}^\mp,\xi_{\bf k}^\mp,-q_0\right)
\end{eqnarray}
\end{widetext}
with $q_0=i\nu_m$. Using the BEC condition $t_\text{pg}^{-1}(0)=0$, we obtain the gap equation
\begin{equation}
\frac{1}{g}=\sum_{\bf k}
\left[\frac{1-2f(E_{\bf k}^-)}{2E_{\bf k}^-}+\frac{1-2f(E_{\bf k}^+)}{2E_{\bf k}^+}\right].
\end{equation}
The number equation $n=\sum_K\text{Tr}\left[\gamma_0 G(K;\mu)\right]$ now becomes
\begin{equation}
n=\sum_{\bf k}\left\{\left[1-\frac{\xi_{\bf k}^-}{E_{\bf k}^-}(1-2f(E_{\bf k}^-))\right]-\left[1-\frac{\xi_{\bf k}^+}{E_{\bf k}^+}(1-2f(E_{\bf k}^+))\right]\right\}.
\end{equation}
While they take the same forms as those in the BCS mean field theory, the excitation gap is replaced by $\Delta$ which contains the contribution
from the uncondensed pairs. The order parameter $\Delta_{\rm sc}$, the pseudogap energy $\Delta_{\rm pg}$, and the chemical potential $\mu$ are
determined by solving together the gap equation, the number equation, and Eq. (\ref{pg1}).

In the nonrelativistic limit with $|\mu-m|, \Delta\ll m$, all the terms including anti-fermion dispersion relations can be safely neglected, and the fermion dispersion relations are well approximated as $\xi_{\bf k}={\bf k}^2/(2m)-(\mu-m)$ and $E_{\bf k}=\sqrt{\xi_{\bf k}^2+\Delta^2}$. Taking into account $|{\bf q}|\ll m$, we obtain
\begin{eqnarray}
\chi_{\text{NR}}(Q)&=&\sum_{\bf k}
\Bigg[\frac{1-f(E_{\bf k})-f(\xi_{{\bf q}-{\bf k}})}{E_{\bf k}+\xi_{{\bf q}-{\bf k}}-q_0}
\frac{E_{\bf k}+\xi_{\bf k}}{2E_{\bf k}}\nonumber\\
&&\ -\frac{f(E_{\bf k})-f(\xi_{{\bf q}-{\bf k}})}{E_{\bf k}-\xi_{{\bf q}-{\bf k}}+q_0}
\frac{E_{\bf k}-\xi_{\bf k}}{2E_{\bf k}}\Bigg],
\end{eqnarray}
which is just the same as the pair susceptibility obtained in the
nonrelativistic theory~\cite{BCSBEC5,G0G,G0G1,G0G2}.

However, solving Eq. (\ref{pg1}) together with the gap and number equations is still complicated. Fortunately, the BEC condition allows us to do
further approximations for the pair propagator $t_{\text{pg}}(Q)$. Using the BEC condition $1-g\chi(0)=0$ for the superfluid phase, we have
\begin{equation}
t_{\text{pg}}(Q)=\frac{1}{\chi(Q)-\chi(0)}.
\end{equation}
The pseudogap contribution is dominated by the gapless pair dispersion in low energy domain. Then we can expand the susceptibility around
$Q=0$ and obtain
\begin{equation}
\label{tpg}
t_{\text{pg}}(Q)\simeq\frac{1}{Z_1q_0+Z_2q_0^2-\xi^2{\bf q}^2},
\end{equation}
where the coefficients $Z_1,Z_2$ and $\xi^2$ are given by
\begin{eqnarray}
Z_1&=&\frac{\partial\chi(Q)}{\partial q_0}\Bigg|_{Q=0},\nonumber\\
Z_2&=&\frac{1}{2}\frac{\partial^2\chi(Q)}{\partial q_0^2}\Bigg|_{Q=0},\nonumber\\
\xi^2&=&-{1\over 2}\frac{\partial^2\chi(Q)}{\partial {\bf q}^2}\Bigg|_{Q=0}.
\end{eqnarray}
Taking the first and second order derivatives with respect to $q_0$, we obtain
\begin{eqnarray}
Z_1&=&\sum_{\bf k}\frac{1}{2E_{\bf k}^-}\left[\frac{1-f(E_{\bf k}^-)-f(\xi_{{\bf k}}^-)}{E_{\bf k}^-+\xi_{\bf k}^-}
+\frac{f(E_{\bf k}^-)-f(\xi_{\bf k}^-)}{E_{\bf k}^--\xi_{\bf k}^-}\right]\nonumber\\
&&+\left(E_{\bf k}^\pm,\xi_{\bf k}^\pm\rightarrow E_{\bf k}^\mp,\xi_{\bf k}^\mp\right),\nonumber\\
Z_2&=&\sum_{\bf k}\frac{1}{2E_{\bf k}^-}\left[\frac{1-f(E_{\bf k}^-)
-f(\xi_{{\bf k}}^-)}{(E_{\bf k}^-+\xi_{\bf k}^-)^2}-\frac{f(E_{\bf k}^-)-f(\xi_{\bf k}^-)}{(E_{\bf k}^--\xi_{\bf k}^-)^2}\right]\nonumber\\
&&+\left(E_{\bf k}^\pm,\xi_{\bf k}^\pm\rightarrow E_{\bf k}^\mp,\xi_{\bf k}^\mp\right).
\end{eqnarray}
Using the identity $(E_{\bf k}^\pm)^2-(\xi_{\bf k}^\pm)^2=\Delta^2$, they can be rewritten as
\begin{widetext}
\begin{eqnarray}
Z_1&=&\frac{1}{2\Delta^2}\left[n-2\sum_{\bf k}\left(f(\xi_{\bf k}^-)-f(\xi_{\bf k}^+)\right)\right],\nonumber\\
Z_2&=&\frac{1}{2\Delta^4}\sum_{\bf k}\left[\frac{(E_{\bf k}^-)^2+(\xi_{\bf k}^-)^2}{E_{\bf k}^-}
\left(1-2f(E_{\bf k}^-)\right)-2\xi_{\bf k}^-\left(1-2f(\xi_{\bf k}^-)\right)\right]
+\left(E_{\bf k}^\pm,\xi_{\bf k}^\pm\rightarrow E_{\bf k}^\mp,\xi_{\bf k}^\mp\right).
\label{z12}
\end{eqnarray}
Taking the second order derivative with respect to ${\bf q}$, we get
\begin{eqnarray}
\xi^2&=&\frac{1}{2}\sum_{\bf k}\Bigg\{\frac{1}{2E_{\bf k}^-}\left[\frac{1-f(E_{\bf
k}^-)-f(\xi_{{\bf k}}^-)}{E_{\bf k}^-+\xi_{\bf k}^-}+\frac{f(E_{\bf k}^-)-f(\xi_{\bf k}^-)}{E_{\bf k}^--\xi_{\bf
k}^-}\right]\frac{\epsilon_{\bf k}^2-{\bf k}^2x^2}{\epsilon_{\bf k}^3}\nonumber\\
&&-\left[\frac{1}{E_{\bf k}^-}\left(\frac{1-f(E_{\bf k}^-)-f(\xi_{{\bf k}}^-)}{(E_{\bf k}^-+\xi_{\bf k}^-)^2}
-\frac{f(E_{\bf k}^-)-f(\xi_{\bf k}^-)}{(E_{\bf k}^--\xi_{\bf k}^-)^2}\right)
+\frac{2f^\prime(\xi_{\bf k}^-)}{\Delta^2}\right]\frac{{\bf k}^2x^2}{\epsilon_{\bf k}^2}\nonumber\\
&&-\left[\frac{1-f(E_{\bf k}^-)-f(\xi_{\bf k}^+)}{E_{\bf k}^-+\xi_{\bf k}^+}
\frac{E_{\bf k}^--\xi_{\bf k}^-}{2E_{\bf k}^-}-\frac{f(E_{\bf k}^-)-f(\xi_{\bf k}^+)}
{E_{\bf k}^--\xi_{\bf k}^+}\frac{E_{\bf k}^-+\xi_{\bf k}^-}{2E_{\bf k}^-}
-\frac{1-2f(E_{\bf k}^-)}{2E_{\bf k}^-}\right]
\frac{\epsilon_{\bf k}^2-{\bf k}^2x^2}{2\epsilon_{\bf k}^4}\Bigg\}+\left(E_{\bf k}^\pm,\xi_{\bf k}^\pm\rightarrow E_{\bf k}^\mp,\xi_{\bf k}^\mp\right)
\end{eqnarray}
\end{widetext}
with $x=\cos\theta$ and $f^\prime(x)=df(x)/dx$. In the superfluid phase the equation (\ref{pg1}) becomes
\begin{equation}
\label{pg2}
\Delta_{\text{pg}}^2=\frac{1}{Z_2}\sum_{\bf q}
\frac{b(\omega_{\bf q}-\nu)+b(\omega_{\bf q}+\nu)}{2\omega_{\bf q}},
\end{equation}
where $b(x)=1/(e^{ x/T}-1)$ is the Bose-Einstein distribution function, and $\omega_{\bf q}$ and $\nu$ are defined as
\begin{equation}
\omega_{\bf q}=\sqrt{\nu^2+c^2{\bf q}^2},\ \ \
\nu=\frac{Z_1}{2Z_2},\ \ c^2=\frac{\xi^2}{Z_2}.
\end{equation}

Without numerical calculations we have the following observations from the above equations.
\\ 1) At zero temperature, the pseudogap $\Delta_{\text{pg}}$ vanishes automatically and the theory reduces to the BCS mean
field
theory~\cite{Abuki,RBCSBEC,RBCSBEC1,RBCSBEC2,RBCSBEC3,RBCSBEC4,RBCSBEC5,RBCSBEC6,RBCSBEC7,RBCSBEC8,RBCSBEC9,RBCSBEC10}.
Therefore such a theory can be called a generalized mean field
theory at finite temperature.
\\ 2) For dilute systems with $k_f\ll m$ or $n\ll m^3$, if the coupling is not strong enough, i.e., the molecule binding energy $E_b\ll 2m$, the theory reduces to its nonrelativistic version~\cite{BCSBEC6}.
\\ 3) If $Z_1q_0$ dominates the propagator $t_\text{pg}$, the pair dispersion is quadratic in $|{\bf q}|$, and therefore the pseudogap behaves as $\Delta_\text{pg}\propto T^{3/4}$ at low temperature. On the other hand, if $Z_2q_0^2$ is the dominant term, the pair dispersion is linear in
$|{\bf q}|$, and the pseudogap behaves as $\Delta_\text{pg}\propto T$ at low temperature. In the following, we will show that the first case occurs in the NBEC region and the second case occurs in the RBEC region.
\\ 4)From the explicit expression of $Z_1$ in Eq.(\ref{z12}),
we find that the quantity in the square brackets can be identified as the total number density $n_{\text{B}}$ of the bound pairs (bosons). Therefore,
we have
\begin{equation}
n_{\text{B}}=Z_1\Delta^2.
\end{equation}
From the relation $\Delta^2=\Delta_{\text{sc}}^2+\Delta_{\text{pg}}^2$, $n_{\rm B}$ can be decomposed into the condensed pair density $n_{\text{sc}}$ and the uncondensed pair density $n_{\text{pg}}$, i.e.,
\begin{equation}
n_{\text{sc}}=Z_1\Delta_{\text{sc}}^2,\ \ \ \ n_{\text{pg}}=Z_1\Delta_{\text{pg}}^2.
\end{equation}
The fraction of the condensed pairs can be defined by
\begin{equation}
P_c=\frac{n_{\text{sc}}}{n/2}=\frac{2Z_1\Delta_{\text{sc}}^2}{n}.
\end{equation}
5) In the weak coupling BCS region, the density $n$ can be well approximated as
\begin{equation}
n\simeq2\sum_{\bf k}\left(f(\xi_{\bf k}^-)-f(\xi_{\bf k}^+)\right),
\end{equation}
which leads to consistently $n_{\text{B}}=0$ in this region. In the strong coupling BEC region, however, almost all fermions form two-body bound
states which results in $n_{\text{B}}\simeq n/2$. At $T=0$, we have $\Delta_\text{pg}=0$, $n_{\text{B}}=n_{\text{sc}}$, and $P_c\simeq1$. At the critical temperature $T=T_c$, the order parameter $\Delta_{\text{sc}}$ vanishes and the uncondensed pair density $n_{\text{pg}}$ becomes dominant.

At the critical temperature $T_c$, the order parameter $\Delta_{\text{sc}}$ vanishes but the pseudogap $\Delta_{\text{pg}}$ in general does not vanish. The transition temperature $T_c$ can be determined by solving the gap and number equations together with Eq.(\ref{pg2}). In general there also exists a limit temperature $T^*$ where the pseudogap becomes small enough. The region $T_c<T<T^*$ is the so-called
pseudogap phase. Above the critical temperature $T_c$, the order parameter $\Delta_{\text{sc}}$ vanishes, and the BEC condition is no longer
valid. As a consequence, the propagator $t_{\rm pg}(Q)$ can be expressed as
\begin{equation}
t_{\text{pg}}(Q)=\frac{1}{\chi(Q)-\chi(0)-Z_0}
\end{equation}
with $Z_0=1/g-\chi(0)\neq0$. As an estimation of $\Delta_{\text{pg}}$ above $T_c$, we still perform the low energy expansion for the susceptibility,
\begin{equation}
t_{\text{pg}}(Q)\simeq\frac{1}{Z_1q_0+Z_2q_0^2-\xi^2|{\bf q}|^2-Z_0}.
\end{equation}
Therefore above $T_c$ the pseudogap equation becomes
\begin{equation}
\label{ppg}
\Delta_{\text{pg}}^2=\frac{1}{Z_2}\sum_{\bf q}
\frac{b(\omega^\prime_{\bf q}-\nu)+b(\omega^\prime_{\bf q}+\nu)}{2\omega^\prime_{\bf q}}
\end{equation}
with
\begin{equation}
\omega^\prime_{\bf q}=\sqrt{\nu^2+\lambda^2+c^2{\bf q}^2},\ \ \ \lambda^2=Z_0/Z_2.
\end{equation}
The equation (\ref{ppg}) together with the number equation determines the pseudogap $\Delta_{\text{pg}}$ and the chemical potential $\mu$ above $T_c$. Since the pair dispersion is no longer gapless, we expect that $Z_0$ increases with increasing temperature and therefore $\Delta_{\text{pg}}$ drops down and approaches zero at some dissociation temperature $T^*$.

In the end of this part, we discuss the thermodynamics of the system. The BCS mean field theory does not include the contribution from the uncondensed bosons which dominate the thermodynamics at strong coupling. In the generalized mean field theory, the total thermodynamic potential $\Omega$ contains both the fermionic and bosonic contributions,
\begin{equation}
\Omega=\Omega_{\text{cond}}+\Omega_{\text{fermion}}+\Omega_{\text{boson}},
\end{equation}
where $\Omega_{\text{cond}}=\Delta_{\text{sc}}^2/g$ is the condensation energy,  $\Omega_{\text{fermion}}$ is the fermionic contribution,
\begin{eqnarray}
\Omega_{\text{fermion}} &=&\sum_{\bf k}
\Bigg\{\left(\xi_{\bf k}^++\xi_{\bf k}^--E_{\bf k}^+-E_{\bf k}^-\right)\\
&&-T\left[\ln{(1+e^{- E_{\bf k}^+/T})}+\ln{(1+e^{- E_{\bf k}^-/T})}\right]\Bigg\},\nonumber
\end{eqnarray}
and $\Omega_{\text{boson}}$ is the contribution from uncondensed pairs,
\begin{equation}
\Omega_{\text{boson}}=\sum_Q\ln[1-g\chi(Q)].
\end{equation}
Under the approximation (\ref{tpg}) for the pair propagator, the bosonic contribution in the superfluid phase can be evaluated as
\begin{equation}
\Omega_{\text{boson}}=T\sum_{\bf q}\left[\ln{(1-e^{- \omega_{\bf q}^+/T})}+\ln{(1-e^{- \omega_{\bf q}^-/T})}\right]
\end{equation}
with $\omega_{\bf q}^\pm=\omega_{\bf q}\pm\nu$.

There exist two limiting cases for the bosonic contribution. If $Z_1q_0$ dominates the pair propagator, the pair dispersion is
quadratic in $|{\bf q}|$. In this case $\Omega_{\text{boson}}$ recovers the thermodynamic potential of a nonrelativistic boson gas,
\begin{equation}
\Omega_{\text{boson}}^{\text{NR}}=T\sum_{\bf q}\ln\left[1-e^{- {\bf q}^2/(2m_{\text B}T)}\right].
\end{equation}
On the other hand, if $Z_2q_0^2$ dominates, the pair dispersion is linear in $|{\bf q}|$. In the case we obtain the thermodynamic potential
for an ultra relativistic boson gas
\begin{equation}
\Omega_{\text{boson}}^{\text{UR}}=2T \sum_{\bf q}\ln\left(1-e^{- c|{\bf q}|/T}\right)
\end{equation}
with $c\rightarrow 1$ for the RBEC region. As we will see below, the former and latter cases correspond to the NBEC and RBEC regions, respectively. The bosons and fermions behave differently in thermodynamics. As is well known, the specific heat $C$ of an ideal boson gas at low temperature is proportional to $T^\alpha$ with $\alpha=3/2$ for nonrelativistic case and $\alpha=3$ for ultra relativistic case. However the BCS mean field theory only predicts an exponential law $C\propto e^{-\Delta_0/T}$ at low temperature.

We now apply the generalized mean field theory to study the BCS-BEC crossover with massive relativistic fermions. We assume here that the
density $n$ satisfies $n<m^3$ or $\zeta<1$. In this case the system is not ultra relativistic and can even be treated nonrelativistically in some parameter region.

From the study in the previous subsection at $T=0$, if the dimensionless coupling $\eta$ varies from $-\infty$ to $+\infty$, the system undergoes two crossovers, the crossover from the BCS state to the NBEC state around $\eta\sim 0$
and the crossover from the NBEC state to the RBEC state around $\eta\sim\zeta^{-1}$. The NBEC state and the RBEC state can be characterized by the molecule binding energy $E_b$. We have $E_b\ll2m$ in the NBEC state and $E_b\sim 2m$ in the RBEC state.
\\ 1) {\bf BCS region.}  In the weak coupling BCS region, there exist no bound pairs in the system. In this case, $Z_1$ is small enough and $Z_2$ dominates the pair dispersion~\cite{BCSBEC5}. We have $\Delta_{\text{pg}}^2\propto1/(Z_2c^3)$, where $c$ can be proven to be approximately equal to the Fermi velocity~\cite{BCSBEC5}. Since $\Delta$ is small in the weak coupling region, we can show that the pseudogap $\Delta_{\rm pg}$ is much smaller than the zero temperature gap $\Delta_0$ and therefore can be safely neglected in this region. Therefore, the BCS mean field theory is good enough at any temperature, and the critical temperature satisfies the well known relation $T_c\simeq0.57\Delta_0$. In the nonrelativistic limit with $\zeta\ll 1$, the anti-fermion degrees of freedom can be ignored and the critical temperature reads~\cite{bcs}
\begin{equation}
T_c=\frac{8e^{\gamma-2}}{\pi}\epsilon_{\rm f}e^{2\eta/\pi},
\end{equation}
where $\gamma$ is Euler's constant. Since the bosonic contribution can be neglected, the specific heat at low temperature behaves as
$C\propto e^{-\Delta_0/T}$.
\\ 2) {\bf NBEC region.}  In the NBEC region we have $\eta>1$ and $\eta\ll \zeta^{-1}$. The molecule binding energy $E_b\ll 2m$ and $|\mu-m|\ll
m$. The system is a nonrelativistical boson gas with effective boson mass $2m$, if $\zeta\ll1$. In this case, the anti-fermion degree of freedom
can be neglected, and we recover the nonrelativistic result~\cite{BCSBEC6}. In this region, the gap $\Delta$ becomes of order of the Fermi kenetic energy $\epsilon_f$. From $Z_1\propto1/\Delta^2$ and $Z_2\propto1/\Delta^4$, $Z_1q_0$ is the dominant term and the pair dispersion becomes quadratic in $|{\bf q}|$. Therefore, the propagator of the uncondensed pairs can be well approximated by
\begin{equation}
t_\text{pg}(q)\simeq\frac{Z_1^{-1}}{q_0-|{\bf q}|^2/\left(2m_\text{B}\right)},
\end{equation}
where the pair mass $m_{\text{B}}$ is given by $m_{\text{B}}=Z_1/(2\xi^2)$. Then we obtain
\begin{equation}
Z_1\Delta_{\text{pg}}^2=\sum_{\bf q}b\left(\frac{|{\bf q}|^2}{2m_{\text{B}}}\right)
=\left(\frac{m_\text{B}T}{2\pi}\right)^{3/2}\zeta\left(\frac{3}{2}\right).
\end{equation}
Since $Z_1\Delta_{\text{pg}}^2$ equals the total boson density $n_{\text{B}}$ at $T=T_c$, we obtain the critical temperature for Bose-Einstein condensation in nonrelativistic boson gas~\cite{kapusta},
\begin{equation}
T_c=\frac{2\pi}{m_{\text B}}\left(\frac{n_{\text B}}{\zeta(\frac{3}{2})}\right)^{2/3}.
\end{equation}
The boson mass $m_{\text B}$ is generally expected to be equal to the boson chemical potential $\mu_{\text B}=2\mu$. In the nonrelativistic limit $\zeta\ll1$, we find $m_{\text{B}}\simeq2m$ and $n_{\rm B}\simeq n/2$. The critical temperature becomes $T_c=0.218\epsilon_{\rm f}$.
Since $Z_1$ dominates the pair dispersion, we have $\Delta_{\rm pg}\propto T^{3/4}$ and $C\propto T^{3/2}$ at low temperature.
\\ 3) {\bf RBEC region.} In this region the molecule binding energy $E_b\rightarrow 2m$ and the chemical potential $\mu\rightarrow 0$. Nonrelativistic limit cannot be reached even for $\zeta\ll1$. Since the bosons with their mass $m_{\text{B}}=2\mu$ become nearly massless in this region, anti-bosons can be easily excited. At $T=T_c$ we have
\begin{equation}
n_{\text{B}}=n_{\text b}-n_{\bar{\text b}}=Z_1\Delta_{\text{pg}}^2,
\end{equation}
where $n_{\text b}$ and $n_{\bar{\text b}}$ are the densities for boson and anti-boson, respectively. Note that $n_{\text b}$ and
$n_{\bar{\text b}}$ are both large, and their difference produces a small density $n_{\text{B}}\simeq n/2$. On the other hand, for
$\mu\rightarrow 0$ we can expand $Z_1$ in powers of $\mu$,
\begin{equation}
Z_1\simeq R\mu+O(\mu^3)=\frac{R}{2}m_{\text B}+ O(\mu^3)
\end{equation}
and hence $Z_2$ dominates the pair dispersion. In this case, the propagator of the uncondensed pairs can be approximated as
\begin{equation}
t_{\text{pg}}(Q)\simeq\frac{Z_2^{-1}}{q_0^2-c^2|{\bf q}|^2}.
\end{equation}
Therefore we obtain
\begin{equation}
Z_2\Delta_{\text{pg}}^2\simeq\sum_{\bf q}\frac{b\left(c|{\bf q}|\right)}{c|{\bf q}|}=\frac{T^2}{12c^3}.
\end{equation}
At low temperature we have $\Delta_{\rm pg}\propto T$. Combining the above equations, we obtain the expression for $T_c$,
\begin{equation}
T_c=\left(\frac{24c^3Z_2}{R}\frac{n_{\text{B}}}{m_{\text{B}}}\right)^{1/2}.
\end{equation}
In the RBEC limit $\mu\rightarrow0$, we find that the above result
recovers the critical temperature for ultra relativistic
Bose-Einstein condensation~\cite{kapusta,haber,haber1},
\begin{equation}
\label{tc}
T_c=\left(\frac{3n_{\text B}}{m_{\text B}}\right)^{1/2}.
\end{equation}
The specific heat at low temperature behaves as to $C\propto T^3$.

\begin{figure}[!htb]
\begin{center}
\includegraphics[width=7.5cm]{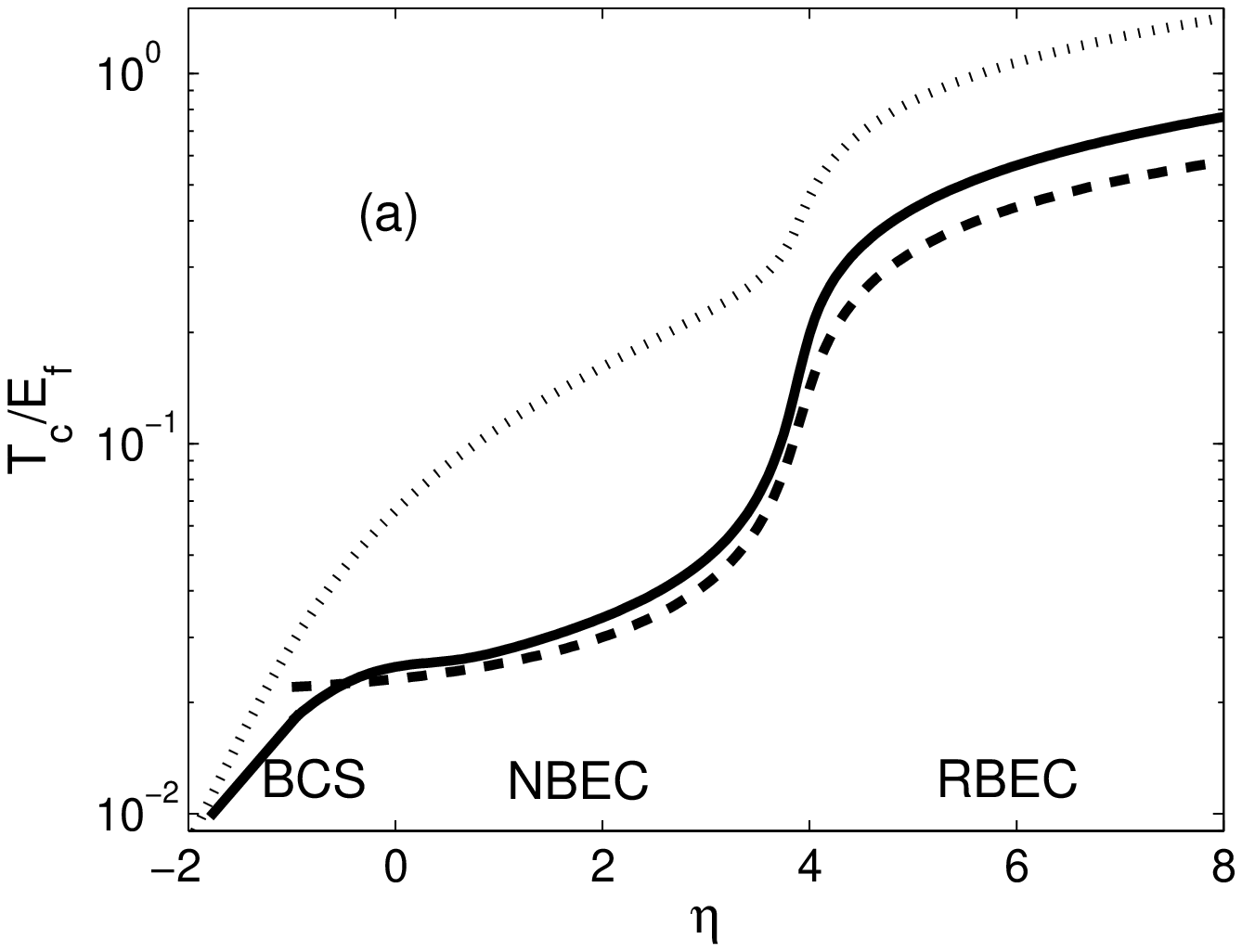}
\includegraphics[width=7.5cm]{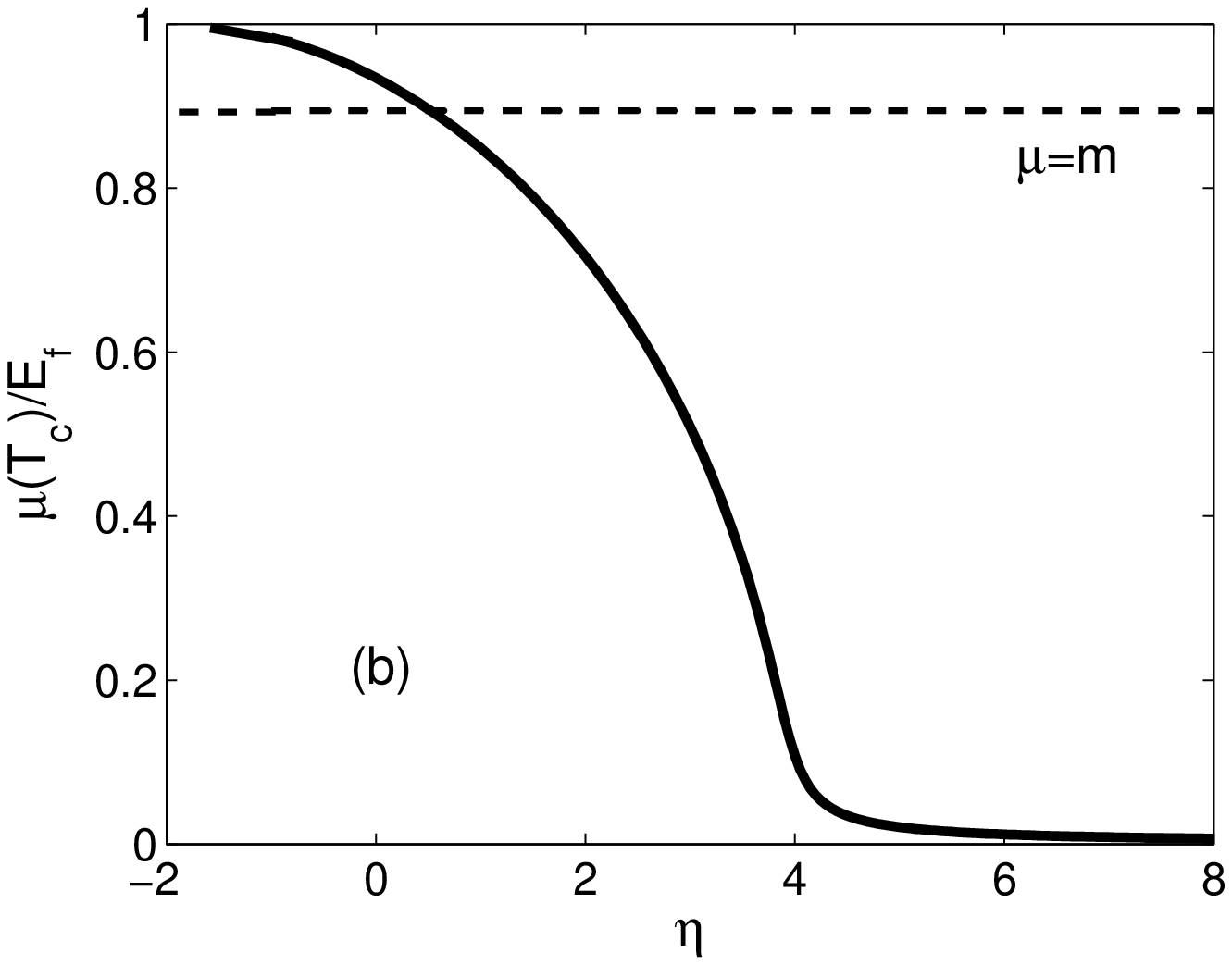}
\includegraphics[width=7.5cm]{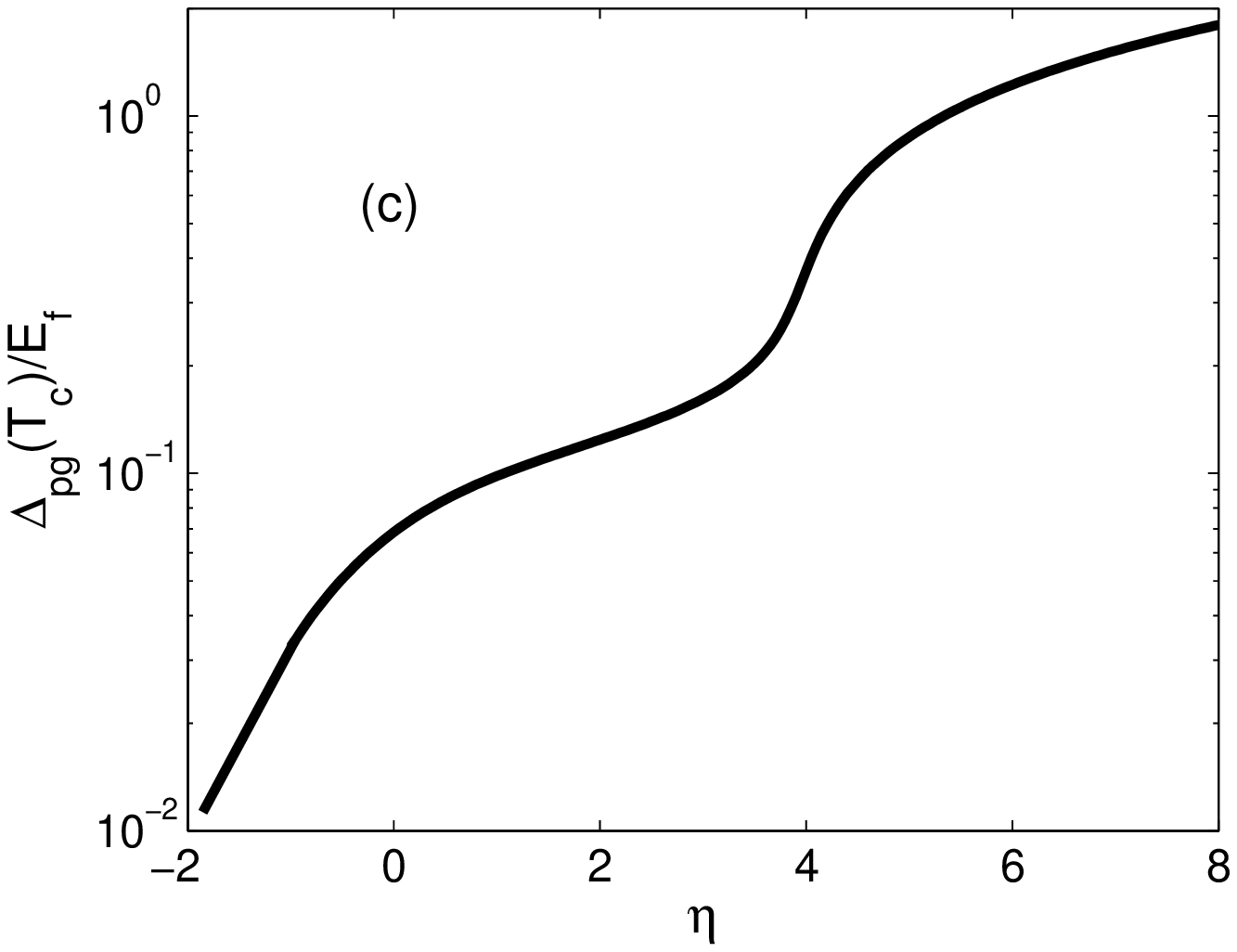}
\caption{The critical temperature $T_c$ (a), chemical potential
$\mu(T_c)$ (b) and pseudogap $\Delta_\text{pg}(T_c)$ (c) as
functions of coupling $\eta$ at $\Lambda/m=10$ and $k_{\rm f}/m=0.5$.
$T_c, \mu$ and $\Delta_\text{pg}$ are all scaled by the Fermi
energy $E_{\rm f}$. The dashed line is the standard critical temperature
for the ideal boson gas in (a) and stands for the position $\mu=m$
in (b), and the dotted line in (a) is the limit temperature $T^*$
where the pseudogap starts to disappear. \label{fig1-5}}
\end{center}
\end{figure}

Now turn to numerical results. In Fig.\ref{fig1-5} we show numerical results for the critical temperature $T_c$, the chemical potential $\mu(T_c)$,
and the pseudogap energy $\Delta_{\text{pg}}(T_c)$ as functions of the dimensionless coupling parameter $\eta$. In the calculations we have set $\Lambda/m=10$ and $k_{\rm f}/m=0.5$. The BCS-NBEC-RBEC crossover can be seen directly from the behavior of the chemical potential $\mu$. In the BCS
region $-\infty<\eta<0.5$, $\mu$ is larger than the fermion mass $m$ and approaches to the Fermi energy $E_{\rm f}$ in the weak coupling limit
$\eta\rightarrow-\infty$. The NBEC region roughly corresponds to $-0.5<\eta<4$ and the NBEC-RBEC crossover occurs at about $\eta\simeq4$.
The critical coupling $\eta\simeq4$ for the NBEC-RBEC crossover is consistent the previous analytical result (\ref{couplingc}).

\begin{figure}[!htb]
\begin{center}
\includegraphics[width=7.5cm]{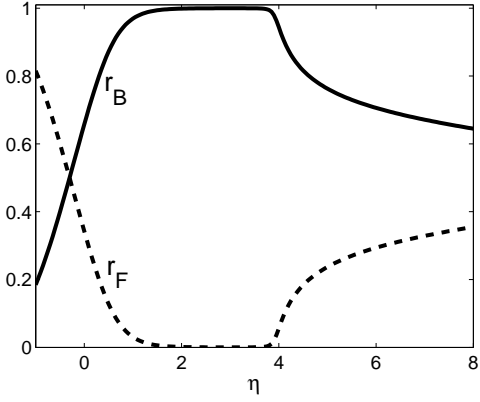}
\caption{The boson number fraction $r_{\text B}$ and the fermion
number fraction $r_{\text F}$ at the critical temperature $T_c$ as
functions of the coupling $\eta$ at $\Lambda/m=10$ and
$k_f/m=0.5$. \label{fig1-6}}
\end{center}
\end{figure}

The critical temperature $T_c$, plotted as the solid line in Fig.\ref{fig1-5}a, shows significant change from the weak to strong coupling. To compare it with the critical temperature for Bose-Einstein condensation, we solve the equation~\cite{kapusta}
\begin{equation}
\int\frac{d^3{\bf q}}{(2\pi)^3}\left[b\left(\epsilon_{\bf q}^{\text B}-\mu_{\text B}\right)
-b\left(\epsilon_{\bf q}^{\text B}+\mu_{\text B}\right)\right]\Big|_{\mu_{\text B}=m_{\text B}}=n_{\text B}
\end{equation}
with $\epsilon_{\bf q}^{\text B}=\sqrt{{\bf q}^2+m_{\text B}^2}$, boson mass $m_{\text B}=2\mu$, and boson density $n_{\text B}=n/2$. The critical temperature obtained from this equation is also shown in Fig.\ref{fig1-5}a as a dashed line. In the weak coupling region $T_c$ is very small and agrees with the BCS mean field theory. In the NBEC region $T_c$ changes smoothly and there is no remarkable difference between the solid and dashed lines. Around the coupling $\eta_c=4$, $T_c$ increases rapidly and then varies smoothly again. In the RBEC region, the critical temperature deviates significantly from the critical temperature for ideal boson gas (dashed line). Note that, $T_c$ is of the order of the Fermi kinetic energy
$\epsilon_{\rm f}\simeq k_{\rm f}^2/(2m)$ in the NBEC region but becomes as large as the Fermi energy $E_f$ in the RBEC region. The pseudogap $\Delta_\text{pg}$ at $T=T_c$, shown in Fig.\ref{fig1-5}c, behaves the same as the critical temperature. To see clearly the pseudogap phase, we also show in Fig.\ref{fig1-5}a the limit temperature $T^*$ as a dotted line. The pseudogap exists in the region between the solid and dotted lines and becomes small enough above the dotted line.

To explain why the critical temperature in the RBEC region deviates remarkably from the result for ideal boson gas, we calculate the boson number fraction $r_{\text B}=n_{\text B}/(n/2)$ and the fermion number fraction $r_{\text F}=1-r_{\text B}$ at $T=T_c$ and show them as functions of the coupling $\eta$ in Fig.\ref{fig1-6}. While $r_{\rm B}\simeq 1$ in the NBEC region, $r_{\rm B}$ is obviously less than $1$ in the RBEC region. This conclusion is consistent with the results from the NSR theory~\cite{Abuki}. In the NBEC region, the binding energy of the molecules is $E_b\simeq1/ma_s^2=2\eta^2\epsilon_{\rm f}$, which is much larger than the critical temperature $T_c\simeq0.2\epsilon_{\rm f}$. In this case the molecules can be safely regarded as point bosons at temperature near $T=T_c$. However, the critical temperature in the RBEC region becomes as large as the Fermi energy $E_{\rm f}$, which is of the order of the molecule binding energy $E_b\simeq2m$. Due to the competition between the condensation and dissociation of composite bosons in hot medium, the bosons in the RBEC region cannot be regarded as point particles and the critical temperature should deviates from the result for ideal boson gas. This may be a general feature of a composite boson system, especially for a system where the condensation temperature $T_c$ is of the order of the molecule binding energy. This phenomenon can also be explained by the competition between free energy and entropy~\cite{Abuki}: in terms of entropy a two-fermion state is more favorable than a one-boson state, but in terms of free energy it is less favorable. Since the condensation temperature $T_c$ in the RBEC region is of the order of $\sqrt{n_{\text B}/m_{\text B}}\sim\sqrt{n/\mu}$,
we conclude that only for a system with sufficiently small value of $k_{\rm f}/m$, the critical temperature for relativistic boson gas can be reached and is much smaller than $2m$.

We now apply the generalized mean field theory to study strong coupling superfluidity/superconductivity in ultra relativistic Fermi systems.
A possible ultra relativistic superfluid/superconductor is color superconducting quark matter which may exist in the core of compact stars.
The high density quark matter corresponds to the ultra relativistic case $n\gg m_0^3$, where $m_0$ is the current quark mass. For light $u$ and $d$ quarks, $m_0\simeq 5$MeV. At moderate baryon density with the quark chemical potential $\mu\sim 400$ MeV, the quark energy gap $\Delta$ due to color superconductivity is of the order of 100 MeV. Since $\Delta/\mu$ is of order $0.1$, the color superconductor is not located in the weak coupling region. As a result, the pseudogap effect is expected to be significant near the critical temperature. To study color superconductivity at $\mu\sim 400$ MeV where perturbative method does not work, we employ the generalized NJL model with four-fermion interaction in the scalar diquark channel. Since the strange quark degree of freedom has no effect for $\mu\sim400$MeV, we restrict us to the two-flavor case.
The Lagrangian density is given by
\begin{eqnarray}
{\cal L} &=&\bar{q}(i\gamma^{\mu}\partial_{\mu}-m_0)q
+G_{\text{s}}\left[\left(\bar{q}q\right)^2+\left(\bar{q}i\gamma_5\tau q\right)^2 \right]\\
&&+G_{\text{d}}\sum_{a=2,5,7}\left(\bar{q} i\gamma^5\tau_2\lambda_a C\bar{q}^{\text T}\right)
\left(q^{\text T}Ci\gamma^5\tau_2\lambda_aq\right),\nonumber
\end{eqnarray}
where $q$ and $\bar{q}$ denote the two-flavor quark fields, $\tau_i$ $(i=1,2,3)$ are the Pauli matrices in flavor space and
$\lambda_a$ $(a=1,2,...,8)$ are the Gell-Mann matrices in color space, and $G_{\text s}$ and $G_{\text d}$ are coupling constants for meson
and diquark channels.

At $\mu\sim400$MeV, the chiral symmetry gets restored and we do not need to consider the possibility of nonzero chiral condensate
$\langle\bar{q}q\rangle$ which generates an effective quark mass $M\gg m$. The order parameter field for color superconductivity is defined as
\begin{equation}
\Phi_a=-2G_{\text d}q^{\text T}C i\gamma^5\tau_2\lambda_aq.
\end{equation}
Nonzero expectation values of $\Phi_a$ spontaneously breaks the color SU(3) symmetry down to a SU(2) subgroup. Due to the color SU(3) symmetry of the Lagrangian, the effective potential depends only on the combination $\Delta_2^2+\Delta_5^2+\Delta_7^2$ with $\Delta_a=\langle\Phi_a\rangle$. Therefore we can choose a specific gauge $\Delta_\text{sc}=\Delta_2\neq 0, \Delta_5=\Delta_7=0$ without loss of generality. In this gauge, the red and green quarks participate in the pairing and condensation, leaving the blue quarks unpaired.

Since the blue quarks do not participate pairing, the generalized mean field theory cannot be directly applied to the color superconducting phase
$T<T_c$. The difficulty here is due to the complicated pairing fluctuations from $\Phi_5$ and $\Phi_7$. However, we can still apply the theory to
study the critical temperature $T_c$ and the pseudogap $\Delta_{\rm pg}$ at $T=T_c$. At and above the critical temperature, the order parameter
$\Delta_{\rm sc}$ vanished and the broken color SU$(3)$ symmetry gets restored. Therefore, all three colors becomes degenerate. At $T=T_c$, we have
$\Delta=\Delta_\text{pg}$. The pair susceptibility $\chi(Q)$ can be derived. It takes the same form as that for the toy U$(1)$ model but there exists a prefactor $N_f(N_c-1)$ ($N_f=2$ and $N_c=3$ are the numbers of flavor and color) due to the existence of flavor and color degrees of freedom.
The gap equation can be obtained from the BEC-like condition $1-4G_{\rm d}\chi(0)=0$.
\begin{figure}[!htb]
\begin{center}
\includegraphics[width=7.5cm]{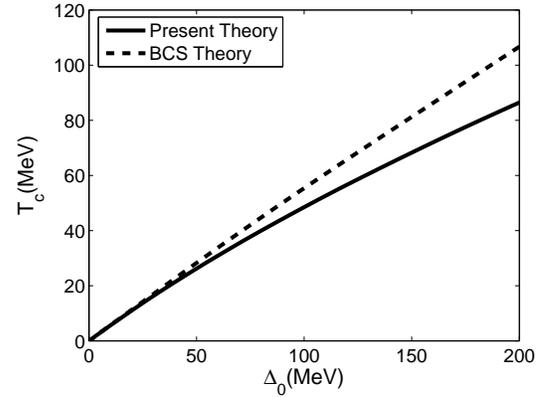}
\caption{The critical temperature $T_c$ for two-flavor
color superconductor as a function of the pairing gap
$\Delta_0$ at zero temperature in the BCS mean field theory
(dashed line) and in the generalized mean field theory
(solid line). \label{fig1-7}}
\end{center}
\end{figure}
\begin{figure}[!htb]
\begin{center}
\includegraphics[width=7.5cm]{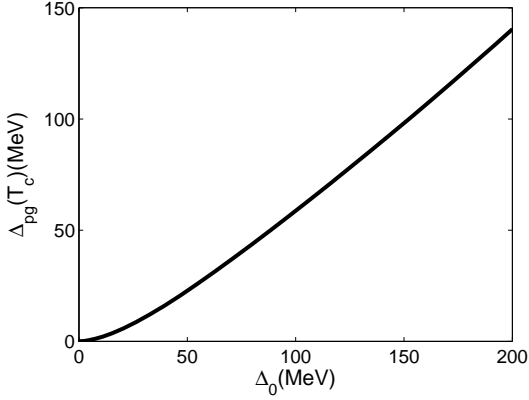}
\caption{The pseudogap $\Delta_{\text{pg}}$ in two-flavor color
superconductor at $T=T_c$ as a function of $\Delta_0$. \label{fig1-8}}
\end{center}
\end{figure}

For numerical calculations, we take the current quark mass $m_0=5$ MeV, the momentum cutoff $\Lambda=650$ MeV, and the quark chemical potential $\mu=400$ MeV. For convenience, we use the pairing gap $\Delta_0$ at zero temperature (obtained by solving the BCS gap equation at $T=0$) to characterize the diquark coupling strength $G_{\text{d}}$. It is generally believed the pairing gap $\Delta_0$ at $\mu\sim400$MeV is of order $100$MeV.

In Fig.\ref{fig1-7} we show the critical temperature $T_c$ as a
function of $\Delta_0$ in the generalized mean field theory and in
the BCS mean field theory. The critical temperature is not strongly
modified by the pairing fluctuations in a wide range of $\Delta_0$.
The difference between the two can reach about $20\%$ for the strong
coupling case $\Delta_0\simeq 200$ MeV. In Fig.\ref{fig1-8}, we show
the pseudogap $\Delta_{\text{pg}}$ at the critical temperature
$T=T_c$. In a wide range of coupling strength, the pseudogap is of
the order of the zero temperature gap $\Delta_0$. Such a behavior
means that the two-flavor color superconductivity at moderate
density ($\mu\sim400$MeV) is likely in the BCS-BEC crossover region
and is similar to the behavior of the pseudogap in
cuprates~\cite{BCSBEC5,G0G,G0G1,G0G2}.

\section {BEC-BCS crossover in two-color QCD and Pion superfluid}
\label{s3}
In Section \ref{s2} we have studied the BCS-BEC crossover with relativistic fermions with a constant mass $m$. In QCD, however, the effective quark mass $M$ generally varies with the temperature and density. For two-flavor QCD, the current masses of light $u$ and $d$ quarks are very small, about $5$MeV. Their effective masses are generated by the nonzero chiral condensate $\langle\bar{q}q\rangle$ which breaks the chiral symmetry of QCD. In the superfluid state, the lightest fermionic quasiparticle has the spectrum
\begin{equation}
E({\bf k})=\sqrt{(\sqrt{{\bf k}^2+M^2}-\mu)^2+\Delta^2}.
\end{equation}
The BEC-BCS crossover occurs when the chemical potential $\mu$ equals the effective mass $M$. However, since $M$ is dynamically generated by the chiral condensate $\langle\bar{q}q\rangle$, it varies with the chemical potential or density. Therefore, there exists interesting interplay between the chiral symmetry restoration and the BEC-BCS crossover. The decreasing of the effective mass $M$ with increasing density lowers the crossover density, and we expect that the BEC-BCS crossover occurs in the nonperturbative region ($\mu\sim \Lambda_{\rm QCD}$) where perturbative QCD does not work. In this section, we study BEC-BCS crossover in two-color QCD (number of color $N_c=2$) at finite baryon chemical potential
$\mu_{\text B}$ and in real QCD at finite isospin chemical potential $\mu_{\text I}$ by using the NJL model.

\subsection {Effective action of two-color QCD at finite ${\text T}$ and $\mu_{\text B}$ }
\label{s3-1}

For vanishing current quark mass $m_0$, two-color QCD possesses an
enlarged flavor symmetry SU$(2N_f)$ ($N_f$ is the number of
flavors), the so-called Pauli-Gursey symmetry~\cite{gur} which
connects quarks and
antiquarks~\cite{QC2D,QC2D1,QC2D2,QC2D3,QC2D4,QC2D5,QC2D6, QL03,
ratti, 2CNJL04}. For $N_f=2$, the flavor symmetry SU$(2N_f)$ is
spontaneously broken down to Sp$(2N_f)$ driven by a nonzero quark
condensate $\langle\bar{q}q\rangle$ and there arise five Goldstone
bosons: three pions and two scalar diquarks. The scalar diquarks are
the lightest baryons that carry baryon number. For nonvanishing
current quark mass, the flavor symmetry is explicitly broken,
resulting in five degenerate pseudo-Goldstone bosons with a small
mass $m_\pi$. At finite baryon chemical potential $\mu_{\text B}$,
the flavor symmetry SU$(2N_f)$ is explicitly broken down to
SU$_{\text L}(N_f)\otimes$SU$_{\text R}(N_f)\otimes$U$_{\text
B}(1)$. Further, a nonzero diquark condensate $\langle qq\rangle$
can form at large chemical potential and breaks spontaneously the
U$_{\text B}(1)$ symmetry. In two-color QCD, the scalar diquarks are
in fact the lightest ``baryons". Therefore, we expect a baryon
superfluid phase with $\langle qq\rangle\neq0$ for $|\mu_{\text
B}|>m_\pi$.

First, we construct a NJL model for two-color two-flavor QCD with the above flavor symmetry. We consider a contact current-current interaction
\begin{eqnarray}
{\cal L}_{\text{int}}=G_{\text c}\sum_{{\text a}=1}^3(\bar{q}\gamma_\mu t_{\text a}q)(\bar{q}\gamma^\mu t_{\text a}q)
\end{eqnarray}
inspired by QCD. Here $t_{\text a}$ (${\text a}=1,2,3$) are the generators of color SU$_{\text c}(2)$ and $G_{\text c}$ is a phenomenological coupling constant. After the Fierz transformation we can obtain an effective NJL Lagrangian density with scalar mesons and color singlet scalar diquarks~\cite{ratti}
\begin{eqnarray}\label{NJL}
{\cal L}_{\text{NJL}}&=&\bar{q}(i\gamma^\mu\partial_\mu-m_0)q
+G\left[(\bar{q}q)^{2}+(\bar{q}i\gamma_{5}\mbox{\boldmath{$\tau$}}q)^{2}\right]\nonumber\\
&&+G(\bar{q}i\gamma_5\tau_2t_2q_c)(\bar{q}_ci\gamma_5\tau_2t_2q),
\end{eqnarray}
where $q_c={\cal C}\bar{q}^{\text{T}}$ and $\bar{q}_c=q^{\text
T}{\cal C}$ are the charge conjugate spinors with ${\cal
C}=i\gamma_0\gamma_2$ and $\tau_{\text i}$ (${\text i}=1,2,3$) are
the Pauli matrices in the flavor space. The four-fermion coupling
constants for the scalar mesons and diquarks are the same,
$G=3G_{\text c}/4$~\cite{ratti}, which ensures the enlarged flavor
symmetry SU$(2N_f)$ of two-color QCD in the chiral limit $m_0=0$.
One can show explicitly that there are five Goldstone bosons (three
pions and two diquarks) driven by a nonzero quark condensate
$\langle\bar{q}q\rangle$.  With explicit chiral symmetry broken
$m_0\neq0$, pions and diquarks are also degenerate, and their mass
$m_\pi$ can be determined by the standard method for the NJL
model~\cite{NJLreview,NJLreview1,NJLreview2,NJLreview3}.

The partition function of the two-color NJL model (\ref{NJL}) at finite temperature $T$ and baryon chemical potential $\mu_{\text B}$ is
\begin{eqnarray}
Z_{\text{NJL}}=\int[d\bar{q}][dq]e^{\int dx\left({\cal L}_{\text{NJL}}+\frac{\mu_{\text B}}{2}\bar{q}\gamma_{0}q\right)},
\end{eqnarray}
The partition function can be bosonized after introducing the auxiliary boson fields
\begin{eqnarray}
\sigma(x)=-2G\bar{q}(x)q(x),\ \ \ \mbox{\boldmath{$\pi$}}(x)=-2G\bar{q}(x)i\gamma_5\mbox{\boldmath{$\tau$}}q(x)
\end{eqnarray}
for mesons and
\begin{eqnarray}
\phi(x)=-2G\bar{q}_c(x)i\gamma_5\tau_2t_2q(x)
\end{eqnarray}
for diquarks.  With the help of the Nambu-Gor'kov representation
$\bar{\Psi} = \left(\begin{array}{cc} \bar{q} & \bar{q}_c\end{array}\right)$, the partition function can be written as
\begin{eqnarray}
{\cal Z}_{\text{NJL}}=\int[d\bar{\Psi}][d\Psi][d\sigma][d\mbox{\boldmath{$\pi$}}][d\phi^\dagger][d\phi]e^{-{\cal A_{\text{eff}}}},
\end{eqnarray}
where the action ${\cal A}_{\text{eff}}$ is given by
\begin{widetext}
\begin{eqnarray}
{\cal A_{\text{eff}}}=\int dx\frac{\sigma^2(x)+\mbox{\boldmath{$\pi$}}^2(x)+|\phi(x)|^2}{4G}
-\frac{1}{2}\int dx\int dx^\prime\bar{\Psi}(x){\bf G}^{-1}(x,x^\prime)\Psi(x^\prime)
\end{eqnarray}
with the inverse quark propagator defined as
\begin{eqnarray}
{\bf G}^{-1}(x,x^\prime)=\left(\begin{array}{cc}\gamma^0(-\partial_{\tau}+\frac{\mu_{\text B}}{2})+i\mbox{\boldmath{$\gamma$}}\cdot\mbox{\boldmath{$\nabla$}}
-\mathcal {M}(x)&-i\gamma_5\phi(x)\tau_2t_2\\
-i\gamma_5\phi^\dagger(x)\tau_2t_2 &
\gamma^0(-\partial_{\tau}-\frac{\mu_{\text
B}}{2})+i\mbox{\boldmath{$\gamma$}}\cdot\mbox{\boldmath{$\nabla$}}-\mathcal
{M}^{\text T}(x)\end{array}\right)\delta(x-x^\prime)
\end{eqnarray}
\end{widetext}
and $\mathcal {M}(x) =m_0+\sigma(x)+ i\gamma_5\mbox{\boldmath{$\tau$}}\cdot\mbox{\boldmath{$\pi$}}(x)$. After integrating out the quarks, we can reduce the partition function to
\begin{eqnarray}
{\cal Z}_{\text{NJL}}&=&\int[d\sigma][d\mbox{\boldmath{$\pi$}}][d\phi^\dagger][d\phi]
e^{-{\cal S}_{\text{eff}}[\sigma,\mbox{\boldmath{$\pi$}},\phi^\dagger,\phi]},
\end{eqnarray}
where the bosonized effective action ${\cal S}_{\text{eff}}$ is given by
\begin{eqnarray}
{\cal S}_{\text{eff}}[\sigma,\mbox{\boldmath{$\pi$}},\phi^\dagger,\phi]&=&\int
dx\frac{\sigma^2(x)+\mbox{\boldmath{$\pi$}}^2(x)+|\phi(x)|^2}{4G}\nonumber\\
&&-\frac{1}{2}\text{Tr}\ln{\bf
G}^{-1}(x,x^\prime).
\end{eqnarray}
Here the trace $\text{Tr}$ is taken in color, flavor, spin, Nambu-Gor'kov and coordinate ($x$ and $x^\prime$) spaces. The thermodynamic potential density of the system is given by $\Omega(T,\mu_{\text B})=-\lim_{V\rightarrow\infty}(T/V)\ln Z_{\text{NJL}}$.

The effective action ${\cal S}_{\text{eff}}$ as well as the thermodynamic potential $\Omega$ cannot be evaluated exactly in the $3+1$ dimensional case. In this work, we firstly consider the saddle point approximation, i.e., the mean-field approximation. Then we investigate the fluctuations around the mean field.

In the mean field approximation, all bosonic auxiliary fields are replaced by their expectation values. Therefore, we set $\langle\sigma(x)\rangle=\upsilon$, $\langle\phi(x)\rangle=\Delta$, and $\langle\mbox{\boldmath{$\pi$}}(x)\rangle=0$. While $\Delta$ can be
set to be real due to the U$_{\rm B}(1)$ symmetry, we do not do this in the formalism. We will show in the following that all physical results
depend only on $|\Delta|^2$. The zeroth order or mean-field effective action reads
\begin{equation}
{\cal
S}_{\text{eff}}^{(0)}=\frac{V}{T}\left[\frac{\upsilon^2+|\Delta|^2}{4G}-\frac{1}{2}\sum_K\text{Trln}{\cal G}^{-1}(K)\right],
\end{equation}
where the inverse of the Nambu-Gor'kov quark propagator ${\cal G}^{-1}(K)$ is given by
\begin{eqnarray}
\left(\begin{array}{cc} (i\omega_n+\frac{\mu_{\text B}}{2})\gamma^0-\mbox{\boldmath{$\gamma$}}\cdot{\bf k}-M & -i\gamma_5\Delta\tau_2 t_2\\
-i\gamma_5\Delta^\dagger\tau_2 t_2 & (i\omega_n-\frac{\mu_{\text
B}}{2})\gamma^0-\mbox{\boldmath{$\gamma$}}\cdot{\bf
k}-M\end{array}\right)\
\end{eqnarray}
with the effective quark mass defined as $M=m_0+\upsilon$.  The
mean-field thermodynamic potential $\Omega_0=(T/V){\cal
S}_{\text{eff}}^{(0)}$ can be evaluated as
\begin{eqnarray}
\Omega_0=\frac{\upsilon^2+|\Delta|^2}{4G}-2N_cN_f\sum_{\bf k}\left[\mathcal {W}(E_{\bf k}^+)+\mathcal {W}(E_{\bf k}^-)\right]
\end{eqnarray}
with the function $\mathcal{W}(E)=E/2+T\ln{(1+e^{-E/T})}$ and the BCS-like quasiparticle dispersion relations
$E_{\bf k}^\pm=\sqrt{(E_{\bf k}\pm\mu_{\text B}/2)^2+|\Delta|^2}$ and $E_{\bf k}=\sqrt{{\bf k}^2+M^2}$. The signs $\mp$ correspond to quasiquark and quasi-antiquark excitations, respectively. The integral over the quark momentum ${\bf k}$ is divergent, and some regularization scheme should be adopted. In this work, we employ a hard three-momentum cutoff $\Lambda$.

The physical values of the variational parameters $M$ (or $\upsilon$) and $\Delta$ should be determined by the saddle point condition
\begin{eqnarray}\label{saddle}
\frac{\delta{\cal S}_{\text{eff}}^{(0)}[\upsilon,\Delta]}{\delta\upsilon}=0,
\ \ \ \ \ \frac{\delta{\cal S}_{\text{eff}}^{(0)}[\upsilon,\Delta]}{\delta \Delta}=0,
\end{eqnarray}
which minimizes the mean-field effective action ${\cal S}_{\text{eff}}^{(0)}$. One can show that the saddle point condition is equivalent to the following Green's function relations
\begin{eqnarray}
\langle\bar{q}q\rangle &=&  \sum_K\text{Tr}{\cal G}_{11}(K)\ ,\nonumber\\
\langle \bar{q}_ci\gamma_5\tau_2t_2q\rangle&=&
\sum_K\text{Tr}\left[{\cal G}_{12}(K)i\gamma_5\tau_2 t_2\right]\ ,
\end{eqnarray}
where the matrix elements of ${\cal G}$ are explicitly given by
\begin{eqnarray}
{\cal G}_{11}(K) &=& {i\omega_n+\xi_{\bf k}^-\over
(i\omega_n)^2-(E_{\bf k}^-)^2}\Lambda_{\bf k}^+\gamma_0+ {i\omega_n-\xi_{\bf k}^+\over (i\omega_n)^2-(E_{\bf k}^+)^2}\Lambda_{\bf k}^-\gamma_0\ ,\nonumber\\
{\cal G}_{22}(K) &=& {i\omega_n-\xi_{\bf k}^-\over
(i\omega_n)^2-(E_{\bf k}^-)^2}\Lambda_{\bf k}^-\gamma_0+ {i\omega_n+\xi_{\bf k}^+\over (i\omega_n)^2-(E_{\bf k}^+)^2}\Lambda_{\bf k}^+\gamma_0\ ,\nonumber\\
{\cal G}_{12}(K) &=& {-i\Delta\tau_2 t_2\over
(i\omega_n)^2-(E_{\bf k}^-)^2}\Lambda_{\bf k}^+\gamma_5+ {-i\Delta\tau_2 t_2\over (i\omega_n)^2-(E_{\bf k}^+)^2}\Lambda_{\bf k}^-\gamma_5\ ,\nonumber\\
{\cal G}_{21}(K) &=& {-i\Delta^\dagger\tau_2 t_2\over
(i\omega_n)^2-(E_{\bf k}^-)^2}\Lambda_{\bf k}^-\gamma_5+
{-i\Delta^\dagger\tau_2 t_2\over (i\omega_n)^2-(E_{\bf
k}^+)^2}\Lambda_{\bf k}^+\gamma_5 ,\nonumber
\end{eqnarray}
with the massive energy projectors
\begin{equation}
\Lambda_{\bf k}^{\pm}= {1\over 2}\left[1\pm{\gamma_0\left(\mbox{\boldmath{$\gamma$}}\cdot{\bf k}+M\right)\over E_{\bf k}}\right]\ .
\end{equation}
Here we have defined the notation $\xi_{\bf k}^\pm=E_{\bf k}\pm\mu_{\text B}/2$.

Next, we consider the fluctuations around the mean field, corresponding to the collective bosonic excitations. Making the field shifts for the auxiliary fields,
\begin{eqnarray}
&&\sigma(x)\rightarrow \upsilon+\sigma(x),\ \
\mbox{\boldmath{$\pi$}}(x)\rightarrow
0+\mbox{\boldmath{$\pi$}}(x),\nonumber\\
&&\phi(x)\rightarrow \Delta+\phi(x), \ \ \phi^\dagger(x)\rightarrow
\Delta^\dagger+\phi^\dagger(x),
\end{eqnarray}
we can express the total effective action as
\begin{eqnarray}
{\cal S}_{\text{eff}}&=&{\cal S}_{\text{eff}}^{(0)}+\int
dx\left(\frac{\sigma^2+\mbox{\boldmath{$\pi$}}^2+|\phi|^2}{4G}+\frac{\upsilon\sigma+\Delta\phi^\dagger+\Delta^\dagger\phi}{2G}\right)\nonumber\\
&-&\frac{1}{2}\text{Tr}\ln{\left[\mathbbold{1}+\int dx_1{\cal
G}(x,x_1)\Sigma(x_1,x^\prime)\right]}.
\end{eqnarray}
Here ${\cal G}(x,x^\prime)$ is the Fourier transformation of ${\cal G}(K)$, and $\Sigma(x,x^\prime)$ is defined as
\begin{eqnarray}
\Sigma(x,x^\prime)&=&\left(\begin{array}{cc}-\sigma(x)-
i\gamma_5\mbox{\boldmath{$\tau$}}\cdot\mbox{\boldmath{$\pi$}}(x)&-i\gamma_5\phi(x)\tau_2t_2\\-i\gamma_5\phi^\dagger(x)\tau_2t_2&
-\sigma(x)- i\gamma_5\mbox{\boldmath{$\tau$}}^{\text
T}\cdot\mbox{\boldmath{$\pi$}}(x)\end{array}\right)\nonumber\\
&\times&\delta(x-x^\prime).
\end{eqnarray}
With the help of the derivative expansion
\begin{eqnarray}
\text{Tr}\ln{\left[\mathbbold{1}+{\cal
G}\Sigma\right]}=\sum_{n=1}^\infty\frac{(-1)^{n+1}}{n}\text{Tr}[{\cal
G}\Sigma]^n,
\end{eqnarray}
we can calculate the effective action in powers of the fluctuations $\sigma(x),\mbox{\boldmath{$\pi$}}(x),\phi(x),\phi^\dagger(x)$.

The first-order effective action ${\cal S}_{\text{eff}}^{(1)}$ which includes linear terms of the fluctuations should vanish exactly,
since the expectation value of the fluctuations should be exactly zero. In fact, ${\cal S}_{\text{eff}}^{(1)}$ can be evaluated as
\begin{eqnarray}
{\cal S}_{\text{eff}}^{(1)} &=&\int
dx\Bigg\{\left[\frac{\upsilon}{2G}+\frac{1}{2}\text{Tr}\left({\cal
G}_{11}+{\cal G}_{22}\right)\right]\sigma(x)\nonumber\\
&+&\frac{1}{2}\text{Tr}\left[i\gamma_5\left({\cal
G}_{11}\mbox{\boldmath{$\tau$}}+{\cal
G}_{22}\mbox{\boldmath{$\tau$}}^{\text
T}\right)\right]\cdot\mbox{\boldmath{$\pi$}}(x)\nonumber\\
&+&\left[\frac{\Delta}{2G}+\frac{1}{2}\text{Tr}\left(i\gamma_5\tau_2
t_2{\cal G}_{12}\right)\right]\phi^\dagger(x)\nonumber\\
&+&\left[\frac{\Delta^\dagger}{2G}+\frac{1}{2}\text{Tr}\left(i\gamma_5\tau_2
t_2{\cal G}_{21}\right)\right]\phi(x)\Bigg\}.
\end{eqnarray}
We observe that the coefficient of $\mbox{\boldmath{$\pi$}}(x)$ is automatically zero after taking the trace in Dirac spin space. The
coefficients of $\phi(x),\phi^\dagger(x)$ and $\sigma(x)$ vanish once the quark propagator takes the mean-field form and $M,\Delta$
take the physical values that satisfies the saddle point condition. Therefore, in the present approach, the saddle point condition plays
a crucial role in having vanishing linear terms in the expansion.

The quadratic term ${\cal S}_{\text{eff}}^{(2)}$ or the Gaussian fluctuation corresponds to collective bosonic excitations. Working in the momentum space is convenient. It can be expressed as
\begin{eqnarray}
{\cal S}_{\text{eff}}^{(2)}&=&\frac{1}{2}\sum_Q\Bigg\{\frac{|\sigma(Q)|^2+|\mbox{\boldmath{$\pi$}}(Q)|^2+|\phi(Q)|^2}{2G}\nonumber\\
&&+\frac{1}{2}\sum_K\text{Tr}\left[{\cal G}(K)\Sigma(-Q){\cal
G}(K+Q)\Sigma(Q)\right]\Bigg\}.
\end{eqnarray}
Here $A(Q)$ is the Fourier transformation of the field $A(x)$, and $\Sigma(Q)$ is defined as
\begin{equation}
\Sigma(Q)=\left(\begin{array}{cc}-\sigma(Q)-
i\gamma_5\mbox{\boldmath{$\tau$}}\cdot\mbox{\boldmath{$\pi$}}(Q)&-i\gamma_5\phi(Q)\tau_2t_2\\-i\gamma_5\phi^\dagger(-Q)\tau_2t_2&
-\sigma(Q)- i\gamma_5\mbox{\boldmath{$\tau$}}^{\text
T}\cdot\mbox{\boldmath{$\pi$}}(Q)\end{array}\right).
\end{equation}
After taking the trace in the Nambu-Gor'kov space, we find that ${\cal S}_{\text{eff}}^{(2)}$ can be written in the following bilinear form
\begin{eqnarray}
{\cal
S}_{\text{eff}}^{(2)}&=&\frac{1}{2}\sum_Q\left(\begin{array}{ccc}
\phi^\dagger(Q) & \phi(-Q)&
\sigma^\dagger(Q)\end{array}\right){\bf M}(Q)\left(\begin{array}{cc} \phi(Q)\\
\phi^\dagger(-Q)\\ \sigma(Q)
\end{array}\right)\nonumber\\
&&+\frac{1}{2}\sum_Q\left(\begin{array}{ccc} \pi_1^\dagger(Q) &
\pi_2^\dagger(Q)&
\pi_3^\dagger(Q)\end{array}\right){\bf N}(Q)\left(\begin{array}{cc} \pi_1(Q)\\
\pi_2(Q)\\ \pi_3(Q) \end{array}\right).\nonumber\\
\end{eqnarray}

For $\Delta\neq0$, the matrix ${\bf M}$ is non-diagonal and can be expressed as
\begin{equation}
{\bf M}(Q)=\left(\begin{array}{ccc} \frac{1}{4G}+\Pi_{11}(Q)&\Pi_{12}(Q)&\Pi_{13}(Q)\\
\Pi_{21}(Q)&\frac{1}{4G}+\Pi_{22}(Q)&\Pi_{23}(Q)\\
\Pi_{31}(Q)&\Pi_{32}(Q)&\frac{1}{2G}+\Pi_{33}(Q)\end{array}\right).
\end{equation}
The polarization functions $\Pi_{\text{ij}}(Q)$ (${\text i}, {\text j}=1,2,3$) are one-loop susceptibilities composed of the Nambu-Gor'kov quark propagator. They can be expressed as
\begin{eqnarray}
\Pi_{11}(Q)&=&\frac{1}{2}\sum_K\text{Tr}\left[{\cal
G}_{22}(K)\Gamma{\cal G}_{11}(P)\Gamma\right],\nonumber\\
\Pi_{12}(Q)&=&\frac{1}{2}\sum_K\text{Tr}\left[{\cal
G}_{12}(K)\Gamma{\cal G}_{12}(P)\Gamma\right],\nonumber\\
\Pi_{13}(Q)&=&\frac{1}{2}\sum_K\text{Tr}\left[{\cal
G}_{12}(K)\Gamma{\cal G}_{11}(P)+{\cal G}_{22}(K)\Gamma{\cal
G}_{12}(P)\right],\nonumber\\
\Pi_{21}(Q)&=&\frac{1}{2}\sum_K\text{Tr}\left[{\cal
G}_{21}(K)\Gamma{\cal G}_{21}(P)\Gamma\right],\nonumber\\
\Pi_{22}(Q)&=&\frac{1}{2}\sum_K\text{Tr}\left[{\cal
G}_{11}(K)\Gamma{\cal G}_{22}(P)\Gamma\right],\nonumber\\
\Pi_{23}(Q)&=&\frac{1}{2}\sum_K\text{Tr}\left[{\cal
G}_{11}(K)\Gamma{\cal G}_{21}(P)+{\cal G}_{21}(K)\Gamma{\cal
G}_{22}(P)\right],\nonumber\\
\Pi_{31}(Q)&=&\frac{1}{2}\sum_K\text{Tr}\left[{\cal G}_{21}(K){\cal
G}_{11}(P)\Gamma+{\cal G}_{22}(K){\cal
G}_{21}(P)\Gamma\right],\nonumber\\
\Pi_{32}(Q)&=&\frac{1}{2}\sum_K\text{Tr}\left[{\cal G}_{11}(K){\cal
G}_{12}(P)\Gamma+{\cal G}_{12}(K){\cal G}_{22}(P)\Gamma\right],\nonumber\\
\Pi_{33}(Q)&=&\frac{1}{2}\sum_K\text{Tr}\big[{\cal G}_{11}(K){\cal
G}_{11}(P)+{\cal G}_{22}(K){\cal G}_{22}(P)\nonumber\\
&&+{\cal G}_{12}(K){\cal
G}_{21}(P)+{\cal G}_{21}(K){\cal
G}_{12}(P)\big],
\end{eqnarray}
with $P=K+Q$, $\Gamma=i\gamma_5\tau_2t_2$, where the trace is taken in color, flavor and spin spaces. Using the fact
${\cal G}_{22}(K,\mu_{\text B})={\cal G}_{11}(K,-\mu_{\text B})$ and ${\cal G}_{21}(K,\mu_{\text B})={\cal G}_{12}^\dagger(K,-\mu_{\text B})$,
we can show
\begin{eqnarray}
&&\Pi_{22}(Q)=\Pi_{11}(-Q),\ \ \ \ \Pi_{12}(Q)=\Pi_{21}^\dagger(Q),\nonumber\\
&&\Pi_{13}(Q)=\Pi_{31}^\dagger(Q)=\Pi_{23}^\dagger(-Q)=\Pi_{32}(-Q).
\end{eqnarray}
Therefore, only five of the polarization functions are independent. At $T=0$, their explicit form is shown in Appendix \ref{app}. For general case, we can show $\Pi_{12}\propto\Delta^2$ and $\Pi_{13}\propto M\Delta$. Therefore, in the normal phase with $\Delta=0$, the matrix ${\bf M}$ recovers the diagonal form. The off-diagonal elements $\Pi_{13}$ and $\Pi_{23}$ represents the mixing between the sigma meson and the diquarks.
At large chemical potentials where the chiral symmetry is approximately restored, $M\rightarrow m_0$, this mixing effect can be safely neglected.

On the other hand, the matrix ${\bf N}$ of the pion sector is diagonal and proportional to the identity matrix, i.e.,
\begin{equation}
{\bf N}_{\text{ij}}(Q)=
\delta_{\text{ij}}\left[\frac{1}{2G}+\Pi_{\pi}(Q)\right], \ \ \
\text{i,j}=1,2,3.
\end{equation}
This means that pions are eigen mesonic excitations even in the superfluid phase. The polarization function $\Pi_\pi(Q)$ is given by
\begin{eqnarray}\label{pionpo}
\Pi_{\pi}(Q)&=&\frac{1}{2}\sum_K\text{Tr}\bigg[{\cal
G}_{11}(K)i\gamma_5{\cal G}_{11}(P)i\gamma_5+{\cal
G}_{22}(K)i\gamma_5{\cal G}_{22}(P)i\gamma_5\nonumber\\
&&-{\cal G}_{12}(K)i\gamma_5{\cal G}_{21}(P)i\gamma_5-{\cal
G}_{21}(K)i\gamma_5{\cal G}_{12}(P)i\gamma_5\bigg].
\end{eqnarray}
Its explicit form at $T=0$ is given in Appendix \ref{app}. We find that $\Pi_{\pi}(Q)$ and $\Pi_{33}(Q)$ is different only up to a term proportional to $M^2$. Therefore, at high density with $\langle\bar{q}q\rangle\rightarrow 0$, the spectra of pions and sigma meson become nearly degenerate, which represents the approximate restoration of chiral symmetry.

The U$_{\text B}(1)$ baryon number symmetry is spontaneously broken by the nonzero diquark condensate $\langle qq\rangle$ in the superfluid phase (even for $m_0\neq0$), resulting in one Goldstone boson. In our model, this is ensured by the condition $\det{\bf M}(Q=0)=0$. From the explicit
form of the polarization functions given in Appendix \ref{app}, we find that this condition holds if and only if the saddle point condition (\ref{saddle}) for $\upsilon$ and $\Delta$ is satisfied.

\subsection {Vacuum and model parameter fixing}
\label{s3-2}
For a better understanding of our derivation in the following, it is
useful to review the vacuum state at $T=\mu_{\text B}=0$. In the
vacuum, it is evident that $\Delta=0$ and the mean-field effective
potential $\Omega_{\text{vac}}$ can be evaluated as
\begin{eqnarray}
\Omega_{\text{vac}}(M)=\frac{(M-m_0)^2}{4G}-2N_cN_f\sum_{\bf
k}E_{\bf k}.
\end{eqnarray}
The physical value of $M$, denoted by $M_*$, satisfies the saddle
point condition $\partial\Omega_{\text{vac}}/\partial M=0$ and
minimizes $\Omega_{\text{vac}}$.

The meson and diquark excitations can be obtained from ${\cal
S}_{\text{eff}}^{(2)}$, which in the vacuum can be expressed as
\begin{eqnarray}
&&{\cal S}_{\text{eff}}^{(2)}=
-\frac{1}{2}\int\frac{d^4Q}{(2\pi)^4}\bigg[\sigma(-Q){\cal
D}_\sigma^{*-1}(Q)\sigma(Q)\nonumber\\
&&\ \ \ \ \ \ \ \ \ \ \ +\sum_{i=1}^3\pi_i(-Q){\cal
D}_{\pi}^{*-1}(Q)\pi_i(Q)\nonumber\\
&&\ \ \ \ \ \ \ \ \ \ \ +\sum_{i=1}^2\phi_i(-Q){\cal
D}_{\phi}^{*-1}(Q)\phi_i(Q)\bigg],
\end{eqnarray}
where $\phi_1,\phi_2$ are the real and imaginary parts of $\phi$,
respectively. The inverse propagators in vacuum can be expressed in
a symmetrical form \cite{NJLreview,NJLreview1,NJLreview2,NJLreview3}
\begin{eqnarray}
{\cal D}^{*-1}_{l}(Q)&=&\frac{1}{2G}+\Pi_{l}^*(Q),\ \ \ \
l=\sigma,\pi,\phi\nonumber\\
\Pi_{l}^*(Q)&=&2iN_cN_f(Q^2-\epsilon_l^2)I(Q^2)\nonumber\\
&&-4iN_cN_f\int\frac{d^4K}{(2\pi)^4}\frac{1}{K^2-M_*^2}
\end{eqnarray}
with $\epsilon_\sigma=2M_*$ and $\epsilon_{\pi}=\epsilon_\phi=0$, where
the function $I(Q^2)$ is defined as
\begin{eqnarray}
I(Q^2)=\int\frac{d^4K}{(2\pi)^4}\frac{1}{(K_+^2-M_*^{2})(K_-^2-M_*^{2})}
\end{eqnarray}
with $K_\pm=K\pm Q/2$. Keeping in mind that $M_*$ satisfies the
saddle point condition, we find that the pions and diquarks are
Goldstone bosons in the chiral limit, corresponding to the
symmetry breaking pattern SU$(4)\rightarrow$Sp$(4)$. Using the gap
equation of $M_*$, we find that the masses of mesons and diquarks
can be determined by the equation
\begin{eqnarray}\label{mesonmass}
m_{l}^2=-\frac{m_0}{M_*}\frac{1}{4iGN_cN_fI(m_{l}^2)}+\epsilon_{l}^2.
\end{eqnarray}
Since the $Q^2$ dependence of the function $I(Q^2)$ is very weak, we
find $m_\pi^2\sim m_0$ and $m_\sigma^2\simeq 4M_*^2+m_\pi^2$.

Since pions and diquarks are deep bound states, their propagators
can be well approximated by ${\cal D}_\pi^*(Q)\simeq -g_{\pi
qq}^2/(Q^2-m_\pi^2)$ with $g_{\pi qq}^{-2}\simeq -2iN_cN_f
I(0)$~\cite{NJLreview,NJLreview1,NJLreview2,NJLreview3}. The pion
decay constant $f_\pi$ can be determined by the matrix element of
the axial current,
\begin{eqnarray}
iQ_\mu f_\pi\delta_{\text{ij}}&=&-\frac{1}{2}\text{Tr}\int\frac{d^4K}{(2\pi)^4}\left[\gamma_\mu
\gamma_5\tau_{\text i}{\cal G}(K_+)g_{\pi
qq}\gamma_5\tau_{\text j}{\cal G}(K_-)\right]\nonumber\\
&=&2N_cN_fg_{\pi qq}M_*Q_\mu I(Q^2)\delta_{\text{ij}}.
\end{eqnarray}
Here ${\cal G}(K)=(\gamma^\mu K_\mu-M_*)^{-1}$. Therefore, the pion decay
constant can be expressed as
\begin{eqnarray}\label{piondecay}
f_\pi^2\approx-2iN_cN_fM_*^{2}I(0).
\end{eqnarray}
Finally, together with (\ref{mesonmass}) and (\ref{piondecay}), we
recover the well-known Gell-Mann--Oakes--Renner relation $m_\pi^2
f_\pi^2=-m_0\langle\bar{q}q\rangle_0$~\cite{gell}.

\begin{table}[b!]
\begin{center}
\begin{tabular}{|c| c c c | c c c |}
\hline
&&&&&&\\[-3mm]
 Set & $\Lambda$\ \ \ & $G\Lambda^2$ \ \ \ & $m_0$ & $\langle\bar{u}u\rangle_0^{1/3}$\ \ \ & $M_*$\ \ \ & $m_\pi$
\\[1mm]
\hline
&&&&&&\\[-3mm]
1 & 657.9 & 3.105 & 4.90 & -217.4 & 300 & 133.6 \\
2 & 583.6 & 3.676 & 5.53 & -209.1 & 400 & 134.0 \\
3 & 565.8 & 4.238 & 5.43 & -210.6 & 500 & 134.2 \\
4 & 565.4 & 4.776 & 5.11 & -215.1 & 600 & 134.4 \\
\hline
\end{tabular}
\end{center}
\caption{\small Model parameters (3-momentum cutoff $\Lambda$,
coupling constant $G$, and current quark mass $m_0$) and related
quantities (quark condensate $\langle\bar{u}u\rangle_0$, effective
quark mass $M_*$ and pion mass $m_\pi$ in units of MeV) for the
two-flavor two-color NJL model (\ref{NJL}). The pion decay constant
is fixed to be $f_\pi = 75$~MeV.} \label{fit}
\end{table}

There are three parameters in our model, the current quark mass
$m_0$, the coupling constant $G$ and the cutoff $\Lambda$. In
principle they should be determined from the known values of the
pion mass $m_\pi$, the pion decay constant $f_\pi$ and the quark
condensate
$\langle\bar{q}q\rangle_0$~\cite{NJLreview,NJLreview1,NJLreview2,NJLreview3}.
Since two-color QCD does not correspond to our real world, we get
the above values from the empirical values $f_\pi\simeq93$MeV,
$\langle\bar{u}{u}\rangle_0\simeq(250$MeV$)^3$ in the $N_c=3$ case,
according to the relation $f_\pi^2, \langle\bar{q}{q}\rangle_0\sim
N_c$. To obtain the model parameters, we fix the values of the pion
decay constant $f_\pi$ and slightly vary the values of the chiral
condensate $\langle\bar{q}q\rangle_0$ and the pion mass $m_\pi$.
Thus, we can obtain different sets of model parameters corresponding
to different values of effective quark mass $M_*$ and hence
different values of the sigma meson mass $m_\sigma$. Four sets of
model parameters are shown in Table.~\ref{fit}. As we will show in
the following that, the physics near the quantum phase transition
point $\mu_{\text B}=m_\pi$ is not sensitive to different model
parameter sets, since the low energy dynamics is dominated by the
pseudo-Goldstone bosons (i.e., the diquarks). However, at high
density, the physics becomes sensitive to different model parameter
sets corresponding to different sigma meson masses. The predictions
by the chiral perturbation theory should be recovered in the limit
$m_\sigma/m_\pi\rightarrow\infty$.

\subsection {Quantum phase transition and diquark Bose condensation}
\label{s2-3}
Now we begin to study the properties of two-color matter at finite
baryon density. Without loss of generality, we set $\mu_{\text
B}>0$. In this section, we study the two-color baryonic matter in
the dilute limit, which forms near the quantum phase transition
point $\mu_{\text B}=m_\pi$. Since the diquark condensate is
vanishingly small near the quantum phase transition point, we can
make a Ginzburg-Landau expansion for the effective action. As we
will see below, this corresponds to the mean-field theory of weakly
interacting dilute Bose condensates.

Since the diquark condensate $\Delta$ is vanishingly small near the
quantum phase transition, we can derive the Ginzburg-Landau free
energy functional $V_{\text{GL}}[\Delta(x)]$ at $T=0$ for the order
parameter field $\Delta(x)=\langle\phi(x)\rangle$ in the static and
long-wavelength limit. The general form of
$V_{\text{GL}}[\Delta(x)]$ can be written as
\begin{eqnarray}
&&V_{\text{GL}}[\Delta(x)]=\int
dx\Bigg[\Delta^\dagger(x)\left(-\delta\frac{\partial^2}{\partial\tau^2}+
\kappa\frac{\partial}{\partial\tau}-\gamma\mbox{\boldmath{$\nabla$}}^2\right)\Delta(x)\nonumber\\
&&\ \ \ \ \ \ \ \ \ \ \ \ \ \ \ \ \ \ \
+\alpha|\Delta(x)|^2+\frac{1}{2}\beta|\Delta(x)|^4\Bigg],
\end{eqnarray}
where the coefficients $\alpha,\beta,\gamma,\delta,\kappa$ should be
low energy constants which depend only on the vacuum
properties. The calculation is somewhat similar to the derivation
of Ginzburg-Landau free energy of a superconductor from the
microscopic BCS theory~\cite{nao}, but for our case there is a
difference in that we have another variational parameter, i.e., the
effective quark mass $M$ which should be a function of $|\Delta|^2$
determined by the saddle point condition.

In the static and long-wavelength
limit, the coefficients $\alpha,\beta$ of the potential terms can be
obtained from the effective action ${\cal S}_{\text{eff}}$ in the
mean-field approximation. At $T=0$, the mean-field effective action
reads ${\cal S}_{\text{eff}}^{(0)}=\int dx\Omega_0$, where the mean-field
 thermodynamic potential is given by
\begin{eqnarray}
\Omega_0(|\Delta|^2,M)=\frac{(M-m_0)^2+|\Delta|^2}{4G}-N_cN_f\sum_{\bf
k}(E_{\bf k}^++E_{\bf k}^-).
\end{eqnarray}
The Ginzburg-Landau coefficients $\alpha,\beta$ can be obtained via
a Taylor expansion of $\Omega_0$ in terms of $|\Delta|^2$,
\begin{equation}
\Omega_0=\Omega_{\text{vac}}(M_*)+\alpha|\Delta|^2+\frac{1}{2}\beta|\Delta|^4+O(|\Delta|^6),
\end{equation}
where $\Omega_{\text{vac}}(M_*)$ is the vacuum contribution which
should be subtracted. One should keep in mind that the
effective quark mass $M$ is not a fixed parameter, but an implicit function of
$|\Delta|^2$ determined by the gap equation
$\partial\Omega_0/\partial M=0$.

For convenience, we define $y\equiv|\Delta|^2$. The Ginzburg-Landau
coefficient $\alpha$ is defined as
\begin{eqnarray}
\alpha&=&\frac{d\Omega_0(y,M)}{dy}\Bigg|_{y=0}\nonumber\\
&=&\frac{\partial\Omega_0(y,M)}{\partial y}\Bigg|_{y=0}
+\frac{\partial\Omega_0(y,M)}{\partial M}\frac{dM}{d
y}\Bigg|_{y=0}\nonumber\\
&=&\frac{\partial\Omega_0(y,M)}{\partial y}\Bigg|_{y=0},
\end{eqnarray}
where the indirect derivative term vanishes due to the saddle point
condition for $M$. After some simple algebra, we get
\begin{equation}
\alpha=\frac{1}{4G}-N_cN_f\sum_{\bf k}\frac{E_{\bf k}^*}{E_{\bf
k}^{*2}-\mu_{\text B}^2/4}
\end{equation}
with $E_{\bf k}^*=\sqrt{{\bf k}^2+M_*^2}$. We can make the above
expression more meaningful using the pion mass equation in the same
three-momentum regularization
scheme\cite{NJLreview,NJLreview1,NJLreview2,NJLreview3},
\begin{equation}
\frac{1}{4G}-N_cN_f\sum_{\bf k}\frac{E_{\bf k}^*}{E_{\bf
k}^{*2}-m_\pi^2/4}=0.
\end{equation}
We therefore obtain a $G$-independent result
\begin{equation}
\alpha=\frac{1}{4}N_cN_f(m_\pi^2-\mu_{\text B}^2)\sum_{\bf
k}\frac{E_{\bf k}^*}{(E_{\bf k}^{*2}-m_\pi^2/4)(E_{\bf
k}^{*2}-\mu_{\text B}^2/4)}.
\end{equation}
From the fact that $m_\pi\ll 2M_*$ and $\beta>0$ (see below), we see
clearly that a second order quantum phase transition takes place
exactly at $\mu_{\text B}=m_\pi$. Therefore, the Ginzburg-Landau free energy
is meaningful only near the quantum phase transition point, i.e.,
$|\mu_{\text B}-m_\pi|\ll m_\pi$, and $\alpha$ can be further
simplified as
\begin{equation}\label{alpha}
\alpha\simeq(m_\pi^2-\mu_{\text B}^2){\cal J},
\end{equation}
where the factor ${\cal J}$ is defined as
\begin{eqnarray}
{\cal J}=\frac{1}{4}N_cN_f\sum_{\bf k}\frac{E_{\bf k}^*}{(E_{\bf
k}^{*2}-m_\pi^2/4)^2}.
\end{eqnarray}

The coefficient $\beta$ of the quartic term can be evaluated from the
definition
\begin{eqnarray}
\beta&=&\frac{d^2\Omega_0(y,M)}{dy^2}\Bigg|_{y=0}\nonumber\\
&=&\frac{\partial^2\Omega_0(y,M)}{\partial y^2}\Bigg|_{y=0}
+\frac{\partial^2\Omega_0(y,M)}{\partial M\partial y}\frac{dM}{d
y}\Bigg|_{y=0}.
\end{eqnarray}
Notice that the last indirect derivative term does not vanish here
and will be important for us to obtain a correct
diquark-diquark scattering length. The derivative $dM/dy$ can be
analytically derived from the gap equation for $M$. From the fact that
$\partial \Omega_0/\partial M=0$, we obtain
\begin{equation}
\frac{\partial}{\partial y}\left(\frac{\partial
\Omega_0(y,M)}{\partial M}\right)+ \frac{\partial}{\partial
M}\left(\frac{\partial \Omega_0(y,M)}{\partial
M}\right)\frac{dM}{dy}=0.
\end{equation}
Then we get
\begin{equation}
\frac{dM}{dy}= -\frac{\partial^2\Omega_0(y,M)}{\partial M\partial
y}\left(\frac{\partial^2\Omega_0(y,M)}{\partial M^2}\right)^{-1}.
\end{equation}
Then the practical expression for $\beta$ can be written as
\begin{eqnarray}
\beta=\beta_1+\beta_2,
\end{eqnarray}
where $\beta_1$ is the direct derivative term
\begin{equation}
\beta_1=\frac{\partial^2\Omega_0(y,M)}{\partial y^2}\Bigg|_{y=0},
\end{equation}
and $\beta_2$ is the indirect term
\begin{equation}
\beta_2=-\left(\frac{\partial^2\Omega_0(y,M)}{\partial M\partial
y}\right)^2\left(\frac{\partial^2\Omega_0(y,M)}{\partial
M^2}\right)^{-1}\Bigg|_{y=0}.
\end{equation}

Near the quantum phase transition, we can set $\mu_{\text B}=m_\pi$ in $\beta$.
After a simple algebra, the explicit forms of $\beta_1$ and $\beta_2$ can be evaluated as
\begin{eqnarray}
\beta_1=\frac{1}{4}N_cN_f\sum_{e=\pm}\sum_{\bf k}\frac{1}{(E_{\bf
k}^*-em_\pi/2)^3}
\end{eqnarray}
and
\begin{eqnarray}\label{beta2}
\beta_2&=&-\left\{\frac{1}{2}N_cN_f\sum_{e=\pm}\sum_{\bf
k}\frac{M_*}{E_{\bf k}^*}\frac{1}{(E_{\bf
k}^*-em_\pi/2)^2}\right\}^2\nonumber\\
&&\times\left(\frac{m_0}{2GM_*}+2N_cN_f\sum_{\bf
k}\frac{M_*^2}{E_{\bf k}^{*3}}\right)^{-1}.
\end{eqnarray}
The $G$-dependent term $m_0/(2GM_*)$ in (\ref{beta2}) can be
approximated as $m_\pi^2 f_\pi^2/M_*^2$ by using the relation
$m_\pi^2f_\pi^2=-m_0\langle\bar{q}{q}\rangle_0$.

The kinetic terms in the Ginzburg-Landau free energy can be derived from the inverse of the
diquark propagator. In the general case with $\Delta\neq0$, there exists mixing between the diquarks and the sigma meson.
However, approaching the quantum phase transition point, $\Delta\rightarrow 0$, the
problem is simplified. After the analytical continuation
$i\nu_m\rightarrow \omega+i0^+$, the inverse of the diquark
propagator at $\mu_{\text B}= m_\pi+0^+$ can be evaluated as
\begin{eqnarray}\label{dipro}
{\cal D}_{\text d}^{-1}(\omega,{\bf q})=\frac{1}{4G}+\Pi_{\text
d}(\omega,{\bf q}),
\end{eqnarray}
where the polarization function $\Pi_{\text d}(\omega,{\bf q})$ is
given by
\begin{eqnarray}
&&\Pi_{\text d}(\omega,{\bf q})=N_cN_f\sum_{\bf k}\frac{E^*_{{\bf
k}}+E^*_{{\bf k}+{\bf
q}}}{(\omega+\mu_{\text B})^2-(E^*_{{\bf k}}+E^*_{{\bf k}+{\bf q}})^2}\nonumber\\
&&\ \ \ \ \ \ \ \ \ \ \ \ \ \ \ \times\left(1+\frac{{\bf
k}\cdot({\bf k}+{\bf q})+M_*^2}{E^*_{{\bf k}}E^*_{{\bf k}+{\bf
q}}}\right).
\end{eqnarray}

In the static and long-wavelength limit ($\omega,|{\bf
q}|\rightarrow 0$), the coefficients $\kappa,\delta,\gamma$ can be
determined by the Taylor expansion ${\cal D}_{\text
d}^{-1}(\omega,{\bf q})={\cal D}_{\text d}^{-1}(0,{\bf
0})-\delta\omega^2-\kappa\omega+\gamma{\bf q}^2$. Notice that
$\alpha$ is identical to ${\cal D}_{\text d}^{-1}(0,{\bf 0})$, which
is in fact the Thouless criterion for the superfluid transition. On
the other hand, ${\cal D}_{\text d}^{-1}(\omega,{\bf q})$ can be
related to the pion propagator in the vacuum, i.e., ${\cal D}_{\text
d}^{-1}(\omega,{\bf q})=(1/2){\cal D}_\pi^{*-1}(\omega+\mu_{\text
B},{\bf q})$. In the static and long-wavelength limit and for
$\mu_{\text B}\rightarrow m_\pi\ll 2M_*$ it can be well approximated
as\cite{NJLreview,NJLreview1,NJLreview2,NJLreview3}
\begin{eqnarray}\label{diproa}
{\cal D}_{\text d}^{-1}(\omega,{\bf q})\simeq-{\cal
J}\left[(\omega+\mu_{\text B})^2-{\bf q}^2-m_\pi^2\right],
\end{eqnarray}
where ${\cal J}$ is the factor defined in (\ref{alpha}) which describes the coupling strength of the quark-quark-diquark interaction
${\cal J}\simeq g_{\pi qq}^{-2}/2$ and with which we obtain
$\delta\simeq\gamma\simeq{\cal J}$ and $\kappa\simeq2\mu_{\text B}{\cal J}$.

\begin{table}[b!]
\begin{center}
\begin{tabular}{|c|c|c|c|c|}
  \hline
  Set & 1 & 2 & 3 & 4 \\
  \hline
  $a_{\text{dd}}$ from Eq. (\ref{add})& 0.0631 & 0.0635 & 0.0637 & 0.0639 \\
  \hline
  $a_{\text{dd}}$ from Eq. (\ref{adda}) & 0.0624 & 0.0628 & 0.0630 & 0.0633 \\
  \hline
\end{tabular}
\end{center}
\caption{\small The values of diquark-diquark scattering length
$a_{\text{dd}}$ (in units of $m_\pi^{-1}$) for different model
parameter sets.} \label{scatt}
\end{table}

We now show how the Ginzburg-Landau free energy can be reduced to
the theory describing weakly repulsive Bose condensates, i.e., the
Gross-Pitaevskii free energy~\cite{GP01,GP02}.

First, since the Bose condensed matter is indeed dilute, let us consider the
nonrelativistic version with $\omega\ll m_\pi$ and neglecting the kinetic
term $\propto \partial^2/\partial\tau^2$. To this end,
we define the nonrelativistic chemical potential $\mu_{\text d}$
for diquarks, $\mu_{\text d}=\mu_{\text B}-m_\pi$. The coefficient $\alpha$ can
be further simplified as
\begin{eqnarray}
\alpha&\simeq&-\mu_{\text d}(2m_\pi{\cal J}).
\end{eqnarray}
Then the Ginzburg-Landau free energy can be reduced to the
Gross-Pitaevskii free energy of a dilute repulsive Bose gas, if we
define a new condensate wave function $\psiup(x)$ as
\begin{eqnarray}
\psiup(x)=\sqrt{2m_\pi{\cal J}}\Delta(x).
\end{eqnarray}
The resulting Gross-Pitaevskii free energy is given by
\begin{eqnarray}\label{fgp}
V_{\text{GP}}[\psiup(x)]&=&\int
dx\bigg[\psiup^\dagger(x)\left(\frac{\partial}{\partial\tau}-\frac{\mbox{\boldmath{$\nabla$}}^2}{2m_\pi}\right)\psiup(x)\nonumber\\
&&-\mu_{\text
d}|\psiup(x)|^2+\frac{1}{2}g_0|\psiup(x)|^4\bigg]
\end{eqnarray}
with $g_0=4\pi a_{\text{dd}}/m_\pi$. The repulsive diquark-diquark
interaction is characterized by a positive scattering length
$a_{\text{dd}}$ defined as
\begin{eqnarray}\label{add}
a_{\text{dd}}=\frac{\beta}{16\pi m_\pi}{\cal J}^{-2}.
\end{eqnarray}
Keep in mind that the scattering length obtained here is at the mean-field level. We will discuss the possible
beyond-mean-field corrections later. Therefore, near the quantum phase transition where the density $n$ satisfies $na_{\text{dd}}^3\ll1$,
the system is indeed a weakly interacting Bose condensate~\cite{Bose01}.

Even though we have shown that the Ginzburg-Landau free energy is indeed a
Gross-Pitaevskii version near the quantum phase transition, a key
problem is whether the obtained diquark-diquark scattering length
is quantitatively correct. The numerical calculation for (\ref{add}) is straightforward.
The obtained values of $a_{\text{dd}}$ for the four model parameter sets are shown in Table
\ref{scatt}. We can also give an analytical expression based on the
formula of the pion decay constant in the three-momentum cutoff
scheme,
\begin{eqnarray}
f_\pi^2=N_cM_*^2\sum_{\bf k}\frac{1}{E_{\bf k}^{*3}}.
\end{eqnarray}
According to the fact $m_\pi\ll 2M_*$, $\beta$ and ${\cal J}$
can be well approximated as
\begin{eqnarray}
&&\beta\simeq\frac{f_\pi^2}{M_*^2}-\frac{(2f_\pi^2/M_*)^2}{m_\pi^2f_\pi^2/M_*^2+4f_\pi^2}\simeq\frac{f_\pi^2m_\pi^2}{4M_*^4},\nonumber\\
&&{\cal J}\simeq\frac{f_\pi^2}{2M_*^2}.
\end{eqnarray}
There, the diquark-diquark scattering length $a_{\text{dd}}$ in the
limit $m_\pi/(2M_*)\rightarrow 0$ is only related to the pion mass
and decay constant,
\begin{eqnarray}\label{adda}
a_{\text{dd}}=\frac{m_\pi}{16\pi f_\pi^2}.
\end{eqnarray}
The values of $a_{\text{dd}}$ for the four model parameter sets
according to the above expression are also listed in Table
\ref{scatt}. The errors are always about $1\%$ comparing with the
exact numerical results, which means that the expression
(\ref{adda}) is a good approximation for the diquark-diquark
scattering length. The error should come from the finite value of
$m_\pi/(2M^*)$. We can obtain a correction in powers of
$m_\pi/(2M^*)$~\cite{schulze}, but it is obviously small, and its
explicit form is not shown here.

The result $a_{\text{dd}}\propto m_\pi$ is universal for the
scattering lengths of the pseudo-Goldstone bosons. Eventhough the SU$(4)$ flavor symmetry
is explicitly broken in presence of a nonzero quark mass, a descrete symmetry $\phi_1,\phi_2\leftrightarrow\pi_1,\pi_2$
holds exactly for arbitrary quark mass. This also means that the partition function of two-color QCD has a descrete
symmetry $\mu_{\text B}\leftrightarrow\mu_{\text I}$ \cite{QL03}. Because of
this descrete symmetry of two-color QCD, the analytical
expression (\ref{adda}) of $a_{\text{dd}}$ (which is in fact the
diquark-diquark scattering length in the $B=2$ channel) should be identical
to the pion-pion scattering length at tree level in the $I=2$
channel which was first obtained by Weinberg many years ago
\cite{pipi}. Therefore, the mean-field theory can describe not only
the quantum phase transition to a dilute diquark condensate but also
the effect of repulsive diquark-diquark interaction.

\begin{figure*}
\begin{center}
\includegraphics[width=7.5cm]{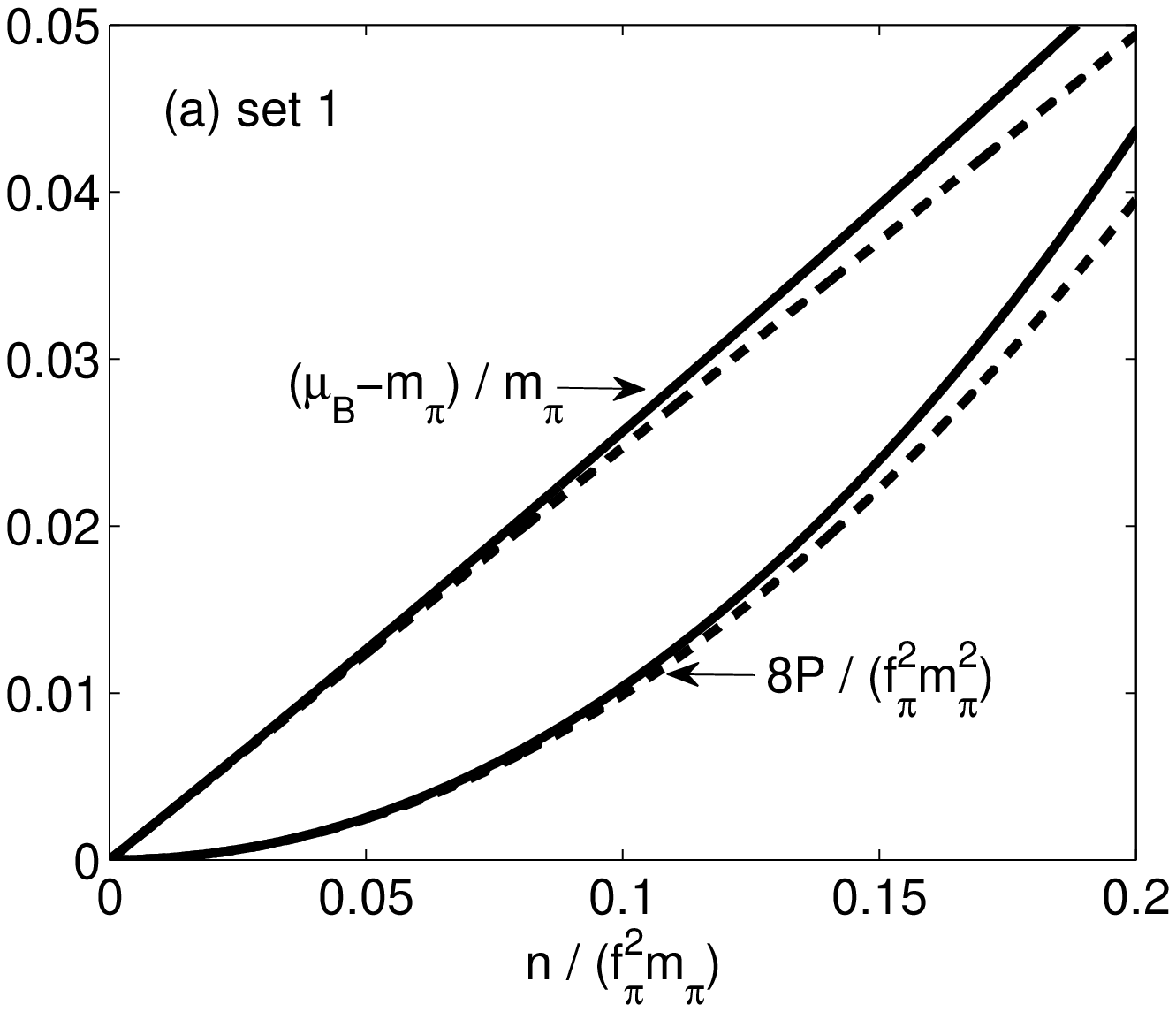}
\includegraphics[width=7.5cm]{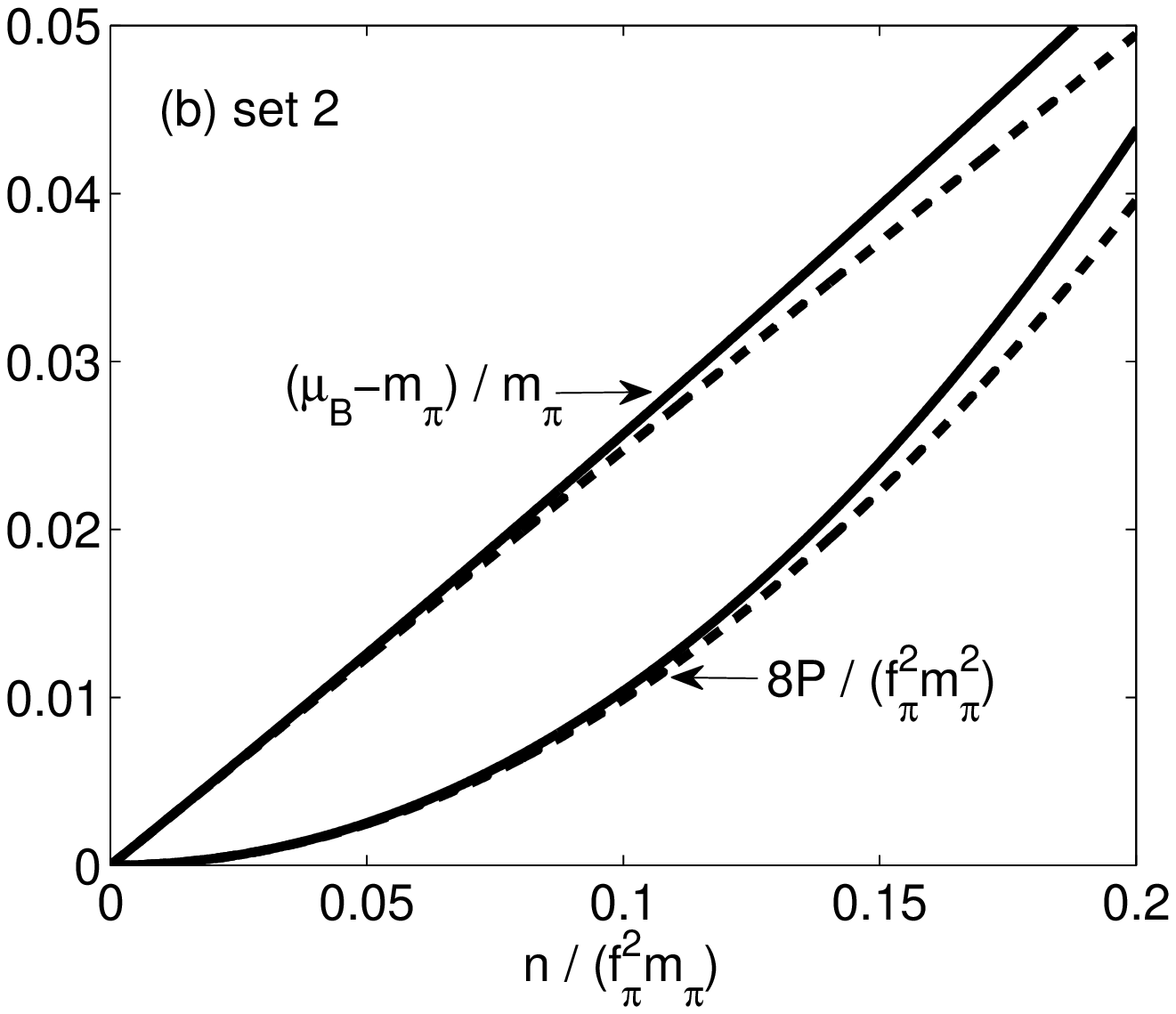}
\includegraphics[width=7.5cm]{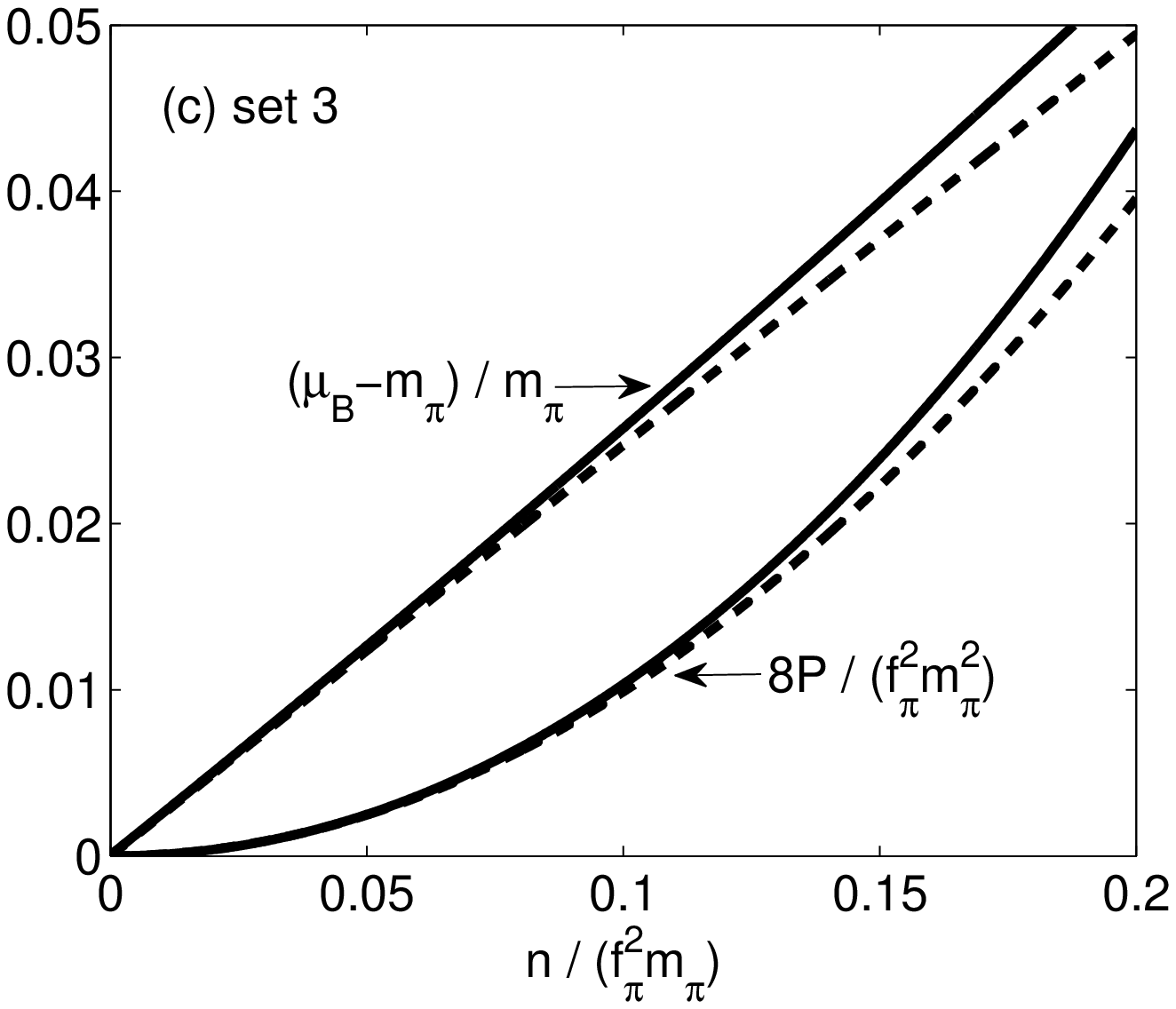}
\includegraphics[width=7.5cm]{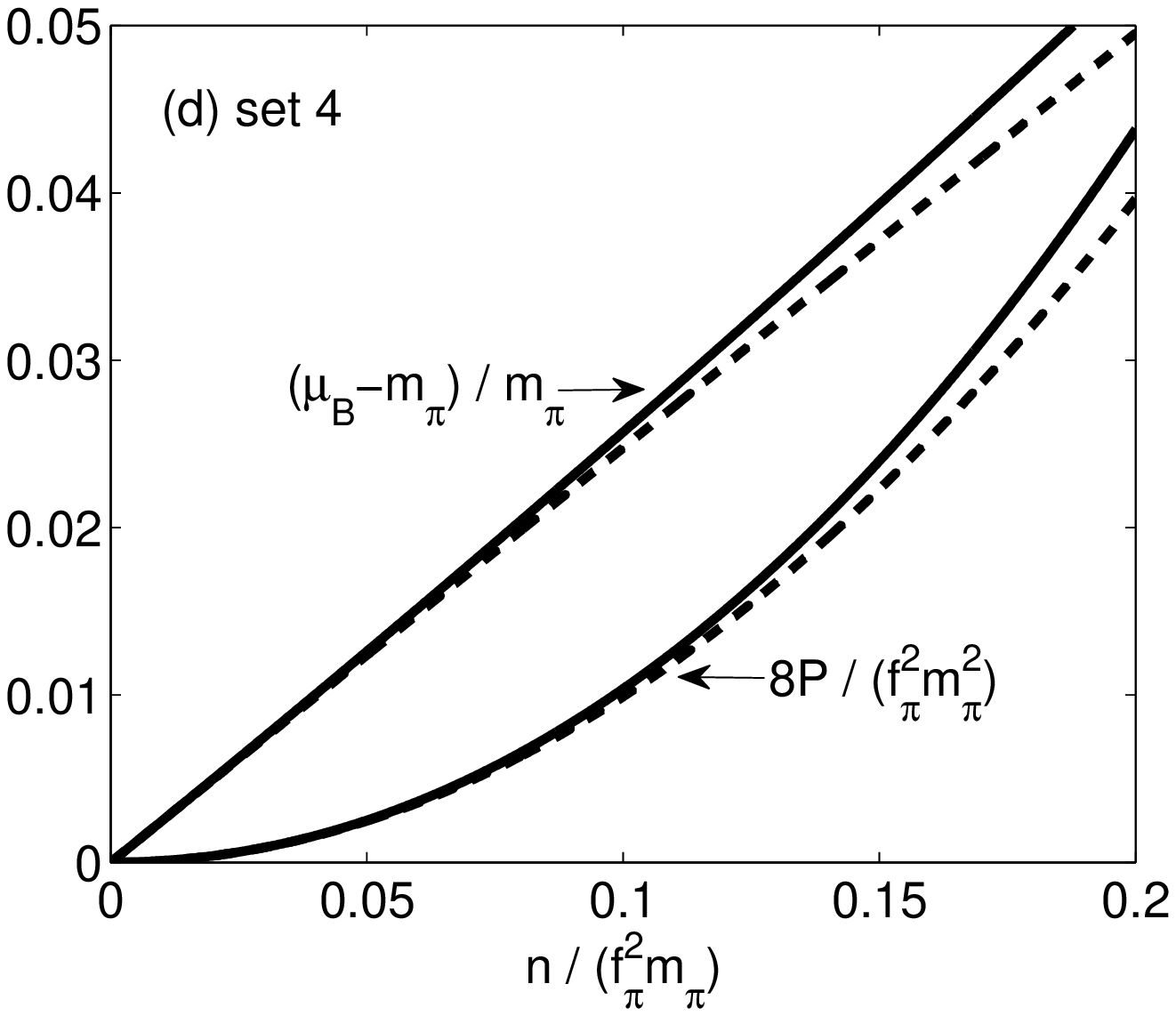}
\caption{The baryon chemical potential $\mu_{\text B}$ and the
pressure $P$ as functions of the baryon density $n$ for different
model parameter sets. The solid lines correspond to the direct mean-field
 calculation and the dashed lines are given by (\ref{eos}).
\label{fig1}}
\end{center}
\end{figure*}

The mean-field equations of state of the dilute diquark condensate are thus determined by the
Gross-Pitaevskii free energy (\ref{fgp}). Minimizing
$V_{\text{GP}}[\psiup(x)]$ with respect to a uniform condensate
$\psiup$, we find that the physical minimum is given by
\begin{eqnarray}\label{chemi}
|\psiup_0|^2=\frac{\mu_{\text d}}{g_0},
\end{eqnarray}
and the baryon density is $n=|\psiup_0|^2$. Using the thermodynamic
relations, we therefore get the well-known results at $T=0$ for the pressure
$P$, the energy density ${\cal E}$ and the chemical potential
$\mu_{\text B}$ in terms of the baryon density $n$~\cite{Bose01},
\begin{eqnarray}\label{eos}
&&P(n)=\frac{2\pi a_{\text{dd}}}{m_\pi}n^2,\nonumber\\
&&{\cal E}(n)=m_\pi n+\frac{2\pi a_{\text{dd}}}{m_\pi}n^2,\nonumber\\
&&\mu_{\text B}(n)=m_\pi+\frac{4\pi a_{\text{dd}}}{m_\pi}n.
\end{eqnarray}
We can examine the above results through a direct
numerical calculation with the mean-field thermodynamic potential.
Since all the thermodynamic quantities are relative to the vacuum, we subtract the vacuum contribution from the pressure, $P=-(\Omega_0(n)-\Omega_0(0))$, and the
baryon density reads $n=-\partial\Omega_0/\partial\mu_{\text B}$. In
Fig.\ref{fig1} we show the numerical results for the pressure and
the chemical potential as functions of the density for the four
model parameter sets. At low enough density, the equations of state
are indeed consistent with the results (\ref{eos}) with the
scattering length given by (\ref{add}). It is evident that the
results at low density are not sensitive to different model
parameter sets, since the physics at low density should be dominated
by the pseudo-Goldstone bosons.

Further, since our treatment is only at the mean-field level, the
Lee-Huang-Yang corrections~\cite{Bose03} which are proportional to
$(na_{\text{dd}}^3)^{1/2}$ are absent in the equations of state. To
obtain such corrections, it is necessary to go beyond the mean
field, and there is also a beyond-mean-field correction to the
scattering length $a_{\text{dd}}$~\cite{Unitary,HU,Diener}.

We can also consider a relativistic version of the Gross-Pitaevskii free energy via
defining the condensate wave function
\begin{eqnarray}
\Phi(x)=\sqrt{{\cal J}}\Delta(x).
\end{eqnarray}
In this case, the Ginzburg-Landau free energy is reduced to a
relativistic version of the Gross-Pitaevskii free energy,
\begin{eqnarray}\label{RGP}
V_{\text{RGP}}[\Phi(x)]&=&\int
dx\bigg[\Phi^\dagger(x)\left(-\frac{\partial^2}{\partial\tau^2}+2\mu_{\text
B}\frac{\partial}{\partial\tau}
-\mbox{\boldmath{$\nabla$}}^2\right)\Phi(x)\nonumber\\
&&+(m_\pi^2-\mu_{\text
B}^2)|\Phi(x)|^2+\frac{\lambda}{2}|\Phi(x)|^4\bigg].
\end{eqnarray}
The self-interacting coupling $\lambda=\beta {\cal J}^{-2}$ is now
dimensionless and can be approximated by $\lambda\simeq
m_\pi^2/f_\pi^2$. For realistic values of $m_\pi$ and $f_\pi$, we
find $\lambda\sim O(1)$. In this sense, the Bose condensate is not
weakly interacting, except for the low density limit
$na_{\text{dd}}^3\ll1$. One should keep in mind that this result
cannot be applied to high density, since it is valid only near the
quantum phase transition point.

An ideal Bose-Einstein condensate is not a superfluid. In presence
of weak repulsive interactions among the bosons, a Goldstone mode
which has a linear dispersion at low energy appears, and
the condensate becomes a superfluid according to Landau's criterion
$\min_{\bf q}[\omega({\bf q})/|{\bf q}|]>0$. The Goldstone mode
which is also called the Bogoliubov mode here should have a
dispersion given by \cite{Bose01}
\begin{eqnarray}\label{bog}
\omega({\bf q})=\sqrt{\frac{{\bf q}^2}{2m_\pi}\left(\frac{{\bf
q}^2}{2m_\pi}+\frac{8\pi a_{\text{dd}}n}{m_\pi}\right)},\ \ \ \
|{\bf q}|\ll m_\pi.
\end{eqnarray}

Since the Gross-Pitaevskii free energy obtained above is at the
classical level, to study the bosonic collective excitations we
should consider the fluctuations around the mean field. The propagator of the bosonic
collective modes is given by ${\bf M}^{-1}(Q)$ and ${\bf
N}^{-1}(Q)$. The Bogoliubov mode corresponds to the lowest
excitation obtained from the equation $\det{\bf M}(\omega,{\bf
q})=0$. With the explicit form of the matrix elements of ${\bf M}$
in the superfluid phase, we can analytically show $\det{\bf
M}(0,{\bf 0})=0$ which ensures the Goldstone's theorem. In fact, for
$(\omega,{\bf q})=(0,{\bf 0})$, we find $\det{\bf M}=({\bf
M}_{11}^2-|{\bf M}_{12}|^2){\bf M}_{33}+2|{\bf M}_{13}|^2(|{\bf
M}_{12}|-{\bf M}_{11})$. Using the saddle point condition for
$\Delta$, we can show that ${\bf M}_{11}(0,{\bf 0})=|{\bf
M}_{12}(0,{\bf 0})|$ and hence the Goldstone's theorem holds in the
superfluid phase. Further, we may obtain an analytical expression of
the velocity of the Bogoliubov mode via a Taylor expansion for ${\bf
M}(\omega,{\bf q})$ around $(\omega,{\bf q})=(0,{\bf 0})$ like those
done in \cite{BCSBEC3}. Such a calculation for our case is more complicated due to the mixing between the sigma
meson and diquarks, and it cannot give the full dispersion
(\ref{bog}).

On the other hand, considering $\Delta\rightarrow 0$ near the quantum
phase transition point, we can expand the matrix elements of ${\bf
M}$ in powers of $|\Delta|^2$. The advantage of such an expansion is
that it can not only give the full dispersion (\ref{bog}) but also
link the meson properties in the vacuum. Formally, we can write down
the following expansions,
\begin{eqnarray}\label{bexp}
&&{\bf M}_{11}(\omega,{\bf q})={\cal D}_{\text d}^{-1}(\omega,{\bf
q})+|\Delta|^2A(\omega,{\bf q})+O(|\Delta|^4),\nonumber\\
&&{\bf M}_{22}(\omega,{\bf q})={\cal D}_{\text d}^{-1}(-\omega,{\bf
q})+|\Delta|^2A(-\omega,{\bf q})+O(|\Delta|^4),\nonumber\\
&&{\bf M}_{12}(\omega,{\bf q})={\bf M}_{21}^\dagger(\omega,{\bf
q})=\Delta^2B(\omega,{\bf q})+O(|\Delta|^4),\nonumber\\
&&{\bf M}_{13}(\omega,{\bf q})={\bf M}_{31}^\dagger(\omega,{\bf
q})=\Delta H(\omega,{\bf q})+O(|\Delta|^3),\nonumber\\
&&{\bf M}_{23}(\omega,{\bf q})={\bf M}_{32}^\dagger(\omega,{\bf
q})=\Delta^\dagger H(-\omega,{\bf q})+O(|\Delta|^3),\nonumber\\
&&{\bf M}_{33}(\omega,{\bf q})={\cal D}_\sigma^{*-1}(\omega,{\bf
q})+O(|\Delta|^2).
\end{eqnarray}
Notice that the effective quark mass $M$ is regarded as a function
of $|\Delta|^2$ as we have done in deriving the Ginzburg-Landau free
energy. Since we are interested in the dispersion in the low energy
limit, i.e., $\omega,|{\bf q}|\ll m_\pi$, we can approximate the
coefficients of the leading order terms as their values at
$(\omega,{\bf q})=(0,{\bf 0})$,
\begin{eqnarray}
&&A(\omega,{\bf q})\simeq A(-\omega,{\bf q})\simeq A(0,{\bf 0})\equiv A_0,\nonumber\\
&&B(\omega,{\bf q})\simeq B(0,{\bf 0})\equiv B_0,\nonumber\\
&&H(\omega,{\bf q})\simeq H(-\omega,{\bf q})\simeq H(0,{\bf
0})\equiv H_0.
\end{eqnarray}
Further, from $m_\sigma\gg m_\pi$, we can approximate the inverse
sigma propagator ${\cal D}_\sigma^{*-1}(\omega,{\bf q})$ as its
value at $(\omega,{\bf q})=(0,{\bf 0})$. Therefore, the dispersion of
the Goldstone mode in the low energy limit can be determined by the
following equation,
\begin{eqnarray}\label{dispersion}
&&\det\left(\begin{array}{ccc} {\cal D}_{\text d}^{-1}(\omega,{\bf
q})+|\Delta|^2A_0&\Delta^2B_0&\Delta H_0\\
\Delta^{\dagger 2}B_0&{\cal D}_{\text d}^{-1}(-\omega,{\bf
q})+|\Delta|^2A_0&\Delta^\dagger H_0\\
\Delta^\dagger H_0&\Delta H_0&{\cal D}_\sigma^{*-1}(0,{\bf
0})\end{array}\right)\nonumber\\
&&=0.
\end{eqnarray}

Now we can link the coefficients $A_0,B_0,H_0$ and ${\cal
D}_\sigma^{-1}(0,{\bf 0})$ to the derivatives of the mean-field
thermodynamic potential $\Omega_0$ and its Ginzburg-Landau
coefficients. Firstly, using the explicit form of ${\bf M}_{12}$, we
find
\begin{eqnarray}
|{\bf M}_{12}(0,{\bf 0})|=|\Delta|^2\beta_1\Longrightarrow
B_0=\beta_1.
\end{eqnarray}
Secondly, using the fact that
\begin{eqnarray}
{\bf M}_{11}(0,{\bf 0})-|{\bf M}_{12}(0,{\bf
0})|=\frac{\partial\Omega_0}{\partial |\Delta|^2},
\end{eqnarray}
and together with the definition for $A(\omega,{\bf q})$,
\begin{eqnarray}
A(\omega,{\bf q})&=&\frac{d{\bf M}_{11}(y,M)}{d
y}\Bigg|_{y=0}\nonumber\\
&=&\frac{\partial{\bf M}_{11}(y,M)}{\partial
y}\Bigg|_{y=0}+\frac{\partial{\bf M}_{11}(y,M)}{\partial M}\frac{d
M}{dy}\Bigg|_{y=0},
\end{eqnarray}
we find the following exact relation:
\begin{eqnarray}
A_0=\beta+B_0=\beta+\beta_1.
\end{eqnarray}
On the other hand, we have the following relations for $H_0$ and
${\cal D}_\sigma^{*-1}(0,{\bf 0})$,
\begin{eqnarray}
H_0&=&\frac{\partial^2\Omega_0(y,M)}{\partial M\partial
y}\Bigg|_{y=0},\nonumber\\
{\cal D}_\sigma^{*-1}(0,{\bf
0})&=&\frac{\partial^2\Omega_0(y,M)}{\partial M^2}\Bigg|_{y=0}.
\end{eqnarray}
One can check the above results from the explicit forms of ${\bf
M}_{13}$ and ${\bf M}_{33}$ in Appendix \ref{app} directly. Thus we
have
\begin{eqnarray}
-\frac{H_0^2}{{\cal D}_\sigma^{*-1}(0,{\bf 0})}=\beta_2.
\end{eqnarray}

According to the above relations, Eq. (\ref{dispersion}) can be
reduced to
\begin{eqnarray}
&& 3\beta^2|\Delta|^4+2\beta|\Delta|^2[{\cal D}_{\text
d}^{-1}(\omega,{\bf
q})+{\cal D}_{\text d}^{-1}(-\omega,{\bf q})]\nonumber\\
+&&{\cal D}_{\text d}^{-1}(\omega,{\bf q}){\cal D}_{\text
d}^{-1}(-\omega,{\bf q})=0.
\end{eqnarray}
It is evident that only the coefficient $\beta$ appears in the final
equation. Further, in the nonrelativistic limit $\omega,|{\bf
q}|\ll m_\pi$ and near the quantum phase transition point, ${\cal
D}_{\text d}^{-1}(\omega,{\bf q})$ can be approximated as
\begin{eqnarray}
{\cal D}_{\text d}^{-1}(\omega,{\bf q})\simeq -2m_\pi {\cal
J}\left(\omega-\frac{{\bf q}^2}{2m_\pi}+\mu_{\text d}\right).
\end{eqnarray}
Together with the mean-field results for the chemical potential
$\mu_{\text d}=g_0|\psiup_0|^2=\beta|\Delta|^2/(2m_\pi {\cal J})$
and for the baryon density $n=|\psiup_0|^2$, we finally get the
Bogoliubov dispersion (\ref{bog}).

We here emphasize that the mixing between the sigma meson
and the diquarks, denoted by the terms $\Delta H_0$ and
$\Delta^\dagger H_0$, plays an important role in recovering the
correct Bogoliubov dispersion. Even though we do get this
dispersion, we find that the procedure is quite different to the standard
theory of weakly interacting Bose gas
\cite{Bose01,GP01,GP02}. There, the elementary excitation is
given only by the diquark-diquark sectors, i.e.,
\begin{equation}
\det\left(\begin{array}{cc} {\bf M}_{11}(Q)&{\bf M}_{12}(Q)\\
{\bf M}_{21}(Q)&{\bf M}_{22}(Q)\end{array}\right)=0\ \ \ \ \ \ \ \Longrightarrow\nonumber
\end{equation}
\begin{eqnarray}
&& \det\left(\begin{array}{cc} -\omega+\frac{{\bf q}^2}{2m_\pi}-\mu_{\text d}+2g_0|\psiup_0|^2&g_0|\psiup_0|^2\\
g_0|\psiup_0|^2&\omega+\frac{{\bf q}^2}{2m_\pi}-\mu_{\text
d}+2g_0|\psiup_0|^2\end{array}\right)\nonumber\\
&&=0.
\end{eqnarray}
But in our case, we cannot get the correct Bogoliubov excitation if
we simply set $H_0=0$ and consider only the diquark-diquark sector.
In fact, this requires $A_0=2B_0=2\beta$ which is not true in our
case.

One can also check how the momentum dependence of $A,B,H$ and ${\cal
D}_\sigma^{*-1}$ modifies the dispersion. This needs direct
numerical solution of the equation $\det{\bf M}(\omega,{\bf q})=0$.
We have examined that for $|\mu_{\text B}-m_\pi|$ up to $0.01m_\pi$,
the numerical result agrees well with the Bogoliubov formula
(\ref{bog}). However, at higher density, a significant deviation is
observed. This is in fact a signature of BEC-BCS crossover which
will be discussed later.

Up to now we have studied the properties of the dilute Bose
condensate induced by a small diquark condensate $\langle
qq\rangle$. The chiral condensate $\langle\bar{q}{q}\rangle$ will also be
modified in the medium. In such a dilute Bose condensate, we can
study the response of the chiral condensate to the baryon density
$n$.

\begin{figure*}
\begin{center}
\includegraphics[width=7.5cm]{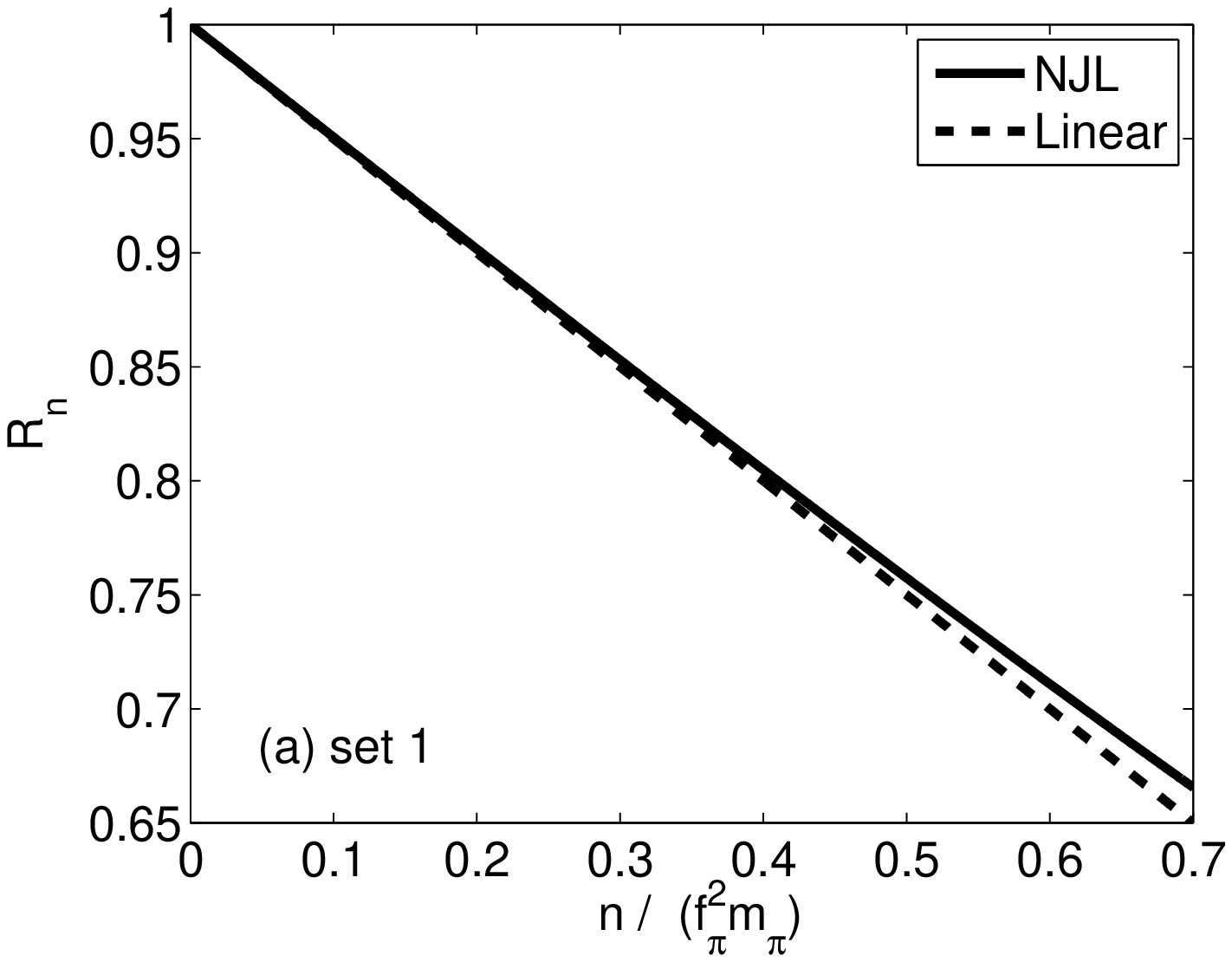}
\includegraphics[width=7.5cm]{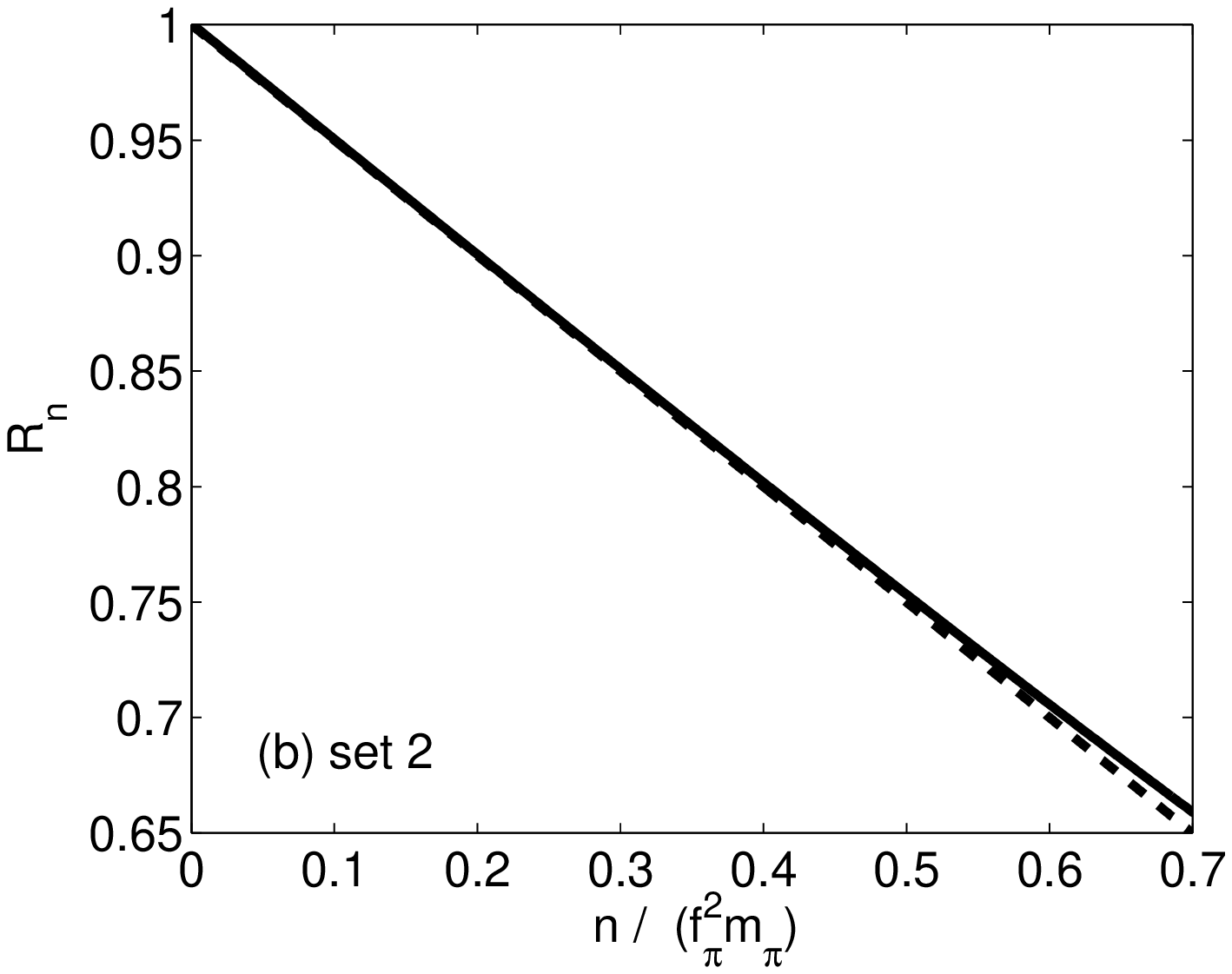}
\includegraphics[width=7.5cm]{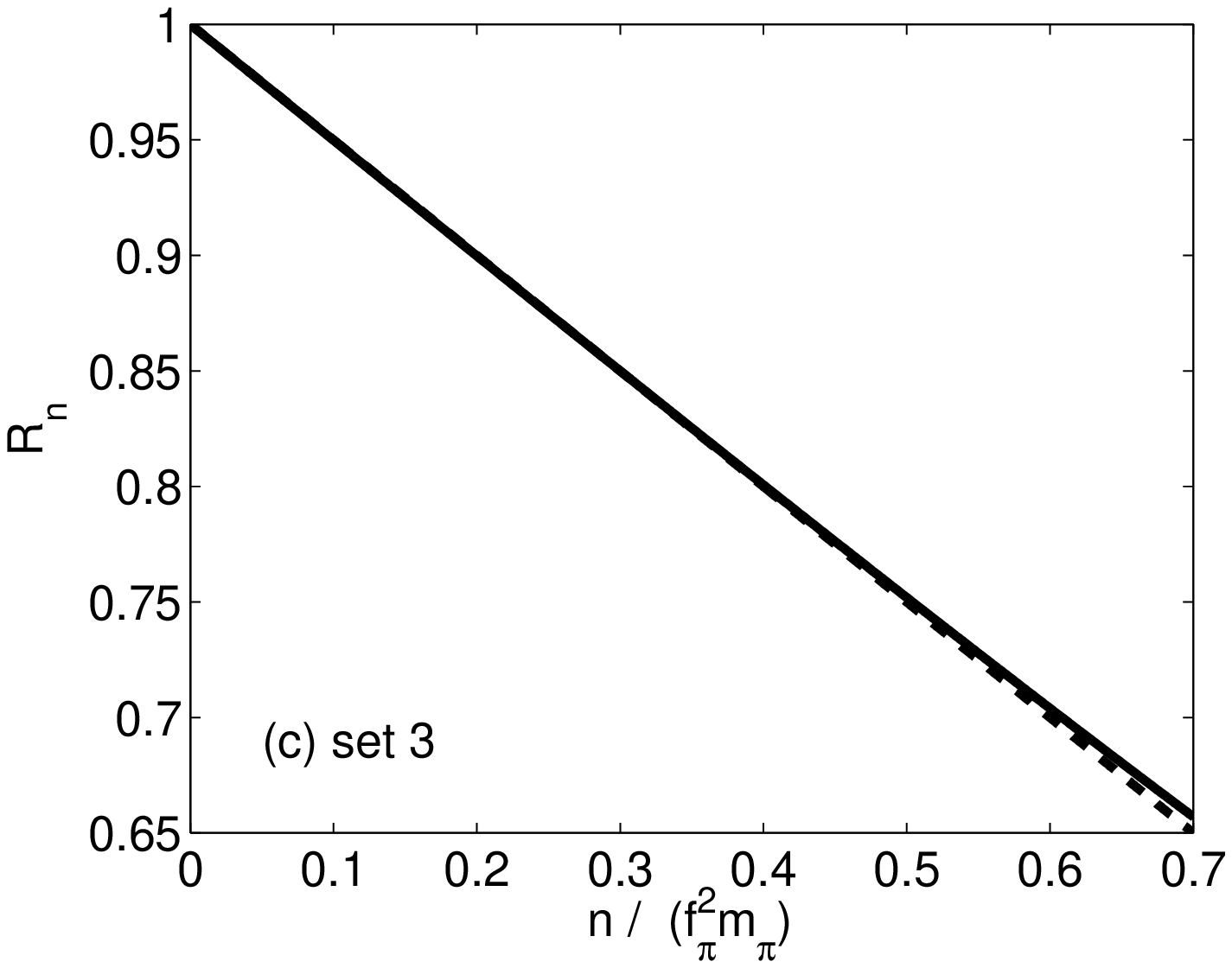}
\includegraphics[width=7.5cm]{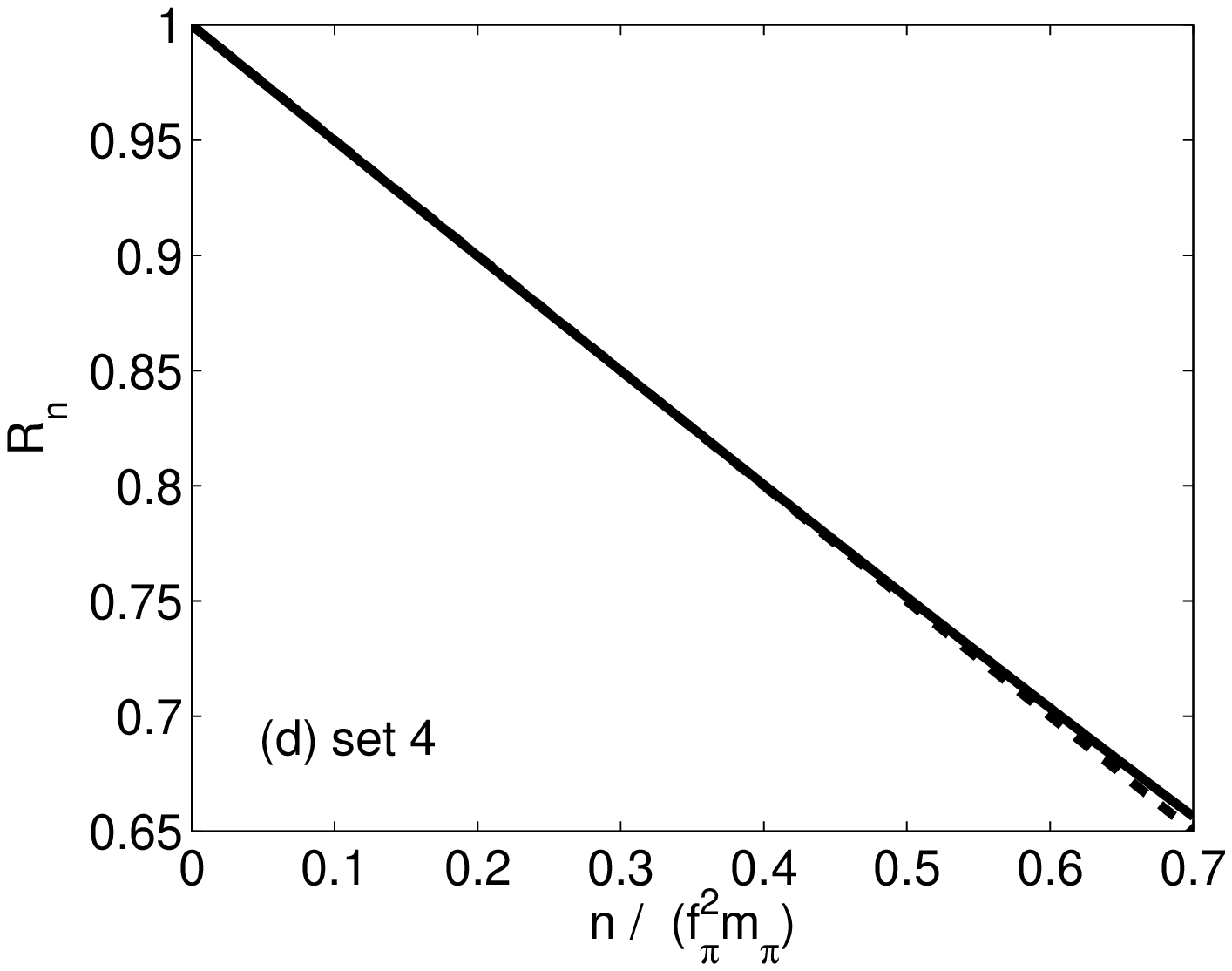}
\caption{The ratio
$R_n=\langle\bar{q}{q}\rangle_n/\langle\bar{q}{q}\rangle_0$ as a
function of $n/(f_\pi^2m_\pi)$ for different model parameter sets.
The dashed line is the linear behavior given by (\ref{linear}).
\label{fig2}}
\end{center}
\end{figure*}

To this end, we expand the effective quark mass $M$ in terms of
$y=|\Delta|^2$. We have
\begin{equation}
M-M_*=\frac{dM}{dy}\bigg|_{y=0}y+O(y^2)
\end{equation}
The expansion coefficient can be approximated as
\begin{eqnarray}
\frac{dM}{dy}\bigg|_{y=0}&\simeq&
-\frac{2f_\pi^2/M_*}{m_\pi^2f_\pi^2/M_*^2+4f_\pi^2}\nonumber\\
&=&-\frac{1}{2M_*}\left[1+O\left(\frac{m_\pi^2}{4M_*^2}\right)\right].
\end{eqnarray}
Using the definition of the effective quark mass,
$M=m_0-2G\langle\bar{q}{q}\rangle$, we find
\begin{equation}
\frac{\langle\bar{q}{q}\rangle_n}{\langle\bar{q}{q}\rangle_0}=1-\frac{|\Delta|^2}{4G\langle\bar{q}{q}\rangle_0
M_*}\simeq1-\frac{|\Delta|^2}{2M_*^2}.
\end{equation}
Since the baryon number density reads $n=|\psiup_0|^2=2m_\pi {\cal
J}|\Delta|^2$, using the fact that ${\cal J}\simeq
f_\pi^2/(2M_*^2)$, we obtain to leading order
\begin{equation}\label{linear}
\frac{\langle\bar{q}{q}\rangle_n}{\langle\bar{q}{q}\rangle_0}\simeq1-\frac{n}{2f_\pi^2
m_\pi}.
\end{equation}
This formula is in fact a two-color analogue of the density
dependence of the chiral condensate in the $N_c=3$ case, where we
have \cite{cohen,cohen2}
\begin{equation}
\frac{\langle\bar{q}{q}\rangle_n}{\langle\bar{q}{q}\rangle_0}\simeq1-\frac{\Sigma_{\pi
{\text N}}}{f_\pi^2 m_\pi^2}n
\end{equation}
with $\Sigma_{\pi \text{N}}$ being the pion-nucleon sigma term.
In Fig.\ref{fig2}, we show the numerical results via solving the
mean-field gap equations. One finds that the chiral condensate has a
perfect linear behavior at low density. For large value of $M_*$ (
and hence the sigma meson mass $m_\sigma$), the linear behavior
persists even at higher density.

In fact, the Eq. (\ref{linear}) can be obtained in a model
independent way. Applying the Hellmann-Feynman theorem to a dilute
diquark gas with energy density ${\cal E}(n)$ given by (\ref{eos}),
we can obtain (\ref{linear}) directly. According to the
Hellmann-Feynman theorem, we have
\begin{equation}
2m_0(\langle\bar{q}{q}\rangle_n-\langle\bar{q}{q}\rangle_0)=m_0\frac{d{\cal
E}}{dm_0}.
\end{equation}
The derivative $d{\cal E}/dm_0$ can be evaluated via the chain rule
$d{\cal E}/dm_0=(d{\cal E}/dm_\pi)(dm_\pi/dm_0)$. Together with the
Gell-Mann--Oakes--Renner relation
$m_\pi^2f_\pi^2=-m_0\langle\bar{q}{q}\rangle_0$ and the fact that
$da_{\text {dd}}/dm_\pi\simeq a_{\text{dd}}/m_\pi$, we can obtain to
leading order Eq. (\ref{linear}). Beyond the leading order,
we find that the correction of order $O(n^2)$ vanishes. Thus, the
next-to-leading order correction should be  $O(n^{5/2})$ coming from
the Lee-Huang-Yang correction to the equation of state~\cite{HFiso}.

Finally, we can show analytically that the ``chiral rotation"
behavior \cite{QC2D,QC2D1,QC2D2,QC2D3,QC2D4,QC2D5,QC2D6, QL03,
ratti, 2CNJL04, ISO} predicted by the chiral perturbation theories
is valid in the NJL model near the quantum phase transition. In the
chiral perturbation theories, the chemical potential dependence of
the chiral and diquark condensates can be analytically expressed as
\begin{equation}
\frac{\langle\bar{q}{q}\rangle_{\mu_{\text
B}}}{\langle\bar{q}{q}\rangle_0}=\frac{m_\pi^2}{\mu_{\text B}^2},\ \
\ \ \frac{\langle q{q}\rangle_{\mu_{\text
B}}}{\langle\bar{q}{q}\rangle_0}=\sqrt{1-\frac{m_\pi^4}{\mu_{\text
B}^4}}.\label{chpt}
\end{equation}
Near the phase transition point, we can expand the above formula in
powers of $\mu_{\text d}=\mu_{\text B}-m_\pi$. To leading order, we
have
\begin{equation}
\frac{\langle\bar{q}{q}\rangle_{\mu_{\text
B}}}{\langle\bar{q}{q}\rangle_0}\simeq1-\frac{2\mu_{\text
d}}{m_\pi},\ \ \ \ \frac{\langle q{q}\rangle_{\mu_{\text
B}}}{\langle\bar{q}{q}\rangle_0}\simeq 2\sqrt{\frac{\mu_{\text
d}}{m_\pi}}.
\end{equation}
Using the mean-field result (\ref{chemi}) for the chemical potential
$\mu_{\text d}$, one can easily check that the above relations are
also valid in our NJL model.

In the above studies we focused on the ``physical point" with
$m_0\neq 0$. In the final part of this section, we briefly discuss
the chiral limit with $m_0=0$.

We may naively expect that the results at $m_0\neq0$ can be directly
generalized to the chiral limit via setting $m_\pi=0$. The ground
state is a noninteracting Bose condensate of massless diquarks, due
to $m_\pi=0$ and $a_{\text{dd}}=0$. However, this cannot be true
since many divergences develop due to the vanishing pion mass. In
fact, the conclusion of second order phase transition is not correct
since the Ginzburg-Landau coefficient $\beta$ vanishes. Instead, the
superfluid phase transition is of strongly first order in the chiral
limit~\cite{ratti,ISOother02,ISOother021}.

\begin{figure*}
\begin{center}
\includegraphics[width=7.5cm]{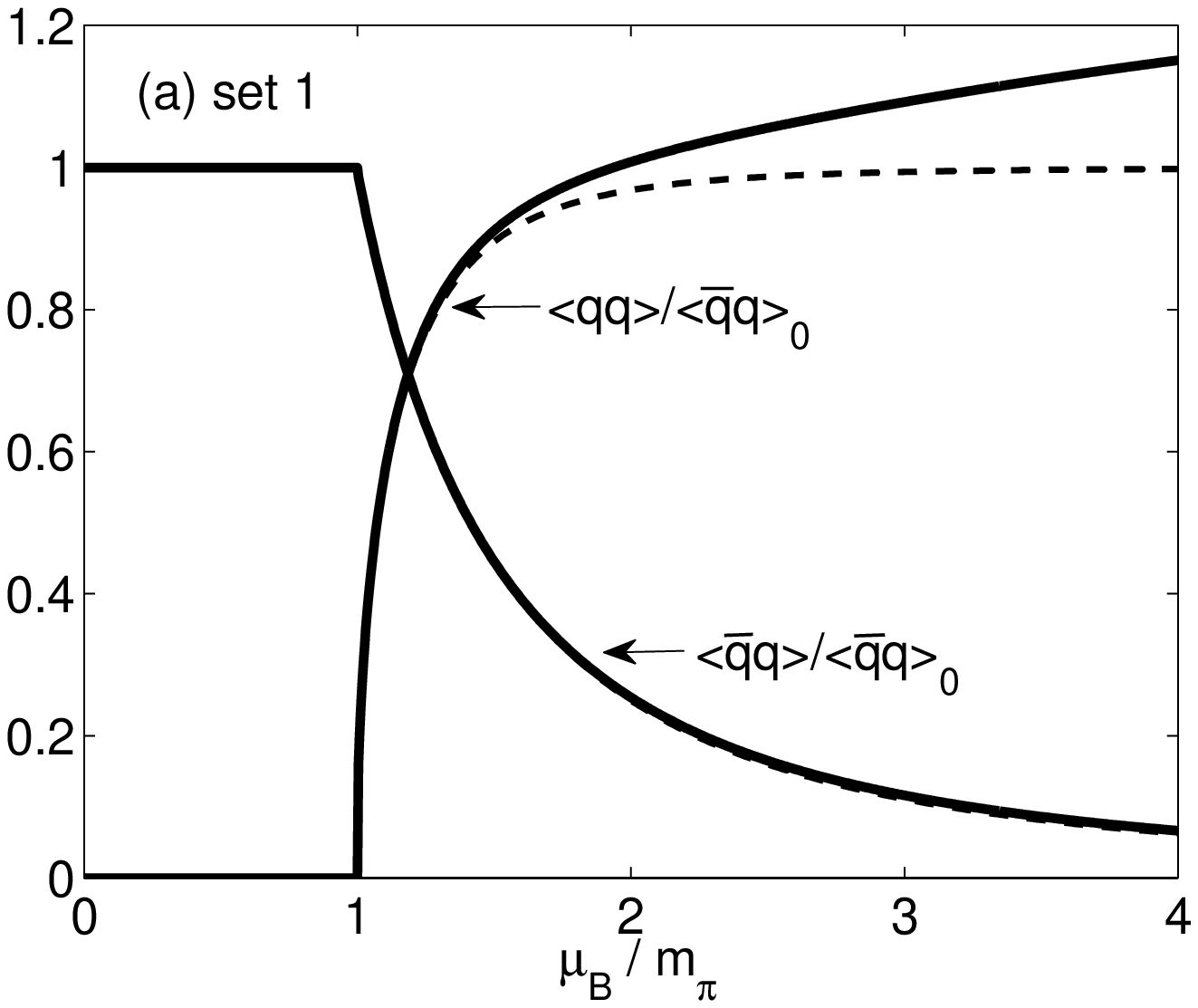}
\includegraphics[width=7.5cm]{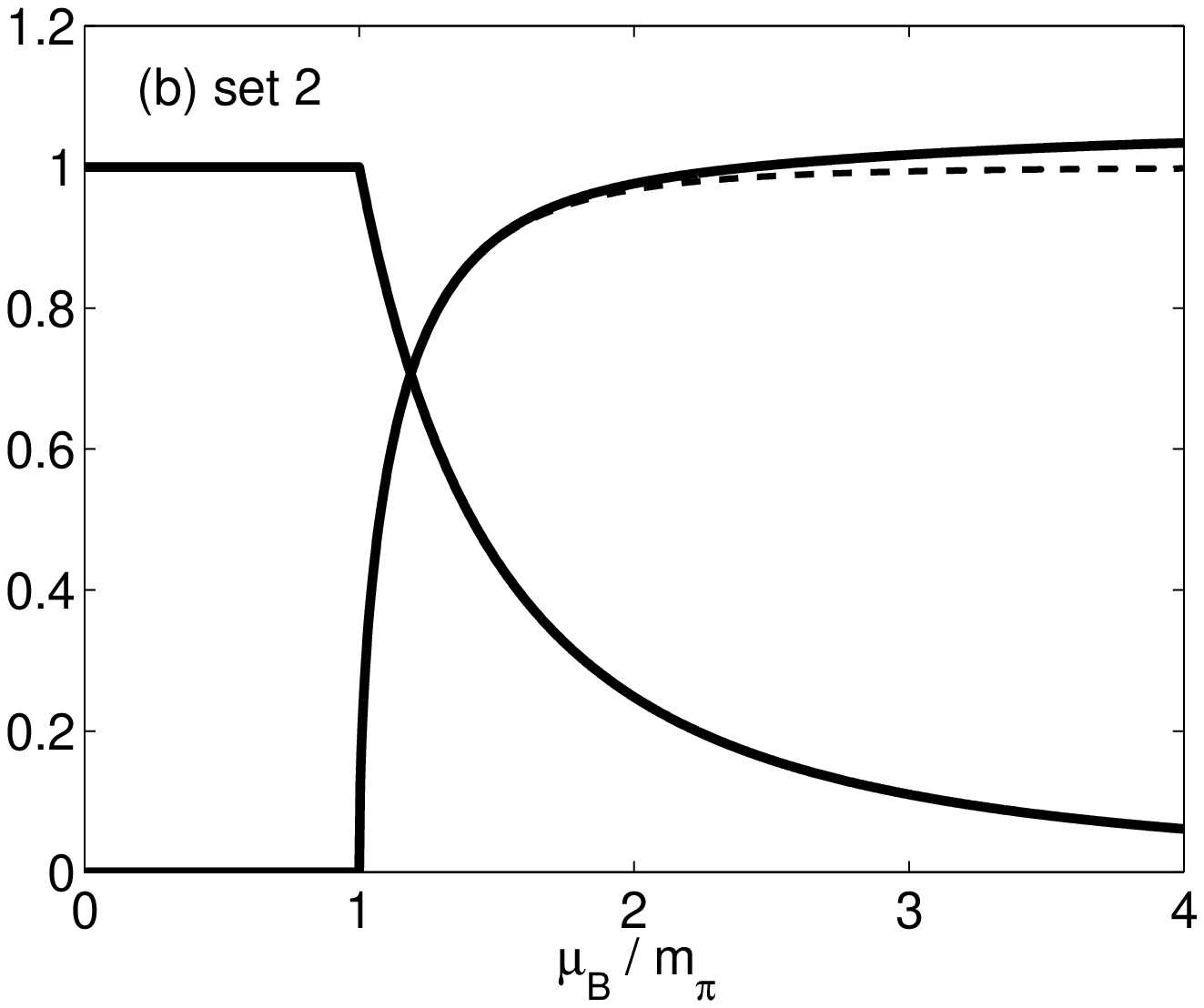}
\includegraphics[width=7.5cm]{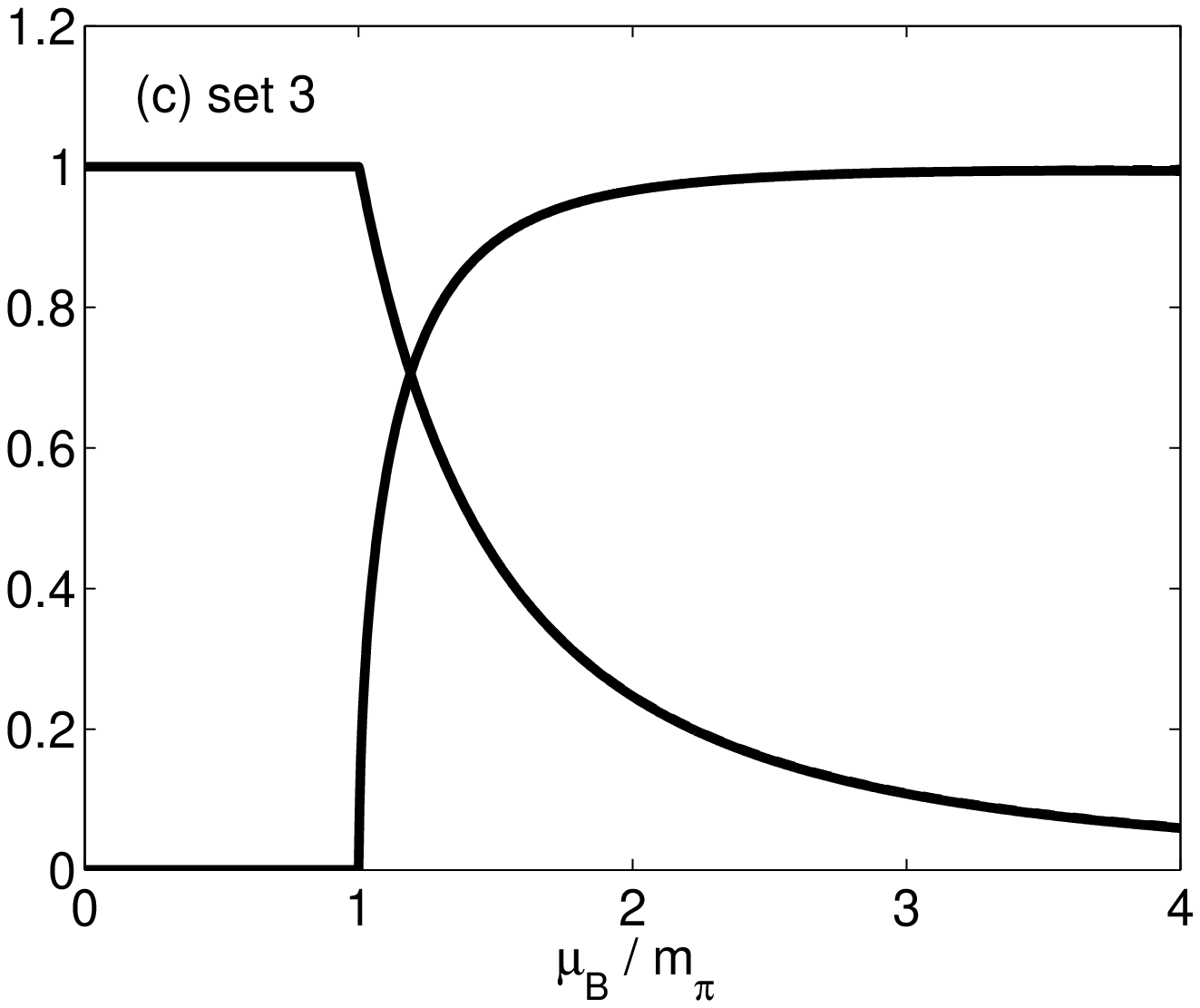}
\includegraphics[width=7.5cm]{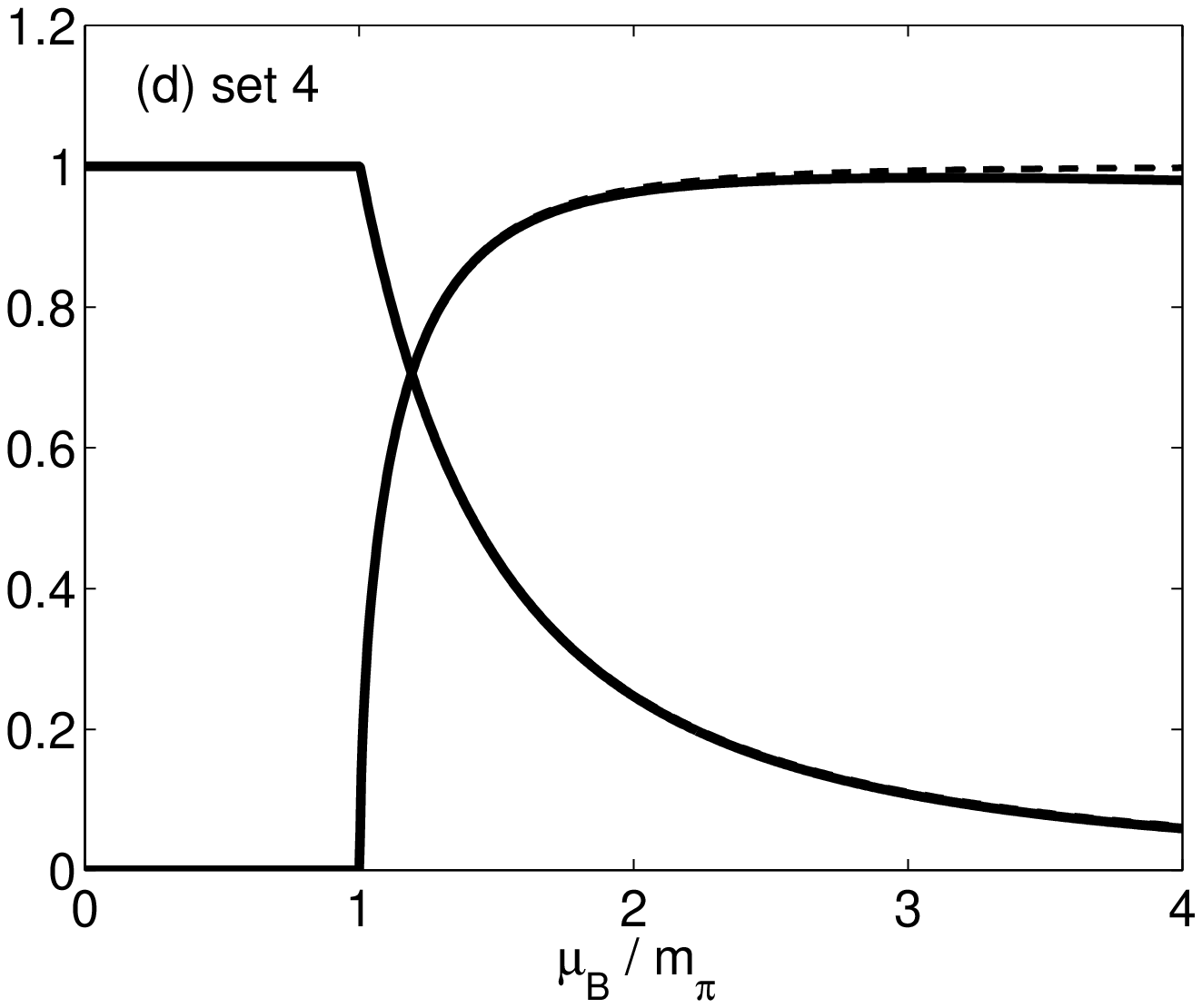}
\caption{The chiral and diquark condensates (in units of
$\langle\bar{q}{q}\rangle_0$) as functions of the baryon chemical
potential(in units of $m_\pi$) for different model parameter sets.
\label{fig3}}
\end{center}
\end{figure*}

In the chiral limit, the effective action in the vacuum should
depend only on the combination
$\sigma^2+\mbox{\boldmath{$\pi$}}^2+|\phi|^2$ due to the exact
flavor symmetry SU$(4)\simeq$ SO$(6)$. The vacuum is chosen to be
associated with a nonzero chiral condensate $\langle\sigma\rangle$
without loss of generality. At zero and at finite chemical
potential, the thermodynamic potential $\Omega_0(M,|\Delta|)$ has
two minima locating at $(M,|\Delta|)=(a,0)$ and
$(M,|\Delta|)=(0,b)$. At zero chemical potential, these two minima
are degenerate due to the exact flavor symmetry. However, at nonzero
chemical potential (even arbitrarily small), the minimum $(0,b)$ has
the lowest free energy. Analytically, we can show $b\rightarrow
M_*$ at $\mu_{\text B}=0^+$. This means that the superfluid phase
transition in the chiral limit is of strongly first order, and takes
place at arbitrarily small chemical potential. Since the effective
quark mass $M$ keeps vanishing in the superfluid phase, a low
density Bose condensate does not exist in the chiral limit.

\subsection {Matter at High Density: BEC-BCS crossover and Mott Transition}
\label{s3-4}
The investigations in the previous subsection are restricted near
the quantum phase transition point $\mu_{\text B}=m_\pi$. Generally
the state of matter at high density should not be a relativistic
Bose condensate described by (\ref{RGP}). In fact, perturbative QCD
calculations show that the matter is a weakly coupled BCS superfluid
at asymptotic
density~\cite{pQCD,pQCD1,pQCD2,pQCD3,pQCD4,pQCD5,pQCD6}. In this
section, we will discuss the evolution of the superfluid matter as
the baryon density increases from the NJL model point of view.

The numerical results for the chiral condensate
$\langle\bar{q}q\rangle$ and diquark condensate $\langle qq\rangle$
are shown in Fig.\ref{fig3}. As a comparison, we also show the
analytical result (\ref{chpt}) predicted by the chiral perturbation
theories (dashed lines). Since both the NJL model and chiral perturbation theories are equivalent realization of chiral symmetry as an effective low-energy theory of QCD, the behavior of the chiral condensate is almost the same in the two cases. However, for the diquark
condensate, there is quantitative difference at large chemical potential. To understand
this deviation, we compare the linear sigma model and its limit of infinite sigma mass. The former is similar to the NJL model with a finite sigma mass, and the latter corresponds to the chiral perturbation theories. We consider the O$(6)$ linear sigma model~\cite{2CNJL04}
\begin{equation}
{\cal L}_{\text{LSM}}=\frac{1}{2}(\partial_\mu
\mbox{\boldmath{$\varphi$}})^2-\frac{1}{2}m^2\mbox{\boldmath{$\varphi$}}^2+\frac{1}{4}\lambda\mbox{\boldmath{$\varphi$}}^4-H\sigma
\end{equation}
with $\mbox{\boldmath{$\varphi$}}=(\sigma, \mbox{\boldmath{$\pi$}},
\phi_1,\phi_2)$ and $m^2<0$. The model parameters $m^2,\lambda,H$
can be determined from the vacuum phenomenology. In this model, we
can show that the chiral and diquark condensates are given by
\begin{eqnarray}
&&\frac{\langle\bar{q}{q}\rangle_{\mu_{\text
B}}}{\langle\bar{q}{q}\rangle_0}=\frac{m_\pi^2}{\mu_{\text
B}^2},\nonumber\\
&&\frac{\langle q{q}\rangle_{\mu_{\text
B}}}{\langle\bar{q}{q}\rangle_0}=\sqrt{1-\frac{m_\pi^4}{\mu_{\text
B}^4}+2\frac{\mu_{\text
B}^2-m_\pi^2}{m_\sigma^2-m_\pi^2}}.\label{lsm}
\end{eqnarray}
In the limit $m_\sigma\rightarrow\infty$, the
above results are indeed reduced to the result (\ref{chpt})
of chiral perturbation theories. However, for finite
values of $m_\sigma$, the difference between the two might be remarkable at large chemical potential.

While the Ginzburg-Landau free energy can be reduced to the
Gross-Pitaevskii free energy near the quantum phase transition
point, it is not the case at arbitrary $\mu_{\text B}$.

\begin{figure}[!htb]
\begin{center}
\includegraphics[width=8.5cm]{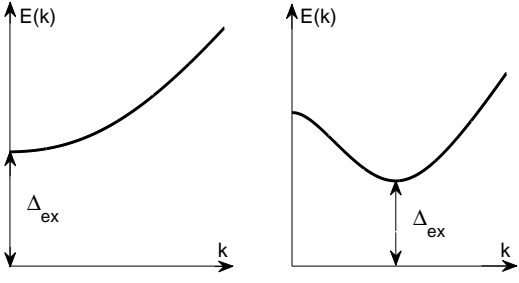}
\caption{A schematic plot of the fermionic excitation spectrum in
the BEC state (left) and the BCS state (right). \label{fig4}}
\end{center}
\end{figure}

When $\mu_{\text B}$ increases, we find that the fermionic
excitation spectra $E_{\bf k}^\pm$ undergo a characteristic change.
Near the quantum phase transition $\mu_{\text B}=m_\pi$, they are
nearly degenerate, due to $m_\pi\ll 2M_*$ and their minima located
at $|{\bf k}|=0$. However, at very large $\mu_{\text B}$, the minimum
of $E_{\bf k}^-$ moves to $|{\bf k}|\simeq \mu_{\text B}/2$ from
$M\rightarrow m_0$. Meanwhile the excitation energy of the
anti-fermion excitation become much larger than that of the fermion
excitation and can be neglected. This characteristic change of the
fermionic excitation spectra takes place when the minimum of the
lowest band excitation $E_{\bf k}^-$ moves from $|{\bf k}|=0$ to
$|{\bf k}|\neq0$, i.e., $\mu_{\text B}/2=M(\mu_{\text B})$. A schematic
plot of this characteristic change is shown in Fig.\ref{fig4}. The
equation $\mu_{\text B}/2=M(\mu_{\text B})$ defines the so-called
crossover point $\mu_{\text B}=\mu_0$ which can be numerically
determined by the mean-field gap equations. The numerical results of
the crossover chemical potential $\mu_0$ for the four model
parameter sets are shown in Table.\ref{crossover}. For reasonable
parameter sets, the crossover chemical potential is in the range
$(1.6-2)m_\pi$. We notice that this crossover chemical potential agrees
with the result from lattice simulation~\cite{LBECBCS}.
\begin{table}[b!]
\begin{center}
\begin{tabular}{|c|c|c|c|c|}
  \hline
  Set & 1 & 2 & 3 & 4 \\
  \hline
  chemical potential $\mu_0$  & 1.65 & 1.81 & 1.95 & 2.07 \\
  \hline
\end{tabular}
\end{center}
\caption{\small The crossover chemical potential $\mu_0$ (in units
of $m_\pi$) for different model parameter sets. } \label{crossover}
\end{table}

In fact, an analytical expression for $\mu_0$ can be achieved
according to the fact that the chiral rotation behavior
$\langle\bar{q}q\rangle_{\mu_{\text B}}/\langle\bar{q}q\rangle_0\simeq m_\pi^2/\mu_{\text B}^2$ is still
valid in the NJL model at large chemical potentials as shown in
Fig.\ref{fig3}. We obtain
\begin{eqnarray}
\frac{\mu_0}{2}\simeq\frac{m_\pi^2}{\mu_0^2}M_*\ \ \Longrightarrow \
\ \mu_0\simeq (2M_* m_\pi^2)^{1/3}. \label{crossmu}
\end{eqnarray}
Using the fact that $m_\sigma\simeq 2M_*$, we find that $\mu_0$ can
be expressed as
\begin{eqnarray}
\frac{\mu_0}{m_\pi}\simeq\left(\frac{m_\sigma}{m_\pi}\right)^{1/3}.
\end{eqnarray}
Thus, in the nonlinear sigma model limit
$m_\sigma/m_\pi\rightarrow\infty$, there should be no BEC-BCS
crossover. On the other hand, this means that the physical prediction
power of the chiral perturbation theories is restricted near the
quantum phase transition point.

\begin{figure*}
\begin{center}
\includegraphics[width=7cm]{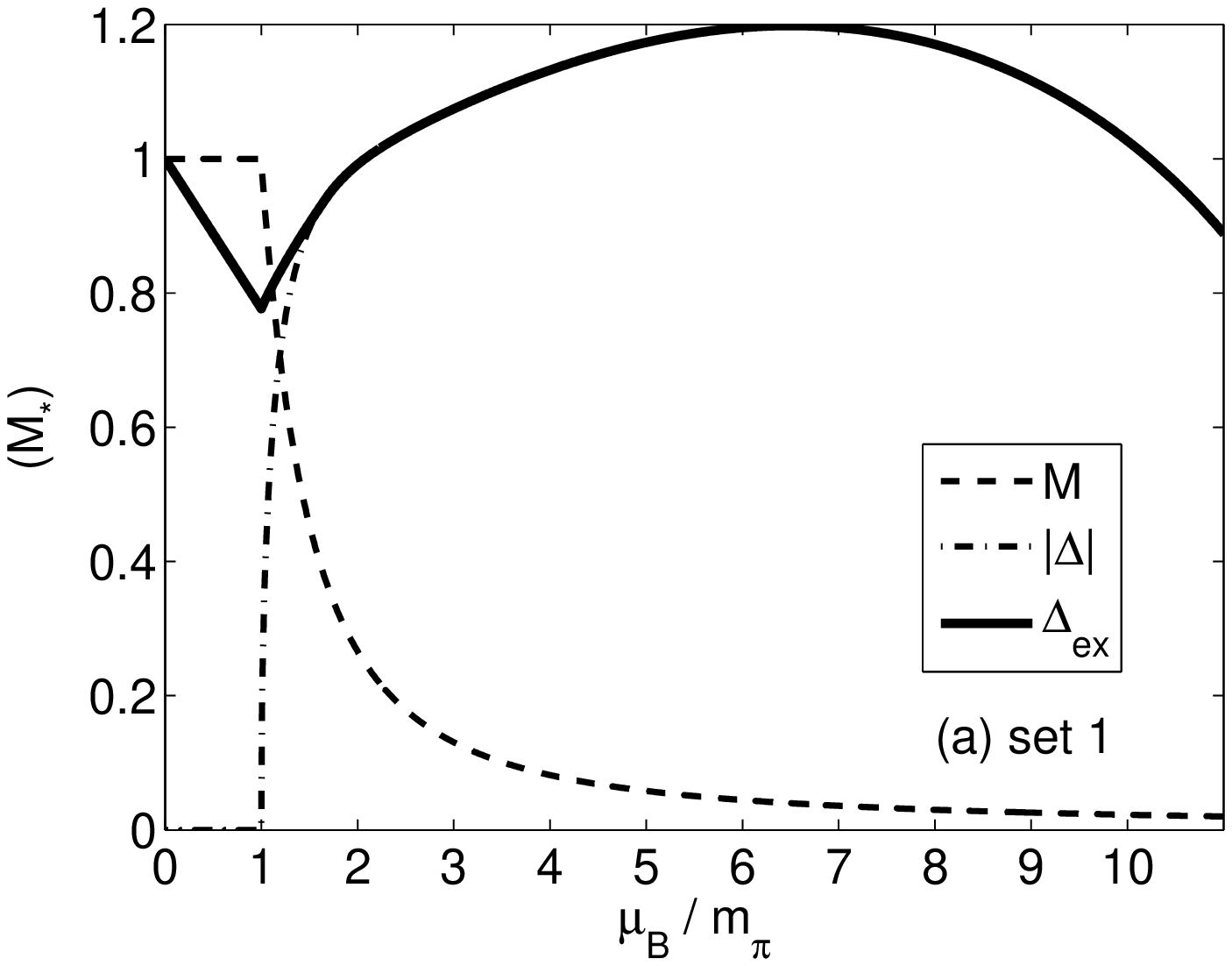}
\includegraphics[width=7cm]{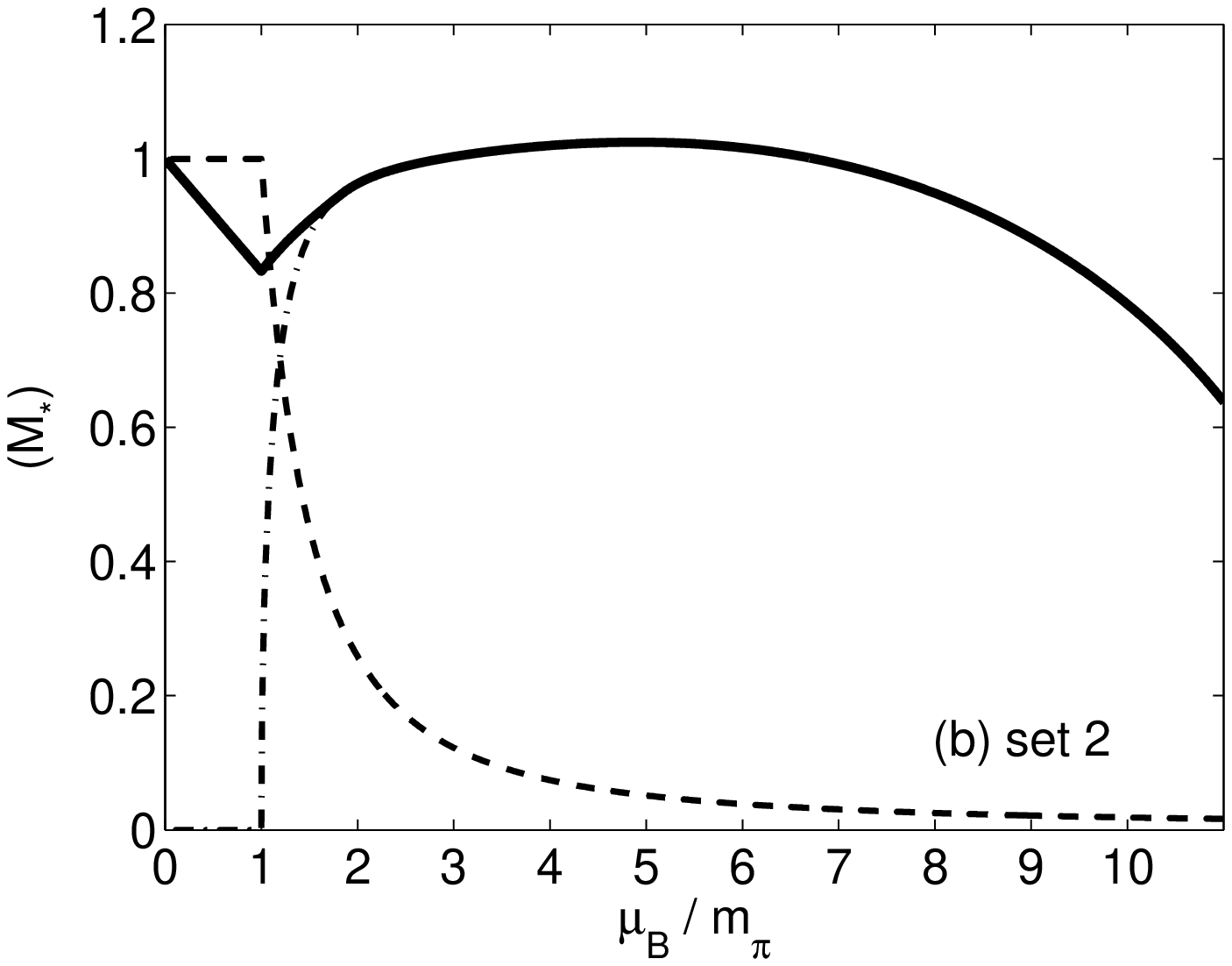}
\includegraphics[width=7cm]{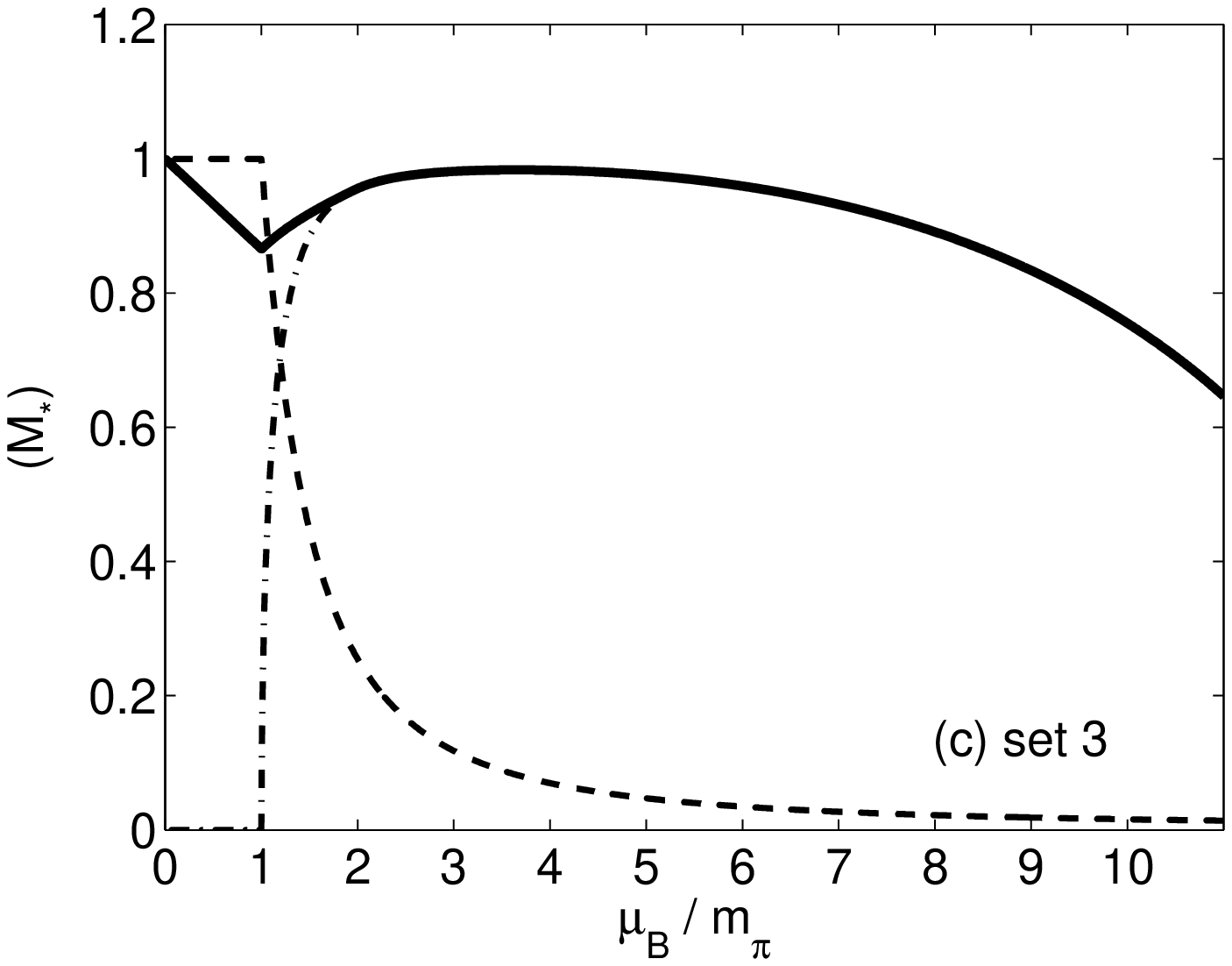}
\includegraphics[width=7cm]{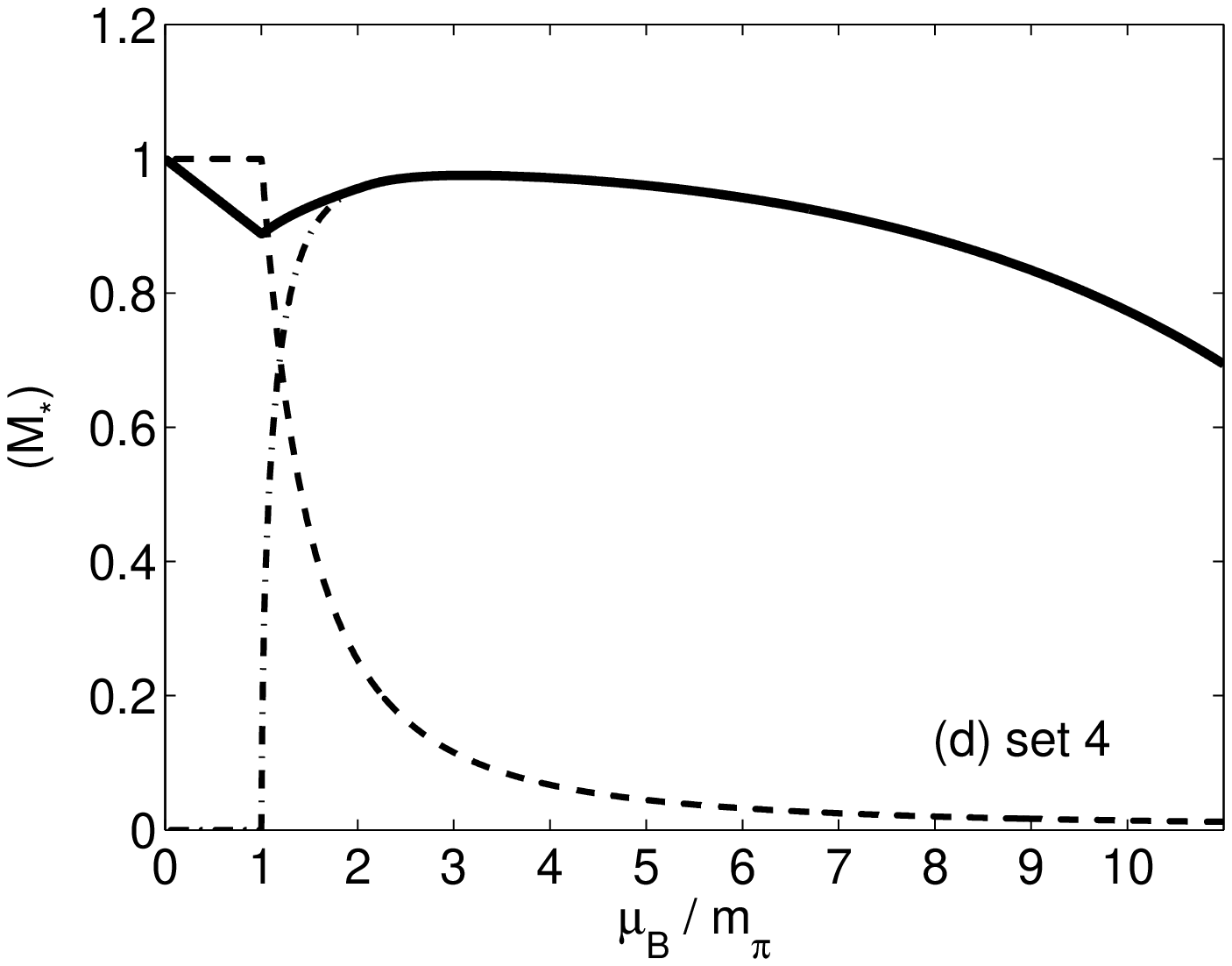}
\caption{The fermionic excitation gap $\Delta_{\text{ex}}$ (in units
of $M_*$) as a function of the baryon chemical potential (in units
of $m_\pi$) for different model parameter sets. The effective quark
mass $M$ and the pairing gap $|\Delta|$ are also shown by dashed and
dash-dotted lines, respectively. \label{fig5}}
\end{center}
\end{figure*}

The fermionic excitation gap $\Delta_{\text{ex}}$ (as shown in
Fig.\ref{fig4}), defined as the minimum of the fermionic excitation
energy, i.e., $\Delta_{\text{ex}}=\min_{\bf k}\{E_{\bf k}^-,E_{\bf
k}^+\}$, can be evaluated as
\begin{equation}
\Delta_{\text{ex}}=
 \left\{ \begin{array}
{r@{\quad,\quad}l}
 \sqrt{(M-\frac{\mu_{\text B}}{2})^2+|\Delta|^2}&
 \mu_{\text B}<\mu_0 \\
 |\Delta| & \mu_{\text B}>\mu_0.
\end{array}
\right.
\end{equation}
It is evident that the fermionic excitation gap is equal to the
superfluid order parameter only in the BCS regime. This is similar
to the BEC-BCS crossover in nonrelativistic systems~\cite{BCSBEC3},
and we find that the corresponding fermion chemical potential $\mu_{\rm n}$
can be defined as $\mu_{\rm n}=\mu_{\text B}/2-M$. The numerical results of
the fermionic excitation gap $\Delta_{\text{ex}}$ for different
model parameter sets are shown in Fig.\ref{fig5}. We find that for a
wide range of the baryon chemical potential, it is of order
$O(M_*)$. The fermionic excitation gap is equal to the pairing gap
$|\Delta|$ only at the BCS side of the crossover, and exhibits a
minimum at the quantum phase transition point.

\begin{figure*}
\begin{center}
\includegraphics[width=7.5cm]{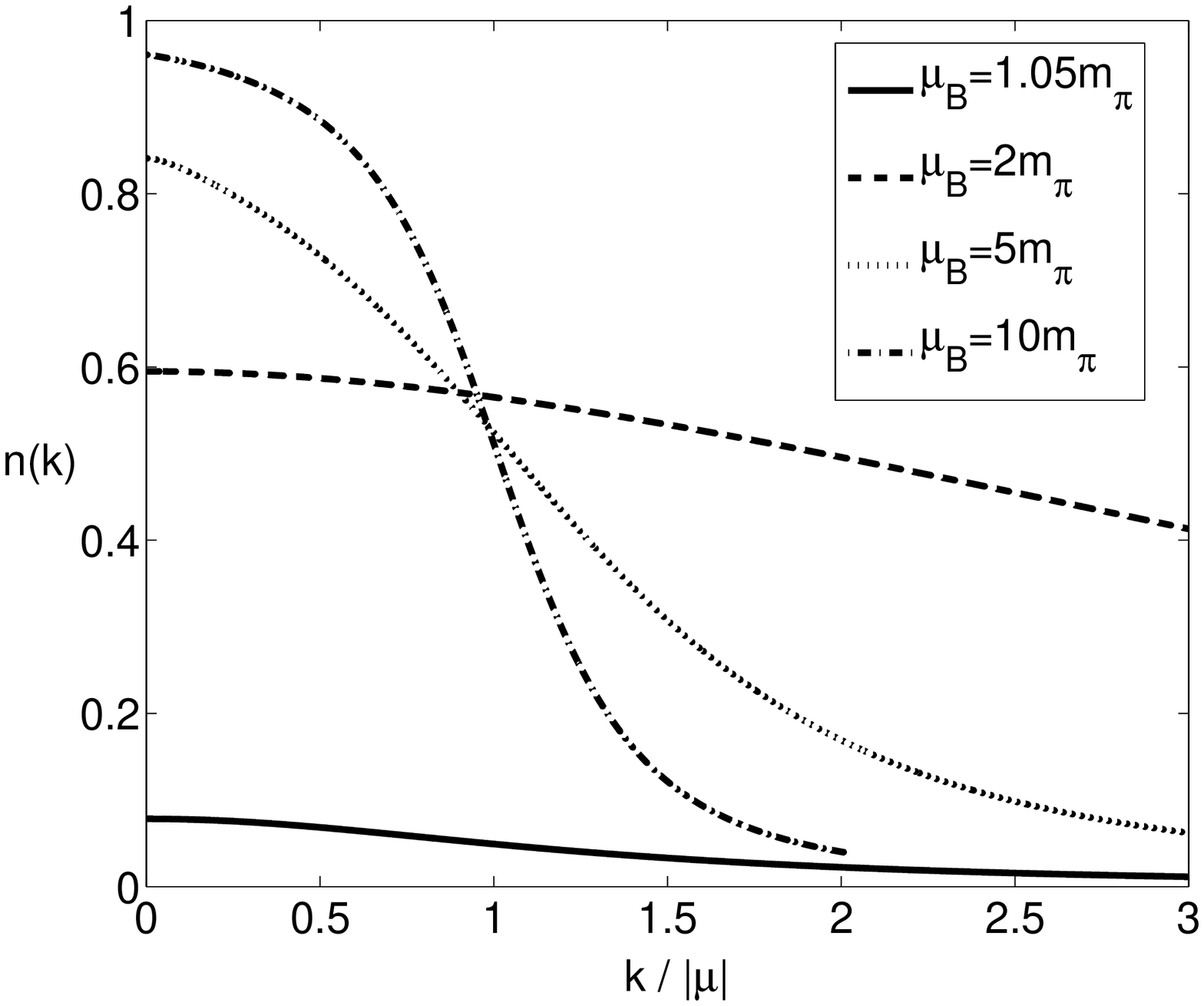}
\includegraphics[width=7.5cm]{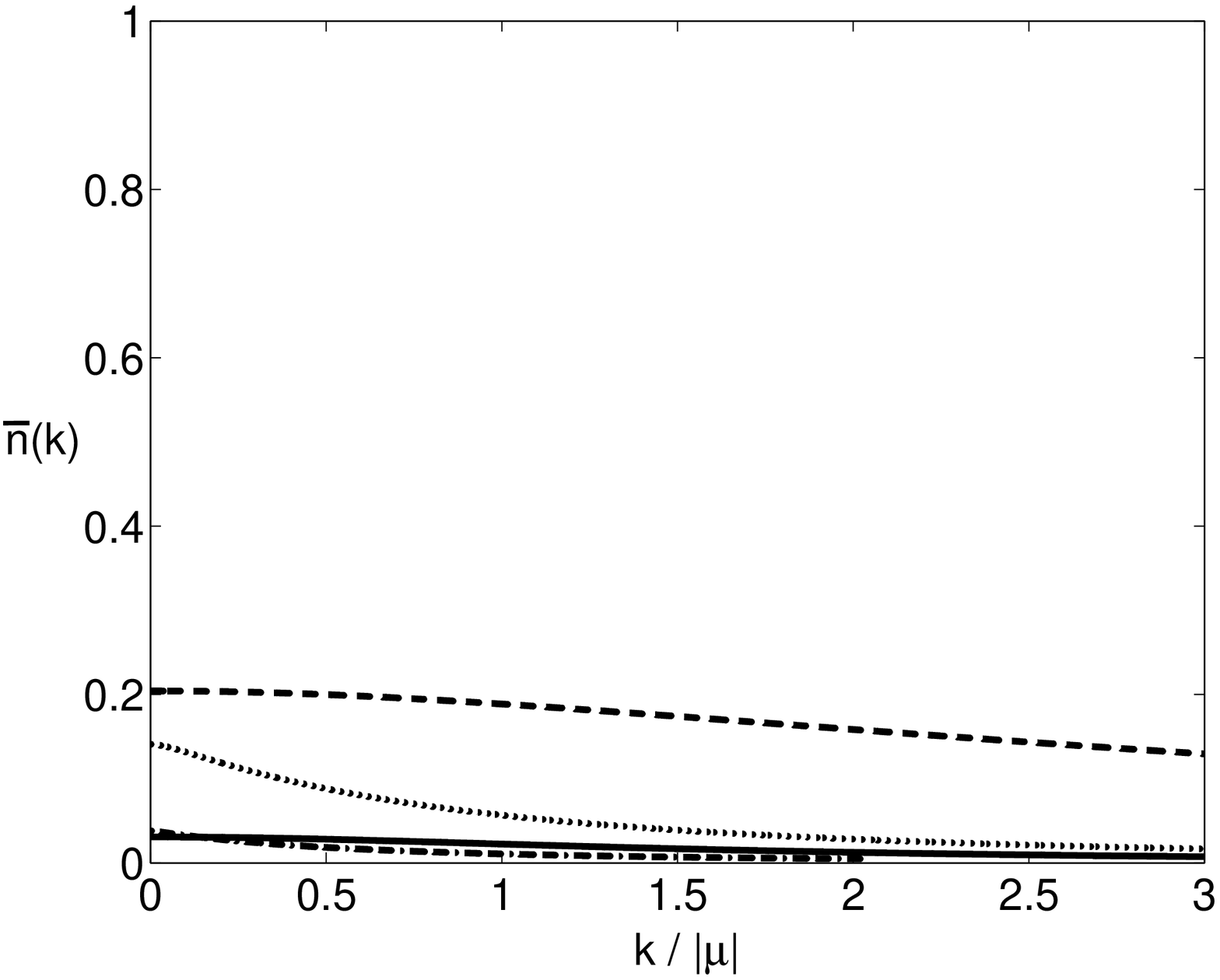}
\caption{The momentum distributions for quarks (upper panel) and
antiquarks (lower panel) for various values of $\mu_{\text B}$. The
momentum is scaled by $|\mu|=|\mu_{\text B}/2-M|$. \label{fig6}}
\end{center}
\end{figure*}

On the other hand, the momentum distributions of quarks (denoted by
$n({\bf k})$) and antiquarks (denoted by $\bar{n}({\bf k})$) can be
evaluated using the quark Green function ${\cal G}_{11}(K)$. We
obtain
\begin{eqnarray}
&&n({\bf k})=\frac{1}{2}\left(1-\frac{\xi_{\bf k}^-}{E_{\bf
k}^-}\right),\ \ \ \ \text{for quarks},\nonumber\\
&&\bar{n}({\bf k})=\frac{1}{2}\left(1-\frac{\xi_{\bf k}^+}{E_{\bf
k}^+}\right),\ \ \ \ \text{for antiquarks}.
\end{eqnarray}
The numerical results for $n({\bf k})$ and $\bar{n}({\bf k})$ (for
model parameter set 1) are shown in Fig.\ref{fig6}. Near the quantum
phase transition point, the quark momentum distribution $n({\bf k})$
is a very smooth function in the whole momentum space. In the
opposite limit, i.e., at large chemical potentials, it approaches
unity at $|{\bf k}|=0$ and decreases rapidly around the effective
``Fermi surface" at $|{\bf k}|\simeq|\mu|$. For the antiquarks, we
find that the momentum distribution $\bar{n}({\bf k})$ exhibits a
nonmonotonous behavior: it is suppressed at both low and high
densities and is visible only at moderate chemical potentials.
However, even at very large chemical potentials, e.g., $\mu_{\text
B}=10m_\pi$, the momentum distribution $n({\bf k})$ does not
approach the standard BCS behavior, which means that the dense matter is
not a weakly coupled BCS superfluid for a wide range of the baryon
chemical potential. Actually, at $\mu_{\text B}\simeq 10m_\pi$, the
ratio $|\Delta|/\mu$ is about $0.5$, which means that the dense
matter is still a strongly coupled superfluid.


The Goldstone mode also undergoes a characteristic change in the
BEC-BCS crossover. Near the quantum phase transition point, i.e., in
the dilute limit, the Goldstone mode recovers the Bogoliubov
excitation of weakly interacting Bose condensates.  In the opposite
limit, we expect the Goldstone mode approaches the
Anderson-Bogoliubov mode of a weakly coupled BCS superfluid, which
takes a dispersion $\omega({\bf q})=|{\bf q}|/\sqrt{3}$ up to the
two-particle continuum $\omega\simeq 2|\Delta|$. In fact, at large
chemical potentials, we can safely neglect the mixing between the
sigma meson and diquarks. The Goldstone boson dispersion is thus
determined by the equation
\begin{equation}
\det\left(\begin{array}{cc} {\bf M}_{11}(Q)&{\bf M}_{12}(Q)\\
{\bf M}_{21}(Q)&{\bf M}_{22}(Q)\end{array}\right)=0 .
\end{equation}
Therefore, at very large chemical potentials where $|\Delta|/\mu$
becomes small enough, the Goldstone mode recovers the Anderson-Bogoliubov
mode of a weakly coupled BCS superfluid.

Finally, we should emphasize that the existence of a smooth
crossover from the Bose condensate to the BCS superfluid depends on
whether there exists a deconfinement phase transition at finite
$\mu_{\text B}$~\cite{LBECBCS,decon} and where it takes place.
Recent lattice calculation predicts a deconfinement crossover which
occurs at a baryon chemical potential larger than that of the
BEC-BCS crossover~\cite{LBECBCS}.

\begin{figure}[!htb]
\begin{center}
\includegraphics[width=7.5cm]{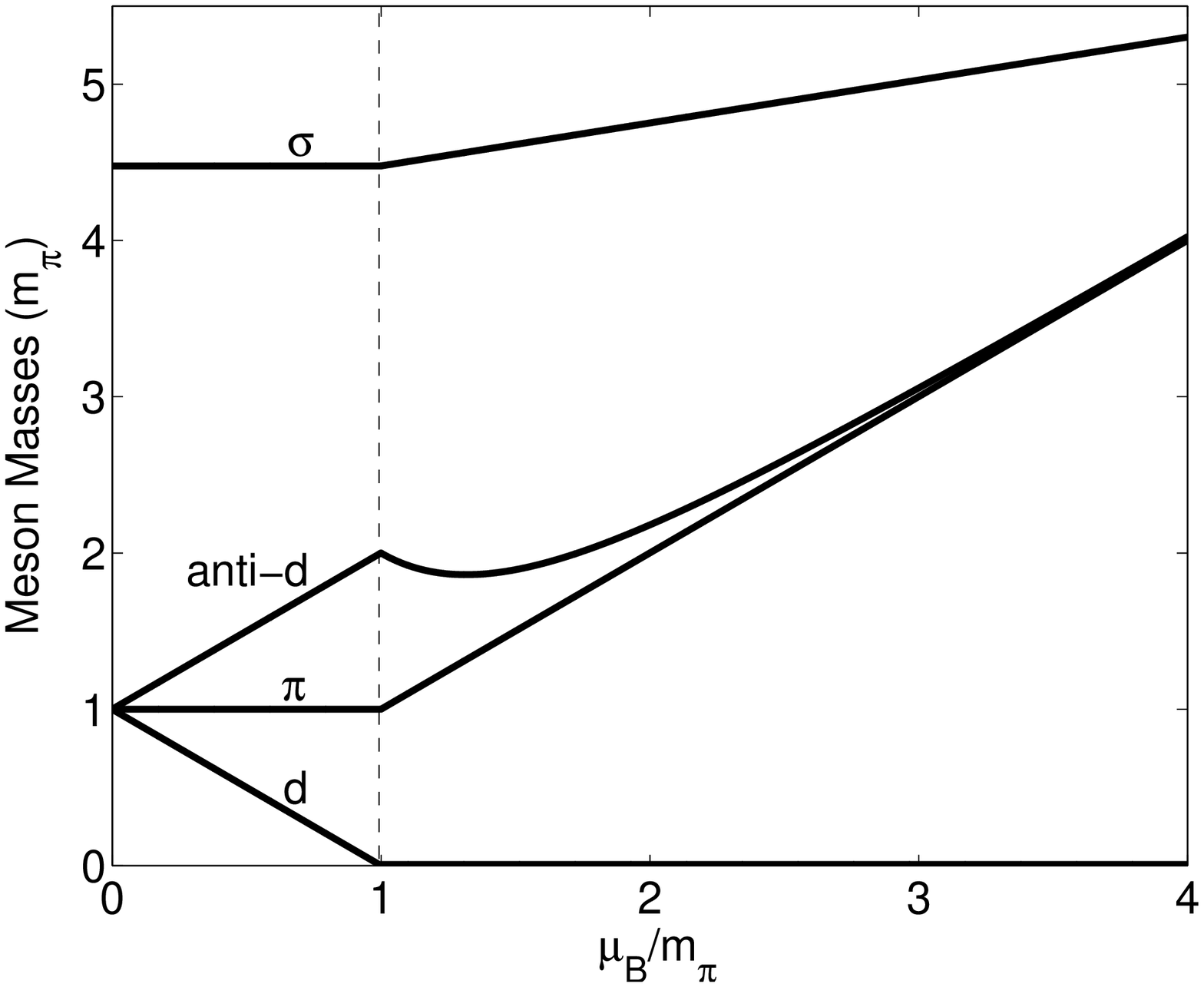}
\caption{The mass spectra of mesons and diquarks (in units of
$m_\pi$) as functions of the baryon chemical potential (in units of
$m_\pi$) for model parameter set 1. For other model parameter sets,
the mass of the heaviest mode is changed but others are almost the
same.\label{fig8}}
\end{center}
\end{figure}

As in real QCD with two quark flavors, we expect the chiral symmetry
is restored and the spectra of sigma meson and pions become
degenerate at high density \cite{note2}. For the two-flavor case and
with vanishing $m_0$, the residue SU$_{\text L}(2)\otimes$SU$_{\text
R}(2)\otimes$U$_{\text B}(1)$ symmetry group at $\mu_{\text B}\neq0$
is spontaneously broken down to Sp$_{\text L}(2)\otimes$Sp$_{\text
R}(2)$ in the superfluid medium with nonzero $\langle qq\rangle$, resulting in one
Goldstone boson. For small nonzero $m_0$, we expect the spectra of
sigma meson and pions become approximately degenerate when the
in-medium chiral condensate $\langle\bar{q}{q}\rangle$ becomes small
enough.

In fact, according to the result
$\langle\bar{q}{q}\rangle_n/\langle\bar{q}{q}\rangle_0\simeq1-n/(2f_\pi^2
m_\pi)$ at low density, we can roughly expect that the chiral
symmetry is approximately restored at $n\sim 2f_\pi^2 m_\pi$. From
the chemical potential dependence of the chiral condensate
$\langle\bar{q}{q}\rangle$ shown in Fig.\ref{fig3}, we find that it
becomes smaller and smaller as the density increases. As a result,
we should have nearly degenerate spectra for the sigma meson and
pions. To show this we need the explicit form of the matrix ${\bf
M}(Q)$ and ${\bf N}(Q)$ given in Appendix \ref{app}. From ${\bf
M}_{13}, {\bf M}_{32}\propto M\Delta$ at high density with
$\langle\bar{q}{q}\rangle\rightarrow 0$, they can be safely
neglected and the sigma meson decouples from the diquarks. The
propagator of the sigma meson is then given by ${\bf
M}_{33}^{-1}(Q)$. From the explicit form of the polarization
functions $\Pi_\sigma(Q)=\Pi_{33}(Q)$ and $\Pi_\pi(Q)$, we can see
that the inverse propagators of the sigma meson and pions differ
from each other in a term proportional to $M^2$. Thus at high
density their spectra are nearly degenerate, and their masses are
given by the equation
\begin{equation}
1-2G\Pi_\pi(\omega,{\bf 0})=0.
\end{equation}
Using the mean-field gap equation for $\Delta$, we find that the solution
is $\omega=\mu_{\text B}$, which means that the meson masses are equal to
$\mu_{\text B}$ at large chemical potentials. In Fig.\ref{fig8}, we
show the chemical potential dependence of the meson and diquark masses determined at zero momentum. We find
that the chiral symmetry is approximately restored at $\mu_{\text
B}\simeq 3m_\pi$, corresponding to $n\simeq 3.5f_\pi^2m_\pi$, where the $\pi$ and "anti-d" start to be degenerate.
In the normal phase with $\mu_{\text B}<m_\pi$, diquark, anti-diquark, $\pi$ and $\sigma$ themselves are eigen modes of the collective excitation
of the system, but in the symmetry breaking phase with $\mu_{\text B}>m_\pi$, except for $\pi$
which is still an eigen mode, diquark, anti-diquark and $\sigma$ are no longer
eigen modes~\cite{ratti,tomas}. However, if we neglect the mixing between $\sigma$ and anti-diquark in the symmetry breaking phase~\cite{hao}, $\sigma$ is still the eigen mode and it becomes degenerate with anti-diquark and $\pi$ at high enough chemical potential. This is clearly shown in Fig.15 of Ref.~\cite{ISOother03}.
\begin{figure*}
\begin{center}
\includegraphics[width=7.5cm]{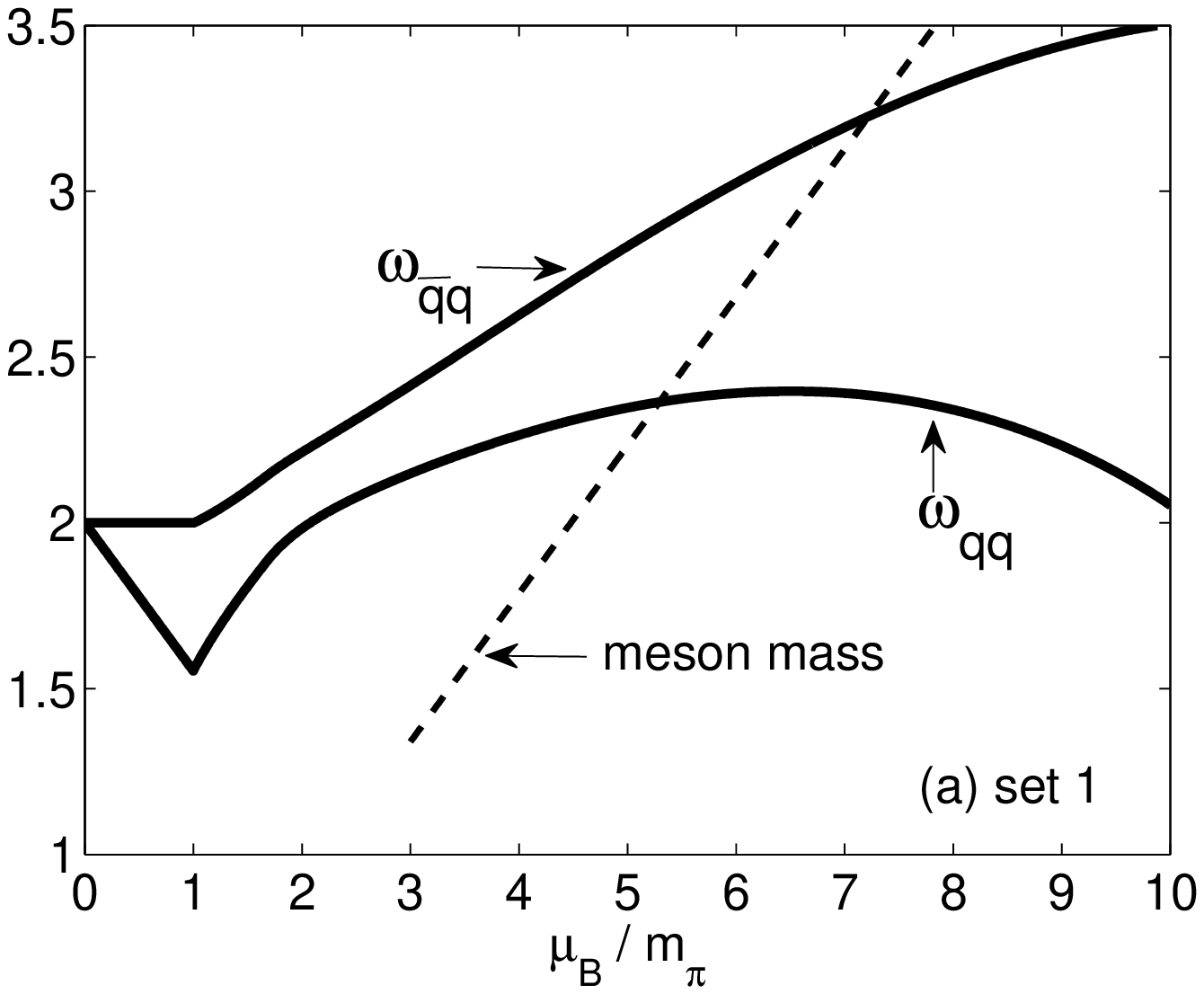}
\includegraphics[width=7.5cm]{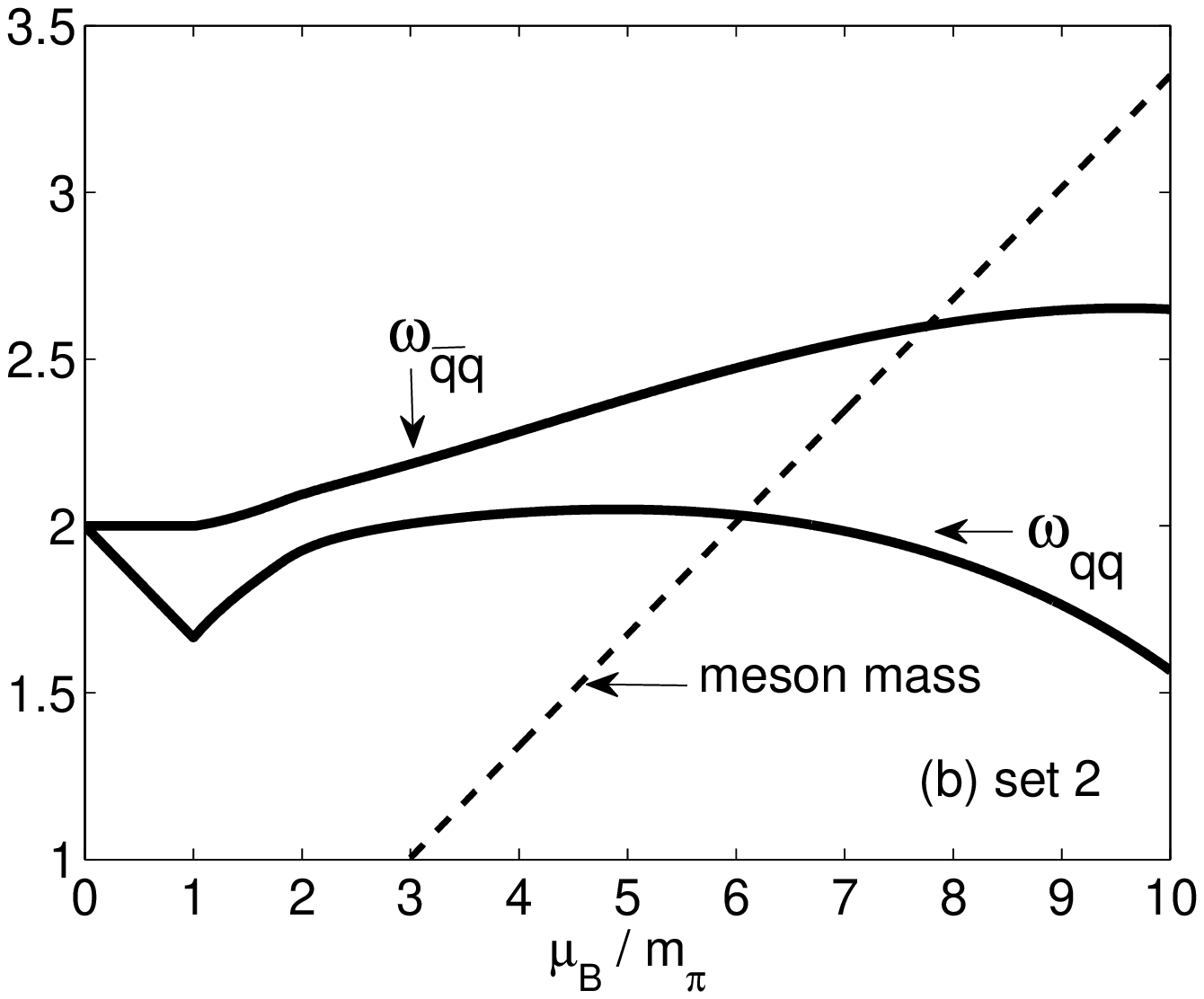}
\includegraphics[width=7.5cm]{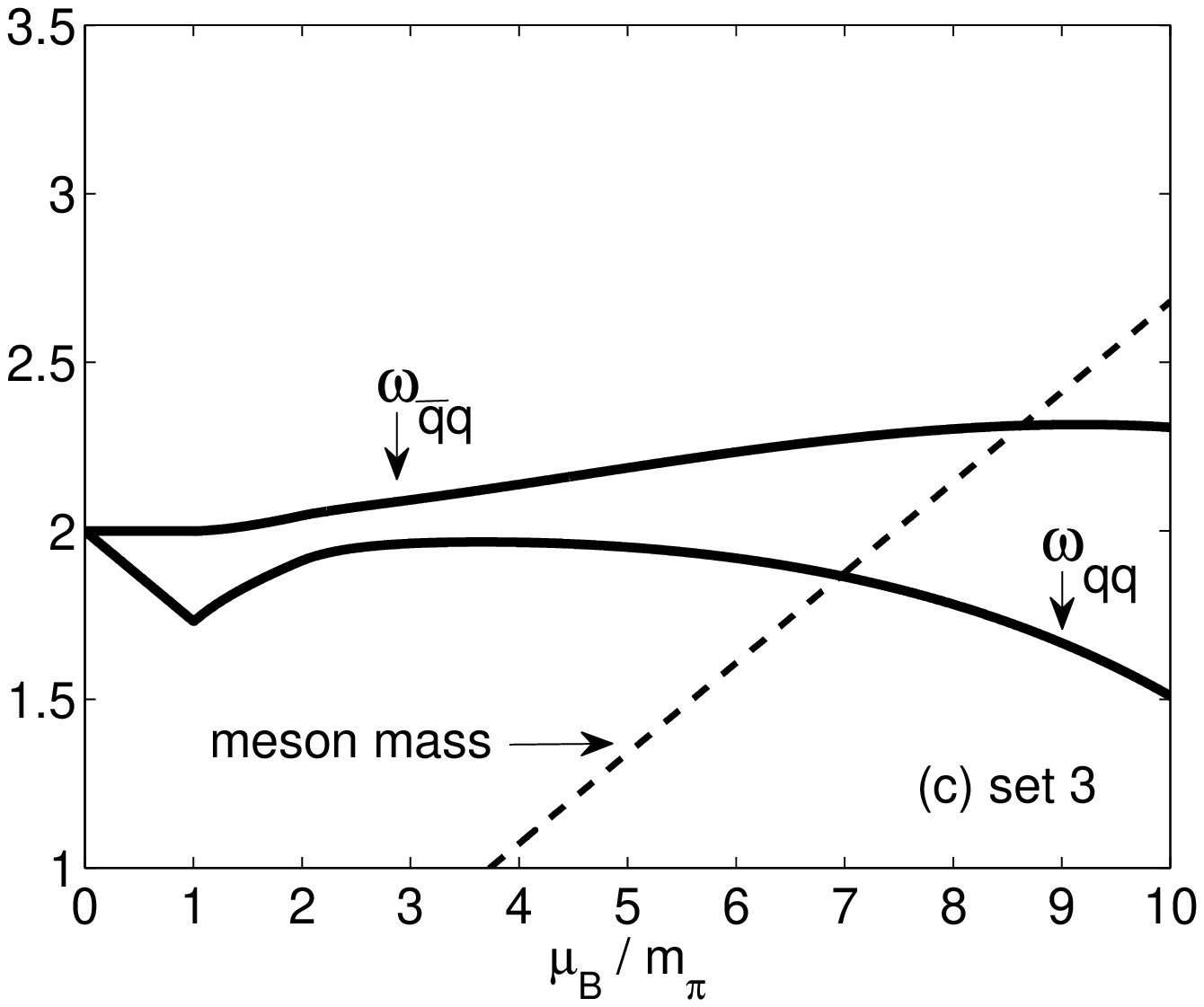}
\includegraphics[width=7.5cm]{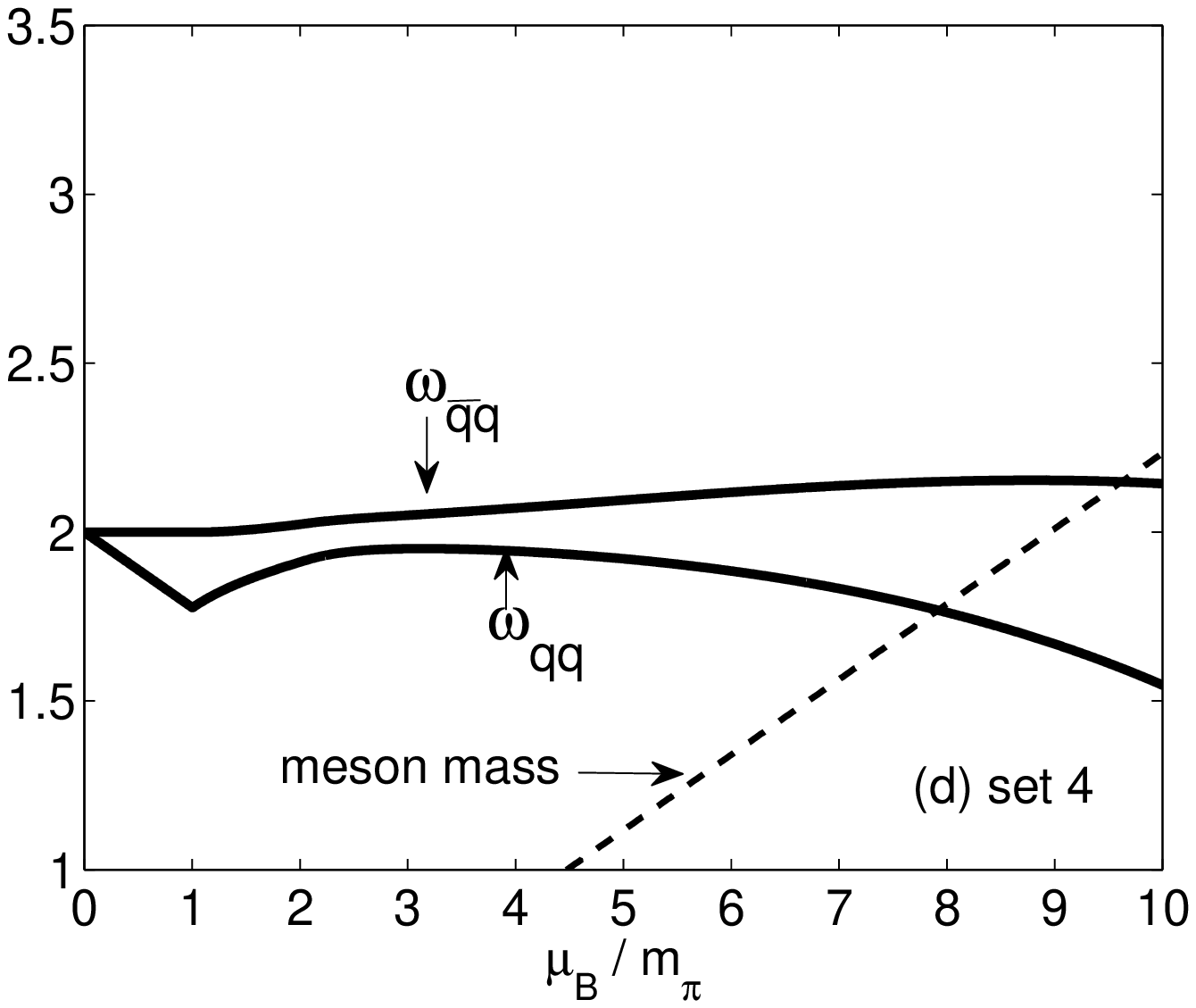}
\caption{The two-particle continua $\omega_{\bar{q}q}$ and
$\omega_{qq}$ (in units of $M_*$) as functions of the baryon
chemical potential (in units of $m_\pi$) for different model
parameter sets. The degenerate mass of pions and sigma meson is
shown by dashed line. \label{fig9}}
\end{center}
\end{figure*}

Even though the deconfinement transition or crossover which
corresponds to the gauge field sector cannot be described in the NJL
model, we can on the other hand study the meson Mott transition
associated with the chiral restoration
\cite{mott,mott1,mott2,precursor,precursor1,precursor2,precursor3,note4}.
The meson Mott transition is defined as the point where the meson
energy becomes larger than the two-particle continuum
$\omega_{\bar{q}q}$ for the decay process $\pi\rightarrow \bar{q}q$
at zero momentum, which means that the mesons are no longer bound
states. The two-particle continuum $\omega_{\bar{q}q}$ is different
at the BEC and the BCS sides. From the explicit form of
$\Pi_\pi(Q)$, we find
\begin{equation}
\omega_{\bar{q}q}=
 \left\{ \begin{array}
{r@{\quad,\quad}l}
 \sqrt{(M-\frac{\mu_{\text
B}}{2})^2+|\Delta|^2}+\sqrt{(M+\frac{\mu_{\text
B}}{2})^2+|\Delta|^2}&
 \mu_{\text B}<\mu_0 \\
 |\Delta|+\sqrt{(M+\frac{\mu_{\text
B}}{2})^2+|\Delta|^2} & \mu_{\text B}>\mu_0.
\end{array}
\right.
\end{equation}
Thus the pions and the sigma meson will undergo a Mott transition
when their masses become larger than the two-particle continuum
$\omega_{\bar{q}q}$, i.e., $\mu_{\text B}>\omega_{\bar{q}q}$. Using
the mean-field results for $\Delta$ and $M$, we can calculate the
two-particle continuum $\omega_{\bar{q}q}$ as a function of
$\mu_{\text B}$, which is shown in Fig.\ref{fig9}. We find that the
Mott transition does occur at a chemical potential $\mu_{\text B}=
\mu_{\text M1}$ which is sensitive to the value of $M_*$. The values
of $\mu_{\text M1}$ for the four model parameter sets are shown in
Table.\ref{mottmu}. For reasonable model parameter sets, the value
of $\mu_{\text M1}$ is in the range $(7-10)m_\pi$. Above this
chemical potential, the mesons are no longer stable bound states and
can decay into quark-antiquark pairs even at zero momentum. We note
that the Mott transition takes place well above the chiral
restoration, in contrast to the pure finite temperature case where
the mesons are dissociated once the chiral symmetry is
restored~\cite{mott,mott1,mott2,precursor,precursor1,precursor2,precursor3}.

\begin{table}[b!]
\begin{center}
\begin{tabular}{|c|c|c|c|c|}
  \hline
  Set & 1 & 2 & 3 & 4 \\
  \hline
  $\mu_{\text M1}$  & 7.22 & 7.76 & 8.63 & 9.62 \\
  \hline
  $\mu_{\text M2}$  & 5.29 & 6.06 & 6.96 & 7.92 \\
  \hline
\end{tabular}
\end{center}
\caption{\small The chemical potentials $\mu_{\text M1}$ and
$\mu_{\text M2}$ (in units of $m_\pi$) for different model parameter
sets.} \label{mottmu}
\end{table}

On the other hand, we find from the explicit forms of the meson
propagators in Appendix \ref{app} that the decay process
$\pi\rightarrow qq$ is also possible at ${\bf q}\neq0$ (even though
$|{\bf q}|$ is small) due to the presence of superfluidity. Thus, we
have another unusual Mott transition in the superfluid phase. Notice
that this process is not in contradiction to the baryon number
conservation law, since the U$_{\text B}(1)$ baryon number symmetry
is spontaneously broken in the superfluid phase. Quantitatively,
this transition occurs when the meson mass becomes larger than the
two-particle continuum $\omega_{qq}$ for the decay process
$\pi\rightarrow qq$ at ${\bf q}=0^+$. In this case, we have
\begin{equation}
\omega_{qq}=
 \left\{ \begin{array}
{r@{\quad,\quad}l}
 2\sqrt{(M-\frac{\mu_{\text B}}{2})^2+|\Delta|^2}&
 \mu_{\text B}<\mu_0 \\
 2|\Delta| & \mu_{\text B}>\mu_0.
\end{array}
\right.
\end{equation}
The two-particle continuum $\omega_{qq}$ is also shown in
Fig.\ref{fig9}. We find that the unusual Mott transition does occur
at another chemical potential $\mu_{\text B}= \mu_{\text M2}$ which
is also sensitive to the value of $M_*$. The values of $\mu_{\text
M2}$ for the four model parameter sets are also shown in
Table.\ref{mottmu}. For reasonable model parameter sets, this value
is in the range $(5-8)m_\pi$. This process can also occur in the 2SC
phase of quark matter in the $N_c=3$ case \cite{ebert}. In the 2SC
phase, the symmetry breaking pattern is SU$_{\text
c}(3)\otimes$U$_{\text B}(1)\rightarrow$SU$_{\text
c}(2)\otimes\tilde{\text U}_{\text B}(1)$ where the generator of the
residue baryon number symmetry $\tilde{\text U}_{\text B}(1)$ is
$\tilde{\text B}={\text B}-2T_8/\sqrt{3}=\text{diag}(0,0,1)$
corresponding to the unpaired blue quarks. Thus the baryon number
symmetry for the paired red and green quarks are broken and our
results can be applied. To show this explicitly, we write down
the explicit form of the polarization function for pions in the 2SC phase \cite{ebert}
\begin{equation}
\Pi_\pi^{2\text{SC}}(Q)=\Pi_\pi^{\text{2-color}}(Q)+\sum_K\text{Tr}[{\cal G}_0(K)i\gamma_5{\cal G}_0(P)i\gamma_5],
\end{equation}
where ${\cal G}_0(K)$ is the propagator for the unpaired blue quarks.
Here $\Pi_\pi^{\text{2-color}}(Q)$ is given by (\ref{pionpo}) (the effective quarks mass $M$ and the pairing gap $\Delta$ should be
given by the $N_c=3$ case of course) and corresponds to the contribution from the paired red and green sectors. The second
term is the contribution from the unpaired blue quarks. Therefore, the unusual decay process is only available for the paired quarks.

\subsection { Beyond-Mean-Field Corrections}
\label{s3-5}
The previous investigations are restricted to the mean-field approximation, even though the bosonic collective
excitations are studied. In this part, we will include the
Gaussian fluctuations in the thermodynamic potential, and thus
really go beyond the mean field. The scheme of going beyond the mean field
is somewhat like those done in the study of finite temperature
thermodynamics of the NJL model \cite{zhuang01,zhuang02}; however,
in this paper we will focus on the beyond-mean-field corrections at
zero temperature, i.e., the pure quantum fluctuations. We will
first derive the thermodynamic potential beyond the mean field which
is valid at arbitrary chemical potential and temperature, and then
briefly discuss the beyond-mean-field corrections near the quantum
phase transition. The numerical calculations are deferred for future
studies.

In Gaussian approximation, the partition function can be
expressed as
\begin{eqnarray}
Z_{\text{NJL}}\simeq\exp{\left(-{\cal
S}_{\text{eff}}^{(0)}\right)}\int[d\sigma][d\mbox{\boldmath{$\pi$}}][d\phi^\dagger][d\phi]e^{-{\cal
S}_{\text{eff}}^{(2)}}
\end{eqnarray}
Integrating out the Gaussian fluctuations, we can express the total
thermodynamic potential as
\begin{equation}
\Omega(T,\mu_{\text B})=\Omega_0(T,\mu_{\text
B})+\Omega_{\text{fl}}(T,\mu_{\text B}),
\end{equation}
where the contribution from the Gaussian fluctuations can be written
as
\begin{eqnarray}
\Omega_{\text{fl}}=\frac{1}{2}\sum_Q\big[\ln\det{\bf
M}(Q)+\ln\det{\bf N}(Q)\big].
\end{eqnarray}

However, there is a problem with the above expression, since it is
actually ill-defined: the sum over the boson Matsubara frequency is
divergent and we need appropriate convergent factors to make it
meaningful. In the simpler case without superfluidity, the
convergent factor is simply given by $e^{i\nu_m 0^+}$
\cite{zhuang01,zhuang02}. In our case, the situation is somewhat
different due to the introduction of the Nambu-Gor'kov spinors. Keep
in mind that in the equal time limit, there are additional factors
$e^{i\omega_n0^+}$ for ${\cal G}_{11}(K)$ and $e^{-i\omega_n0^+}$
for ${\cal G}_{22}(K)$. Therefore, to get the proper convergent
factors for $\Omega_{\text{fl}}$, we should keep these factors when
we make the sum over the fermion Matsubara frequency $\omega_n$ in
evaluating the polarization functions $\Pi_{\text{ij}}(Q)$ and
$\Pi_{\pi}(Q)$.

The problem in the expression of $\Omega_{\text{fl}}$ is thus from
the opposite convergent factors for ${\bf M}_{11}$ and ${\bf
M}_{22}$. From the above arguments, we find that there is a factor
$e^{i\nu_m 0^+}$ for ${\bf M}_{11}$ and $e^{-i\nu_m 0^+}$ for ${\bf
M}_{22}$. Keep in mind that the Matsubara sum $\sum_m$ is converted
to a standard contour integral ($i\nu_m\rightarrow z$). The
convergence for $z\rightarrow+\infty$ is automatically guaranteed by
the Bose distribution function $b(z)=1/(e^{ z/T}-1)$, we thus
should treat only the problem for $z\rightarrow-\infty$. To this
end, we write the first term of $\Omega_{\text{fl}}$ as
\begin{eqnarray}
\sum_Q\ln\det{\bf M}(Q)&=&\sum_Q\bigg[\ln{\bf M}_{11}e^{i\nu_m
0^+}+\ln{\bf M}_{22}e^{-i\nu_m
0^+}\nonumber\\
&&+\ln\left(\frac{\det{\bf M}}{{\bf M}_{11}{\bf
M}_{22}}\right)e^{i\nu_m 0^+}\bigg].
\end{eqnarray}
Using the fact that ${\bf M}_{22}(Q)={\bf M}_{11}(-Q)$, we obtain
\begin{eqnarray}
\sum_Q\ln\det{\bf M}(Q)=\sum_Q\ln\left[\frac{{\bf M}_{11}(Q)}{{\bf
M}_{22}(Q)}\det{\bf M}(Q)\right]e^{i\nu_m 0^+}.
\end{eqnarray}
Therefore, the well-defined form of $\Omega_{\text{fl}}$ is given by
the above formula together with the other term $\sum_Q\ln\det{\bf
N}(Q)$ associated with a factor $e^{i\nu_m 0^+}$.

The Matsubara sum can be written as the contour integral via the
theorem $\sum_m g(i\nu_m)=\oint_{\text C}dz/(2\pi i)b(z)g(z)$, where
${\text C}$ runs on either side of the imaginary $z$ axis, enclosing
it counterclockwise. Distorting the contour to run above and below
the real axis, we obtain
\begin{eqnarray}
\Omega_{\text{fl}}&=&\sum_{\bf
q}\int_{-\infty}^{+\infty}\frac{d\omega}{2\pi}b(\omega)\big[\delta_{\text
M}(\omega,{\bf q})+\delta_{11}(\omega,{\bf
q})\nonumber\\
&&-\delta_{22}(\omega,{\bf q})+3\delta_\pi(\omega,{\bf
q})\big],
\end{eqnarray}
where the scattering phases are defined as
\begin{eqnarray}
&&\delta_{\text M}(\omega,{\bf q})=\text{Im}\ln\det{\bf
M}(\omega+i0^+,{\bf
q}),\nonumber\\
&&\delta_{11}(\omega,{\bf q})=\text{Im}\ln{\bf
M}_{11}(\omega+i0^+,{\bf q}),\nonumber\\
&&\delta_{22}(\omega,{\bf q})=\text{Im}\ln{\bf
M}_{22}(\omega+i0^+,{\bf q}),\nonumber\\
&&\delta_\pi(\omega,{\bf
q})=\text{Im}\ln\left[(2G)^{-1}+\Pi_\pi(\omega+i0^+,{\bf q})\right].
\end{eqnarray}
Keep in mind the pressure of the vacuum should be zero,  the
physical thermodynamic potential at finite temperature and chemical
potential should be defined as
\begin{equation}
\Omega_{\text{phy}}(T,\mu_{\text B})=\Omega(T,\mu_{\text
B})-\Omega(0,0).
\end{equation}

As we have shown in the mean-field theory, at $T=0$, the vacuum
state is restricted in the region $|\mu_{\text B}|<m_\pi$. In this
region, all thermodynamic quantities should keep zero, no matter how
large the value of $\mu_{\text B}$ is. While this should be an
obvious physical conclusion, it is important to check whether our
beyond-mean-field theory satisfies this condition.

Notice that the physical thermodynamic potential is defined as
$\Omega_{\text{phy}}(\mu_{\text B})=\Omega(\mu_{\text
B})-\Omega(0)$, we therefore should prove that the thermodynamic
potential $\Omega(\mu_{\text B})$ stays constant in the region
$|\mu_{\text B}|<m_\pi$. For the mean-field part $\Omega_0$, the
proof is quite easy. Because of the fact that $M_*>m_\pi/2$, the
solution for $M$ is always given by $M=M_*$. Thus $\Omega_0$ keeps
its value at $\mu_{\text B}=0$ in the region $|\mu_{\text
B}|<m_\pi$.

Now we turn to the complicated part $\Omega_{\text{fl}}$. From
$\Delta=0$, all the off-diagonal elements of ${\bf M}$ vanishes,
$\Omega_{\text{fl}}$ is reduced to
\begin{eqnarray}
\Omega_{\text{fl}}&=&\frac{1}{2}\sum_Q\ln\left[\frac{1}{2G}+\Pi_{\sigma}(Q)\right]e^{i\nu_m
0^+}\nonumber\\
&&+\frac{3}{2}\sum_Q\ln\left[\frac{1}{2G}+\Pi_{\pi}(Q)\right]e^{i\nu_m 0^+}\nonumber\\
&&+\sum_Q\ln\left[\frac{1}{4G}+\Pi_{\text d}(Q)\right]e^{i\nu_m
0^+}
\end{eqnarray}
with $\Pi_\sigma(Q)=\Pi_{33}(Q)$, and we should set $\Delta=0$ and
$M=M_*$ in evaluating the polarization functions. First, we can
easily show that the contributions from the sigma meson and pions do
not have explicit $\mu_{\text B}$ dependence and thus keep the same
values as those at $\mu_{\text B}=0$. In fact, since the effective
quark mass $M$ keeps its vacuum value $M_*$ guaranteed by the mean-field
part, all the $\mu_{\text B}$ dependence in
$\Pi_{\sigma,\pi}(Q)$ is included in the Fermi distribution
functions $f(E\pm\mu_{\text B}/2)$. From $M_*>\mu_{\text B}/2$,
they vanish automatically at $T=0$. In fact, from the explicit
expressions for $\Pi_{\sigma,\pi}(Q)$ in Appendix \ref{app}, we can
check that there is no $\mu_{\text B}$ independence in
$\Pi_{\sigma,\pi}(Q)$.

The diquark contribution, however, has an explicit $\mu_{\text
B}$ dependence through the combination $i\nu_m+\mu_{\text B}$ in the
polarization function $\Pi_{\text d}(Q)$. The diquark contribution
(at $T=0$) can be written as
\begin{eqnarray}
\Omega_{\text{d}}&=&-\sum_{\bf
q}\int_{-\infty}^{0}\frac{d\omega}{\pi}\delta_{\text
d}(\omega,{\bf q}),\nonumber\\
\delta_{\text d}(\omega,{\bf
q})&=&\text{Im}\ln\left[(4G)^{-1}+\Pi_{\text d}(\omega+i0^+,{\bf
q})\right].
\end{eqnarray}
Making a shift $\omega\rightarrow\omega-\mu_{\text B}$, and noticing
that fact $\Pi_{\text d}(\omega-\mu_{\text B},{\bf
q})=\Pi_\pi(\omega,{\bf q})/2$, we obtain
\begin{eqnarray}
\Omega_{\text{d}}=-\sum_{\bf q}\int_{-\infty}^{-\mu_{\text
B}}\frac{d\omega}{\pi}\delta_{\pi}(\omega,{\bf q}).
\end{eqnarray}
To show that the above quantity is $\mu_{\text B}$ independent,
we separate it into a pole part and a continuum part. There is a
well-defined two-particle continuum $E_c({\bf q})$ for pions at
arbitrary momentum ${\bf q}$,
\begin{eqnarray}
E_c({\bf q})=\text{min}_{\bf k}\left(E_{\bf k}^*+E_{\bf
k+q}^*\right).
\end{eqnarray}
The pion propagator has two symmetric poles $\pm\omega_\pi({\bf q})$
when ${\bf q}$ satisfies $\omega_\pi({\bf q})<E_c({\bf q})$. Thus in
the region $|\omega|<E_c({\bf q})$, the scattering phase
$\delta_{\pi}$ can be analytically evaluated as
\begin{eqnarray}
\delta_{\pi}(\omega,{\bf
q})=\pi\left[\Theta\left(-\omega-\omega_\pi({\bf
q})\right)-\Theta\left(\omega-\omega_\pi({\bf q})\right)\right].
\end{eqnarray}
From $E_c({\bf q})>\omega_\pi({\bf q})>m_\pi>\mu_{\text B}$, the
thermodynamic potential $\Omega_{\text d}$ can be separated as
\begin{eqnarray}
\Omega_{\text{d}}=\sum_{\bf q}\left[\omega_\pi({\bf q})-E_c({\bf
q})\right]-\sum_{\bf q}\int_{-\infty}^{-E_c({\bf
q})}\frac{d\omega}{\pi}\delta_{\pi}(\omega,{\bf q}),
\end{eqnarray}
which is indeed $\mu_{\text B}$ independent. Notice that in the
first term the integral over ${\bf q}$ is restricted in the region
$|{\bf q}|<q_c$ where $q_c$ is defined as
$\omega_\pi(q_c)=E_c(q_c)$.

In conclusion, we have shown that the thermodynamic potential
$\Omega$ in the Gaussian approximation stays constant in the
vacuum state, i.e., at $|\mu_{\text B}|<m_\pi$ and at $T=0$. All
other thermodynamic quantities such as the baryon number density
keep zero in the vacuum. The subtraction term $\Omega(0,0)$ in the
Gaussian approximation can be expressed as
\begin{eqnarray}
\Omega(0,0)&=&\Omega_{\text{vac}}(M_*)+\frac{5}{2}\sum_{\bf
q}\left[\omega_\pi({\bf
q})-E_c({\bf q})\right]\nonumber\\
&&-\sum_{\bf q}\int_{-\infty}^{-E_c({\bf
q})}\frac{d\omega}{2\pi}\left[\delta_\sigma(\omega,{\bf
q})+5\delta_{\pi}(\omega,{\bf q})\right].
\end{eqnarray}

Now we consider the beyond-mean-field corrections near the quantum
phase transition point $\mu_{\text B}=m_\pi$. Notice that the
effective quark mass $M$ and the diquark condensate $\Delta$ are
determined at the mean-field level, and the beyond-mean-field
corrections are possible only through the equations of state.

Formally, the Gaussian contribution to the thermodynamic
potential $\Omega_{\text{fl}}$ is a function of $\mu_{\text B},M$
and $y=|\Delta|^2$, i.e.,
$\Omega_{\text{fl}}=\Omega_{\text{fl}}(\mu_{\text B},y,M)$. In the
superfluid phase, the total baryon density including the Gaussian
contribution can be evaluated as
\begin{eqnarray}
n(\mu_{\text B})=n_0(\mu_{\text B})+n_{\text{fl}}(\mu_{\text B}),
\end{eqnarray}
where the mean-field part is simply given by $n_0(\mu_{\text
B})=-\partial\Omega_0/\partial\mu_{\text B}$ and the Gaussian
contribution can be expressed as
\begin{eqnarray}
n_{\text{fl}}(\mu_{\text B})=-\frac{\partial
\Omega_{\text{fl}}}{\partial \mu_{\text B}}-\frac{\partial
\Omega_{\text{fl}}}{\partial y}\frac{dy}{d\mu_{\text
B}}-\frac{\partial \Omega_{\text{fl}}}{\partial
M}\frac{dM}{d\mu_{\text B}}.
\end{eqnarray}
The physical values of $M$ and $|\Delta|^2$ should be determined by
their mean-field gap equations. In fact, from the gap equations
$\partial\Omega_0/\partial M=0$ and $\partial\Omega_0/\partial y=0$,
we obtain
\begin{eqnarray}
\frac{\partial^2 \Omega_0}{\partial \mu_{\text B}\partial
M}+\frac{\partial^2\Omega_0}{\partial y\partial
M}\frac{dy}{d\mu_{\text B}}+\frac{\partial^2 \Omega_0}{\partial
M^2}\frac{dM}{d\mu_{\text B}}&=&0,\nonumber\\
\frac{\partial^2 \Omega_0}{\partial \mu_{\text B}\partial
y}+\frac{\partial^2\Omega_0}{\partial y^2}\frac{dy}{d\mu_{\text
B}}+\frac{\partial^2 \Omega_0}{\partial M\partial
y}\frac{dM}{d\mu_{\text B}}&=&0.
\end{eqnarray}
Thus, we can obtain the derivatives $dM/d\mu_{\text B}$ and
$dy/d\mu_{\text B}$ analytically. Finally, $n_{\text{fl}}(\mu_{\text
B})$ is a continuous function of $\mu_{\text B}$ guaranteed by the
properties of second order phase transition, and we have
$n_{\text{fl}}(m_\pi)=0$.

Next we focus on the beyond-mean-field corrections near the quantum
phase transition. Since the diquark condensate $\Delta$ is
vanishingly small, we can expand the Gaussian part
$\Omega_{\text{fl}}$ in powers of $|\Delta|^2$. Notice that
$\mu_{\text B}$ and $M$ can be evaluated as functions of
$|\Delta|^2$ from the Ginzburg-Landau potential and mean-field gap
equations. Thus to order $O(|\Delta|^2)$, the expansion takes the
form
\begin{eqnarray}
\Omega_{\text{fl}}\simeq \eta|\Delta|^2,
\end{eqnarray}
where the expansion coefficient $\eta$ is defined as
\begin{eqnarray}
\eta=\left(\frac{\partial\Omega_{\text{fl}}}{\partial
y}+\frac{\partial\Omega_{\text{fl}}}{\partial\mu_{\text
B}}\frac{d\mu_{\text
B}}{dy}+\frac{\partial\Omega_{\text{fl}}}{\partial
M}\frac{dM}{dy}\right)\Bigg|_{\mu_{\text B}=m_\pi,y=0,M=M_*}.
\end{eqnarray}
Using the definition of $n_{\text{fl}}$, we find that $\eta$ can be
related to $n_{\text{fl}}$ by
\begin{eqnarray}
\eta=n_{\text{fl}}(m_\pi)\frac{d\mu_{\text B}}{dy}\bigg|_{y=0}.
\end{eqnarray}
Therefore, the coefficient $\eta$ vanishes, and the leading order of the
expansion should be $O(|\Delta|^4)$.

As shown above, to leading order, the expansion of $\Omega_{\text{fl}}$ can be formally expressed as
\begin{eqnarray}
\Omega_{\text{fl}}\simeq-\frac{\zeta}{2}\beta|\Delta|^4.
\end{eqnarray}
The explicit form of $\zeta$ is quite complicated and we do not show it here. Notice that the factor
$\zeta$ is in fact $\mu_{\text B}$ independent, thus the total
baryon density to leading order is
\begin{eqnarray}
n=n_0+\zeta\beta|\Delta|^2\frac{d|\Delta|^2}{d\mu_{\text
B}}\bigg|_{\mu_{\text B}=m_\pi}.
\end{eqnarray}
Near the quantum phase transition point, the mean-field contribution
is $n_0=|\psiup_0|^2=2m_\pi {\cal J}|\Delta|^2$ from the
Gross-Pitaevskii free energy. The last term can be evaluated using
the analytical result
\begin{eqnarray}
|\psiup_0|^2=\frac{\mu_{\text d}}{g_0}\Longrightarrow
|\Delta|^2=\frac{2m_\pi{\cal J}}{\beta}\mu_{\text d},
\end{eqnarray}
which is in fact the solution of the mean-field gap equations.
Therefore, to leading order, the total baryon density reads
\begin{eqnarray}
n=(1+\zeta)2m_\pi {\cal J}|\Delta|^2.
\end{eqnarray}
On the other hand, the total pressure $P$ can be expressed as
\begin{eqnarray}
P=(1+\zeta)\frac{\beta}{2}|\Delta|^4.
\end{eqnarray}
Thus we find that the leading order quantum corrections are totally
included in the numerical factor $\zeta$. Setting $\zeta=0$, we
recover the mean-field results obtained previously.

Including the quantum fluctuations, the equations of state shown in
(\ref{eos}) are modified to be
\begin{eqnarray}
P(n)&=&\frac{1}{1+\zeta}\frac{2\pi
a_{\text{dd}}}{m_\pi}n^2,\nonumber\\
\mu_{\text B}(n)&=&m_\pi+\frac{1}{1+\zeta}\frac{4\pi
a_{\text{dd}}}{m_\pi}n.
\end{eqnarray}
This means that, to leading order, the effect of quantum fluctuations is
giving a correction to the diquark-diquark scattering length. The
renormalized scattering length is
\begin{eqnarray}
a_{\text{dd}}^\prime=\frac{a_{\text{dd}}}{1+\zeta}.
\end{eqnarray}
Generally, we have $\zeta>0$ and the renormalized scattering length
is smaller than the mean-field result.

An exact calculation of the numerical factor $\zeta$ can be performed. In this work we will give an analytical estimation of
$\zeta$ based on the fact that the quantum fluctuations are dominated by the gapless Goldstone mode. To this end, we approximate
the Gaussian contribution $\Omega_{\text{fl}}$ as
\begin{eqnarray}
\Omega_{\text{fl}}&\simeq&\frac{1}{2}\sum_Q\ln\bigg[{\cal D}_{\text
d}^{-1}(Q){\cal D}_{\text d}^{-1}(-Q)+3\beta^2|\Delta|^4\nonumber\\
&&+2\beta|\Delta|^2\left({\cal D}_{\text
d}^{-1}(Q)+{\cal D}_{\text d}^{-1}(-Q)\right)\bigg],
\end{eqnarray}
where ${\cal D}_{\text d}^{-1}(Q)$ is given by (\ref{dipro}) and can
be approximated by (\ref{diproa}). Subtracting the value of
$\Omega_{\text{fl}}$ at $\mu_{\text B}=m_\pi$ with $\Delta=0$, and
using the result $\mu_{\text B}=m_\pi+g_0|\psiup_0|^2$ from the
Gross-Pitaevskii equation, we find that $\zeta$ can be evaluated as
\begin{eqnarray}
\zeta=\frac{\beta}{{\cal
J}^2}\left(I_1+I_2\right)\simeq\frac{m_\pi^2}{f_\pi^2}\left(I_1+I_2\right),
\end{eqnarray}
where the numerical factors $I_1$ and $I_2$ are given by
\begin{eqnarray}
I_1&=&\frac{1}{2}\sum_m\sum_{\bf X}\frac{Z_m^2+{\bf
X}^2}{(Z_m^2-{\bf
X}^2)^2-4Z_m^2},\nonumber\\
I_2&=&4\sum_m\sum_{\bf X}\frac{(3Z_m^2-{\bf
X}^2)^2}{\left[(Z_m^2-{\bf X}^2)^2-4Z_m^2\right]^2}.
\end{eqnarray}
Here the dimensionless notations $Z_m$ and ${\bf X}$ are defined as
$Z_m=i\nu_m/m_\pi$ and ${\bf X}={\bf q}/m_\pi$ respectively. Notice
that the integral over ${\bf X}$ is divergent and hence such an
estimation has no prediction power due to the fact that the NJL
model is nonrenormalizable. However, regardless of the numerical
factor $I_1+I_2$, we find $\zeta\propto m_\pi^2/f_\pi^2$. Thus,
the correction should be small for the case $m_\pi\ll 2M_*$.

While the effect of the Gaussian fluctuations at zero temperature is
to give a small correction to the diquark-diquark scattering length
and the equations of state, it can be significant at finite
temperature. In fact, as the temperature approaches the critical
value of superfluidity, the Gaussian fluctuations should dominate.
In this part, we will show that to get a correct critical
temperature in terms of the baryon density $n$, we must go beyond
the mean field. The situation is analogous to the Nozieres--Schmitt-Rink treatment of
molecular condensation in strongly interacting Fermi gases
\cite{BCSBEC1,BCSBEC2}.

The transition temperature $T_c$ is determined by the Thouless
criterion ${\cal D}_{\text d}^{-1}(0,{\bf 0})=0$ which can be shown
to be consistent with the saddle point condition $\delta {\cal
S}_{\text{eff}}/\delta \phi|_{\phi=0}=0$. Its explicit form is a
BCS-type gap equation
\begin{eqnarray}\label{Tc1}
\frac{1}{4G}=N_cN_f\sum_{e=\pm}\int\frac{d^3{\bf
k}}{(2\pi)^3}\frac{1-2f(\xi_{\bf k}^e)}{2\xi_{\bf k}^e}.
\end{eqnarray}
Meanwhile, the effective quark mass $M$ satisfies the mean-field gap
equation
\begin{eqnarray}\label{Tc2}
\frac{M-m_0}{2GM}=N_cN_f\int\frac{d^3{\bf k}}{(2\pi)^3}{1-f(\xi_{\bf
k}^-)-f(\xi_{\bf k}^+)\over E_{\bf k}}.
\end{eqnarray}

To obtain the transition temperature as a function of $n$, we need
the so-called number equation given by $n=-\partial\Omega/\partial
\mu_{\text B}$, which includes both the mean-field contribution
$n_0(\mu_{\text B},T)=2N_f\sum_{\bf k}\left[f(\xi_{\bf
k}^-)-f(\xi_{\bf k}^+)\right]$ and the Gaussian contribution
$n_{\text{fl}}(\mu_{\text B},T)=-\partial\Omega_{\text{fl}}/\partial
\mu_{\text B}$. At the transition temperature with $\Delta=0$,
$\Omega_{\text{fl}}$ can be expressed as
\begin{eqnarray}
\Omega_{\text{fl}}&=&\int\frac{d^3{\bf
q}}{(2\pi)^3}\int_{-\infty}^\infty\frac{d\omega}{2\pi
}b(\omega)\nonumber\\
&&\times\ [2\delta_{\text d}(\omega,{\bf
q})+\delta_{\sigma}(\omega,{\bf q})+3\delta_{\pi}(\omega,{\bf q})],
\end{eqnarray}
where the scattering phases are defined as $\delta_{\text
d}(\omega,{\bf q})=\text{Im}\ln[1/(4G)+\Pi_{\text
d}(\omega+i0^+,{\bf q})]$ for the diquarks,
$\delta_{\sigma}(\omega,{\bf
q})=\text{Im}\ln[1/(2G)+\Pi_\sigma(\omega+i0^+,{\bf q})]$ for the
sigma meson and $\delta_{\pi}(\omega,{\bf
q})=\text{Im}\ln[1/(2G)+\Pi_\pi(\omega+i0^+,{\bf q})]$ for the
pions. Obviously, the polarization functions should take their forms
at finite temperature in the normal phase.

The transition temperature $T_c$ at arbitrary baryon number density
$n$ can be determined numerically via solving simultaneously the gap
and number equations. However, in the dilute limit $n\rightarrow 0$
which we are interested in this section, analytical result can be
achieved. Keep in mind that $T_c\rightarrow 0$ when $n\rightarrow0$,
we find that the Fermi distribution functions $f(\xi_{\bf k}^\pm)$
vanish exponentially (from $M_*-m_\pi/2\gg T_c$) and we obtain
$\mu_{\text B}=m_\pi$ and $M=M_*$ from the gap Eqs. (\ref{Tc1})
and (\ref{Tc2}), respectively. Meanwhile the mean-field contribution
of the density $n_0$ can be neglected and the total density $n$ is
thus dominated by the Gaussian part $n_{\text{fl}}$. When
$T\rightarrow0$ we can show that $\Pi_\sigma(\omega,{\bf q})$ and
$\Pi_\pi(\omega,{\bf q})$ are independent of $\mu_{\text B}$, and
the number equation is reduced to
\begin{eqnarray}
n=-\sum_{\bf q}\int_{-\infty}^\infty\frac{d\omega}{\pi
}b(\omega)\frac{\partial\delta_{\text d}(\omega,{\bf
q})}{\partial\mu_{\text B}}.
\end{eqnarray}
From $T_c\rightarrow0$, the inverse diquark propagator can be
reduced to ${\cal D}_{\text d}^{-1}(\omega,{\bf q})$ in
(\ref{dipro}). Thus the scattering phase $\delta_{\text d}$ can be
well approximated by $\delta_{\text d}(\omega,{\bf
q})=\pi[\Theta(\mu_{\text B}-\epsilon_{\bf
q}-\omega)-\Theta(\omega-\mu_{\text B}-\epsilon_{\bf q})]$ with
$\epsilon_{\bf q}=\sqrt{{\bf q}^2+m_\pi^2}$. Therefore, the number
equation can be further reduced to the well-known equation for ideal
Bose-Einstein condensation,
\begin{eqnarray}
n=\sum_{\bf q}\left[b(\epsilon_{\bf q}-\mu_{\text
B})-b(\epsilon_{\bf q}+\mu_{\text B})\right]\bigg|_{\mu_{\text
B}=m_\pi}.
\end{eqnarray}
Since the above equation is valid only in the low density limit
$n\rightarrow0$, the critical temperature is thus given by the
nonrelativistic result
\begin{eqnarray}
T_c=\frac{2\pi}{m_\pi}\left[\frac{n}{\xi(3/2)}\right]^{2/3}.
\end{eqnarray}
At finite density but $na_{\text{dd}}^3\ll1$, there exists a
correction to $T_c$ which is proportional to
$n^{1/3}a_{\text{dd}}$~\cite{Bose01}. Such a correction is hard to
handle analytically in our model since we should consider
simultaneously the corrections to $M$ and $\mu_{\text B}$, as well
as the contributions from the sigma meson and pions.

\subsection{BCS-BEC crossover in Pion superfluid}
\label{s3-6} The other situation, which can be simulated by lattice
QCD, is quark matter at finite isospin density. The physical
motivation to study QCD at finite isospin density and the
corresponding pion superfluid is related to the investigation of
compact stars, isospin asymmetric nuclear matter and heavy ion
collisions at intermediate energies. In early studies on dense
nuclear matter and compact stars, it has been suggested that charged
pions are condensed at sufficiently high
density~\cite{PiC,PiC1,PiC2,PiC3,PiC4,he36,he361}. The QCD phase
structure at finite isospin chemical potential is recently
investigated in many low energy effective models, such as chiral
perturbation theory, linear sigma model, NJL model, random matrix
method, and ladder
QCD~\cite{ISO,ISOother01,ISOother011,ISOother012,ISOother013,ISOother014,ISOother015,ISOother016,ISOother017,ISOother018,ISOother019,ISOother0110,ISOother0111,ISOother0112,ISOother0113,ISOother0114,ISOother0115,boser,ISOother02,ISOother021,ISOother03,Liso,Liso1,Liso2,Liso3,ran212,ran2121,lad23}.
In this subsection, we review the pion superfluid and the
corresponding BCS-BEC crossover in the frame of two-flavor NJL
model. Since the isospin chemical potential which triggers the pion
condensation is large, $\mu_I\geq m_\pi$, we neglect the diquark
condensation which is favored at large baryon chemical potential and
small isospin chemical potential.

The Lagrangian density of the two-flavor NJL model at quark level is
defined as~\cite{NJLreview,NJLreview1,NJLreview2,NJLreview3}
\begin{equation}
\label{njl}
{\cal L} =
\bar{\psi}\left(i\gamma^{\mu}\partial_{\mu}-m_0+\gamma_0 \mu
\right)\psi +G\Big[\left(\bar{\psi}\psi\right)^2+\left(\bar\psi
i\gamma_5\tau_i\psi\right)^2\Big]
\end{equation}
with scalar and pseudoscalar interactions corresponding to $\sigma$
and $\pi$ excitations, where $m_0$ is the current quark mass, $G$ is
the four-quark coupling constant with dimension GeV$^{-2}$, $\tau_i\
(i=1,2,3)$ are the Pauli matrices in flavor space, and $\mu
=diag\left(\mu_u,\mu_d\right)=diag\left(\mu_B/3+\mu_I/2,\mu_B/3-\mu_I/2\right)$
is the quark chemical potential matrix with $\mu_u$ and $\mu_d$
being the $u$- and $d$-quark chemical potentials and $\mu_B$ and
$\mu_I$ the baryon and isospin chemical potentials.

At zero isospin chemical potential, the Lagrangian density has the
symmetry of $U_B(1)\bigotimes SU_L(2)\bigotimes SU_R(2)$
corresponding to baryon number symmetry, isospin symmetry and chiral
symmetry. At finite isospin chemical potential, the symmetries
$SU_L(2)\bigotimes SU_R(2)$ are firstly explicitly broken down to
$U_L(1)\bigotimes U_R(1)$, and then the nonzero pion condensate leads to
the spontaneous breaking of $U_{I=L+R}(1)$, with pions as the
corresponding Goldstone modes. At $\mu_B=0$, the Fermi surface of $u
(d)$ and anti-$d(u)$ quarks coincide and hence the condensate of $u$
and anti-$d$ is favored at $\mu_I>0$ and the condensate of $d$ and
anti-$u$ quarks is favored at $\mu_I<0$. Finite $\mu_B$ provides a
mismatch between the two Fermi surfaces and will reduce the pion
condensation.

Introducing the chiral and pion condensates
\begin{equation}
\label{condensate1} \sigma = \langle\bar{\psi}\psi\rangle, \ \ \ \ \
\pi =\langle\bar{\psi}i\gamma_5\tau_1\psi\rangle
\end{equation}
and taking them to be real, the quark propagator ${\cal S}$ in mean
field approximation can be expressed as a matrix in the flavor space
\begin{equation}
\label{quarkpropagator}
{\cal S}^{-1}(k)= \left(\begin{array}{cc} \gamma^\mu k_\mu+\mu_u\gamma_0-M_q & 2iG\pi\gamma_5\\
2iG\pi\gamma_5 & \gamma^\mu
k_\mu+\mu_d\gamma_0-M_q\end{array}\right)
\end{equation}
with the effective quark mass $M_q=m_0-2G\sigma$ generated by the
chiral symmetry breaking. By diagonalizing the propagator, the
thermodynamic potential $\Omega(T,\mu_B,\mu_I,M_q,\pi)$ can be
simply expressed as a condensation part plus a summation part of
four quasiparticle
contributions~\cite{ISOother02,ISOother021,ISOother03}. The gap
equations to determine the condensates $\sigma$ (or effective quark
mass $M_q$) and $\pi$ can be obtained by the minimum of the
thermodynamic potential,
\begin{equation}
\label{minimum} {\partial \Omega\over \partial M_q}=0,\ \ \
{\partial \Omega\over \partial \pi}=0.
\end{equation}

In the NJL model, the meson modes are regarded as quantum
fluctuations above the mean field. The two quark scattering via
meson exchange can be effectively expressed at quark level in terms
of quark bubble summation in Random Phase Approximation
(RPA)~\cite{NJLreview,NJLreview1,NJLreview2,NJLreview3}. The quark
bubbles are defined as
\begin{equation}
\label{polarization1} \Pi_{mn}(q) = i\int{d^4 k \over (2\pi)^4} Tr
\left(\Gamma_m^* {\cal S}(k+q)\Gamma_n {\cal S}(k)\right)
\end{equation}
with indices  $m,n=\sigma,\pi_+,\pi_-,\pi_0$, where the trace $Tr =
Tr_C Tr_F Tr_D$ is taken in color, flavor and Dirac spaces, the four
momentum integration is defined as $\int d^4 k/(2 \pi)^4=i T \sum_j
\int d^3{\bf k}/(2 \pi)^3$ with fermion frequency $k_0=i \omega_j=i
(2j+1)\pi T\ (j=0,\pm 1, \pm 2, \cdots)$ at finite temperature $T$,
and the meson vertices are from the Lagrangian density (\ref{njl}),
\begin{equation}
\label{vertex} \Gamma_m = \left\{\begin{array}{ll}
1 & m=\sigma\\
i\gamma_5 \tau_+ & m=\pi_+ \\
i\gamma_5 \tau_- & m=\pi_- \\
i\gamma_5 \tau_3& m=\pi_0\ ,
\end{array}\right.\ \
\Gamma_m^* = \left\{\begin{array}{ll}
1 & m=\sigma\\
i\gamma_5 \tau_-& m=\pi_+ \\
i\gamma_5 \tau_+ & m=\pi_- \\
i\gamma_5 \tau_3& m=\pi_0\ . \\
\end{array}\right.
\end{equation}
Since the quark propagator ${\cal S}$ contains off-diagonal
elements, we must consider all possible channels in the bubble
summation in RPA. Using matrix notation for the meson polarization
function $\Pi(q)$ in the $4\times 4$ meson space, the meson
propagator can be expressed as~\cite{hao}
\begin{equation}
\label{mesonpropagator} {\cal D}(q)={2G\over {1-2G\Pi(q)}}={2G\over
\text {det}\left[1-2G\Pi(q)\right]}{\cal M}(q).
\end{equation}

Since the isospin symmetry is spontaneously broken in the pion
superfluid, the original meson modes $\sigma, \pi_+, \pi_-, \pi_0$
with definite isospin quantum number are no longer the eigen modes
of the Hamiltonian of the system, the new eigen modes
$\overline\sigma, \overline\pi_+, \overline\pi_-, \overline\pi_0$
are linear combinations of the old ones, their masses $M_i
(i=\overline\sigma, \overline\pi_+, \overline\pi_-, \overline\pi_0)$
are determined by poles of the meson propagator at $q_0=M_i$ and
${\bf q=0}$,
\begin{equation}
\label{mass} \text{det} \left[ 1-2G \Pi(M_i, {\bf 0})\right]=0,
\end{equation}
and their coupling constants are defined as the residues of the
propagator at the poles,
\begin{equation}
\label{coupling} g^2_{i q\overline q}={2G\sum_m {\cal
M}_{mm}(M_i,{\bf 0})\over d \text{det}\left[1-2G\Pi(q_0,{\bf
0})\right]/dq_0^2{\big |}_{q_0=M_i}}.
\end{equation}

Since the NJL model is non-renormalizable, we can employ a hard
three momentum cutoff $\Lambda$ to regularize the gap equations for
quarks and pole equations for mesons. In the following numerical
calculations, we take the current quark mass $m_0=5$ MeV, the
coupling constant $G=4.93$ GeV$^{-2}$ and the cutoff $\Lambda=653$
MeV\cite{zhuang02}. This group of parameters correspond to the pion
mass $m_{\pi}=134$ MeV, the pion decay constant $f_{\pi}=93$ MeV and
the effective quark mass $M_q=310$ MeV in the vacuum.

Numerically solving the minimum of thermodynamic potential, the order parameters look like the case in Fig.\ref{fig3}, with replacing $\langle qq \rangle$ by pion condensate and $\mu_B$ by $\mu_I$. The conclusion, which pion superfluid phase transition at zero $\mu_B$ occurs at finite isospin density $\mu_I=m_{\pi}$, can be clearly seen by comparing the gap equation and the polarization function $\Pi_{\pi_+ \pi_+}$. Namely, the phase transition line from normal state to pion superfluid on $T-\mu_I$ plane at $\mu_B=0$ is determined by condition that $1-2 G \Pi_{\pi_+ \pi_+}(q_0=0, {\bf 0})$, and the mass of $\pi_+$ is always zero at phase boundary.

\begin{figure}
\centering
\includegraphics[width=7.3cm]{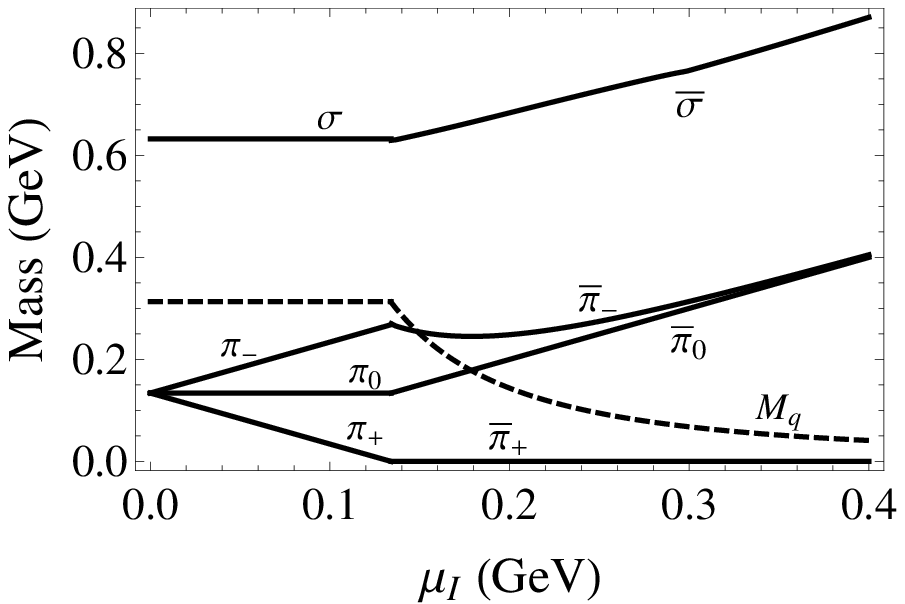}\\
\includegraphics[width=8cm]{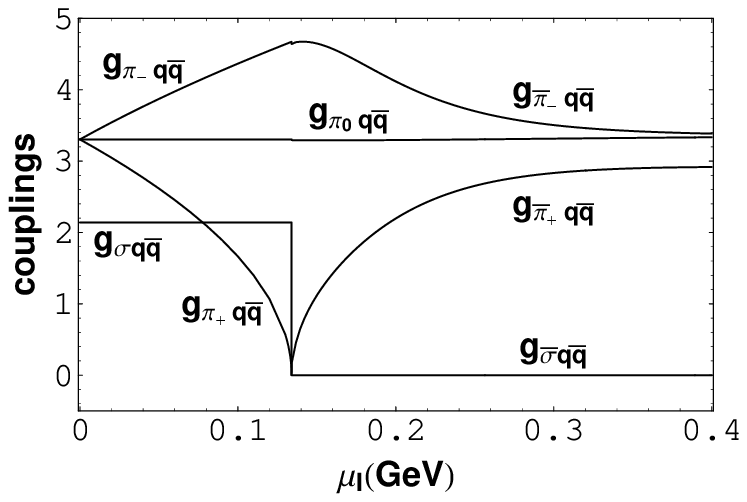}
\caption{(upper) Meson spectra and effective quark mass at $T=\mu_B=0$ as a function of isospin chemical potential $\mu_I$. In normal phase, the meson eigenmode is $\sigma, \pi_0, \pi_+, \pi_-$, and in pion superfluid state, they are denoted by $\overline\sigma, \pi_0, \overline\pi_+,
\overline\pi_-$.~\cite{ISOother03}
(lower) The coupling constants for $\sigma, \pi_0, \pi_+, \pi_-$
in the normal phase and $\overline\sigma, \pi_0, \overline\pi_+,
\overline\pi_-$ in the pion superfluid phase as functions of
$\mu_I$ at $T=\mu_B=0$.~\cite{hao}} \label{figcou}
\end{figure}
The meson mass and the coupling constant $g_{i q\overline q}$ at zero temperature and finite isospin chemical potential are shown in Fig.\ref{figcou}. Note that the meson mass spectrum is similar to what shown in Fig.\ref{fig8}, the "anti-d" and "d" there correspond to the $\pi_+$ and $\pi_-$ here. The condition for a meson to decay into a $q$ and a $\overline q$ is
that its mass lies above the $q-\overline q$ threshold. From the
pole equation (\ref{mass}), the heaviest mode in the pion superfluid
is $\overline\sigma$ and its mass is beyond the threshold value. As
a result, there exists no $\overline\sigma$ meson in the pion
superfluid, and the coupling constant $g_{\overline\sigma q\overline
q}$ drops down to zero at the critical point $\mu^c_I$ and keeps
zero at $\mu_I>\mu^c_I$~\cite{hao}.

\begin{figure}[ht]
\centering
\includegraphics[width=8cm]{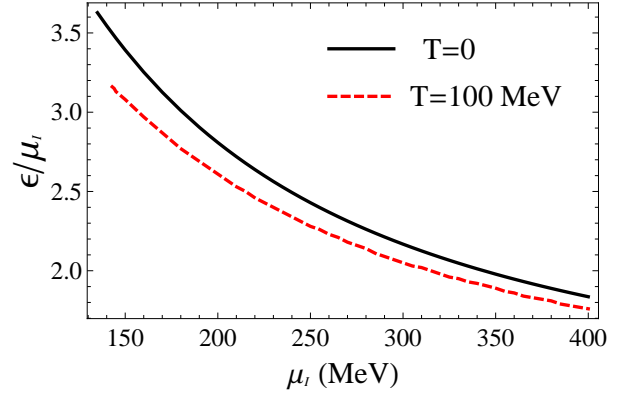}
\caption{The scaled $\overline\pi_+$ binding energy $\epsilon/\mu_I$
as a function of isospin chemical potential $\mu_I$ in the pion
superfluid. The  solid and dashed lines are for $T=0$ and $100$ MeV,
and the baryon chemical potential keeps zero $\mu_B=0$.}
\label{fig4m}
\end{figure}

As discussed in the previous section, there are some characteristic
quantities to describe the BCS-BEC crossover in superfluid or
superconductor~\cite{pion2,pion21,pion22,pion1,pion11}, which are
difficult to be experimentally measured but can be used to confirm
the BCS-BEC crossover picture in pion superfluid. Here we calculate
the scaled binding energy $\epsilon/\mu_I$ as a function of $\mu_I$
in pion superfluid at $T=0$ and $T=100$ MeV, shown in
Fig.\ref{fig4m}. The binding energy of $\overline\pi_+$ is defined
as the the mass difference between $\overline\pi_+$ and the two
quarks, $\epsilon=M_{\overline\pi_+} -M_u-M_{\overline d}$. With
decreasing isospin chemical potential, the binding energy increases,
indicating a BCS-BEC crossover in the pion superfluid. When the
medium becomes hot, the condensate melts and the pairs are gradually
dissociated.

In pion superfluid, the pairs themselves, namely the pion mesons,
are observable objects. We propose to measure the $\pi-\pi$
scattering to probe the properties of the pion condensate and in
turn the BCS-BEC crossover. On one hand, since pions are Goldstone
modes corresponding to the chiral symmetry spontaneous breaking, the
$\pi-\pi$ scattering provides a direct way to link chiral theories
and experimental data and has been widely studied in many chiral
models~\cite{chiralpi1,chiralpi11,schulze,quack,huang}. Note that
pions are also the Goldstone modes of the isospin symmetry
spontaneous breaking, the $\pi-\pi$ scattering should be a sensitive
signature of the pion superfluid phase transition. On the other
hand, the $\pi-\pi$ scattering behaves different according to the
BCS-BEC crossover picture. In the BCS quark superfluid, the large
and overlapped pairs lead to large pair-pair cross section, but the
small and individual pairs in the BEC superfluid interact weakly.

\begin{figure}[hbt]
\centering
\includegraphics[width=8.4cm]{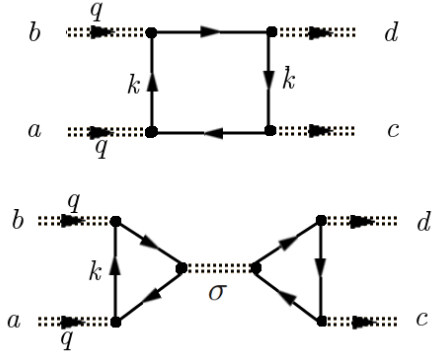}
\caption{The lowest order diagrams for $\pi-\pi$ scattering in the
pion superfluid. The solid and dashed lines are respectively quarks
and mesons (pions or $\sigma$), and the dots denote meson-quark vertices. } \label{fig1m}
\end{figure}

We now study $\pi-\pi$ scattering at finite isospin chemical
potential. To the lowest order in $1/N_c$ expansion, where $N_c$ is
the number of colors, the invariant amplitude ${\cal T}$ is
calculated from the diagrams shown in Fig.\ref{fig1m} for the $s$
channel. Note that the $s$-wave
$\pi-\pi$ scattering calculated~\cite{schulze,quack,huang} in NJL model of $1/N_c$ order is
consistent with the Weinberg limit~\cite{pipi} and the
experimental data~\cite{pocanic} in vacuum. Different from the calculation in normal
state~\cite{schulze,quack,huang} where both the box and $\sigma$-exchange
diagrams contribute, the $\sigma$-exchange diagrams vanish in the
pion superfluid due to the disappearance of the $\overline\sigma$
meson. This greatly simplifies the calculation in the pion
superfluid.

For the calculation in normal matter at $\mu_I=0$, people are
interested in the $\pi$-$\pi$ scattering amplitude with definite
isospin, ${\cal T}_{I=0,1,2}$, which can be measured in experiments
due to isospin symmetry. However, the nonzero isospin chemical
potential breaks down the isospin symmetry and makes the scattering
amplitude ${\cal T}_{I=0,1,2}$ not well defined. In fact, the new
meson modes in the pion superfluid do not carry definite isospin
quantum numbers. Unlike the chiral dynamics in normal matter, where
the three degenerated pions are all the Goldstone modes
corresponding to the chiral symmetry spontaneous breaking, the pion
mass splitting at finite $\mu_I$ results in only one Goldstone mode
$\overline\pi_+$ in the pion superfluid.

The scattering amplitude for any channel of the box diagrams can be
expressed as
\begin{equation}
\label{tstu} i {\cal T}_{s,t,u}(q) = -2 g_{\overline\pi q \overline
q}^4 \int {d^4 k \over (2 \pi)^4}Tr \prod_{l=1}^4\left[\gamma_5 \tau
{\cal S}_l\right]
\end{equation}
with the quark propagators ${\cal S}_1={\cal S}_3={\cal S}(k)$,
${\cal S}_2={\cal S}(k+q)$, and ${\cal S}_4={\cal S}(k-q)$ for the
$s$ and $t$ channels and ${\cal S}_1={\cal S}_3={\cal S}(k+q)$ and
${\cal S}_2={\cal S}_4={\cal S}(k)$ for the $u$ channel. To simplify
the numerical calculation, we consider in the following the limit of
the scattering at threshold $\sqrt{s}=2M_{\overline\pi}$ and
$t=u=0$, where $s, t$ and $u$ are the Mandelstam variables. In this
limit, the amplitude approaches to the scattering length. Note that
the threshold condition can be fulfilled by a simple choice of the
pion momenta, $q_a=q_b=q_c=q_d=q$ and $q^2=M_{\overline\pi}^2=s/4$,
which facilitates a straightforward computation of the diagrams.
Doing the fermion frequency summation over the internal quark lines,
the scattering amplitude for the process of $\overline \pi_+ \ + \
\overline \pi_+ \rightarrow \overline \pi_+ \ + \ \overline \pi_+$
in the pion superfluid is simplified as
\begin{eqnarray}
\label{totalt}
{\cal T}_+ &=& 18 g_{\overline\pi_+ q \overline
q}^4\int {d^3{\bf k}\over (2 \pi)^3}\Bigg\{{1\over
E_+^3}\Big[\Big(f(E_+^-)-f(-E_+^+)\Big)\nonumber\\
&&-E_+\Big(
f'(E_+^-)+f'(-E_+^+)\Big)\Big]\nonumber\\
&&+{1\over
E_-^3}\Big[\Big(f(E_-^-)-f(-E_-^+)\Big)\nonumber\\
&&-E_-\Big(f'(E_-^-)+f'(-E_-^+)\Big)\Big]\Bigg\},
\end{eqnarray}
where $E_\pm^\mp=E_\pm \mp\mu_B/3$ are the energies of the four
quasiparticles with
$E_\pm=\sqrt{\left(E\pm\mu_I/2\right)^2+4G^2\pi^2}$ and
$E=\sqrt{{\bf k}^2+M_q^2}$, $f(x)$ is the Fermi-Dirac distribution
function $f(x)=\left(e^{x/T}+1\right)^{-1}$, and $f'(x)=df/dx$ is
the first order derivative of $f$. For the scattering amplitude
outside the pion superfluid, one should consider both the box and
$\sigma$-exchange diagrams. The calculation is straightforward.

\begin{figure}[hbt]
\centering
\includegraphics[width=7.5cm]{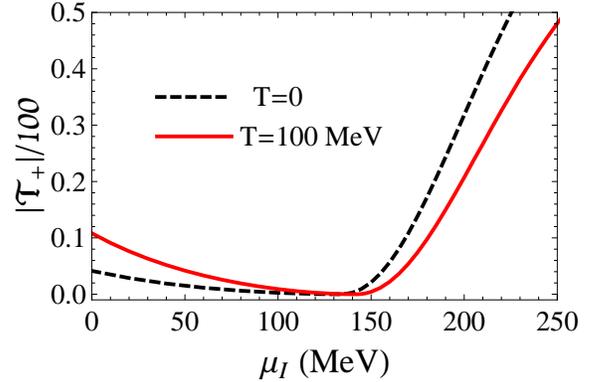}
\caption{ (Color online) The scaled scattering amplitude ${\cal
T}_+$ as a function of isospin chemical potential $\mu_I$ at two
values of temperature $T$. } \label{fig2m}
\end{figure}

In Fig.\ref{fig2m}, we plot the
scattering amplitude $|{\cal T}_+|$ as a function of isospin
chemical potential $\mu_I$ at two temperatures $T=0$ and $T=100$
MeV, keeping baryon chemical potential $\mu_B=0$. The normal matter with $\mu_I<\mu_I^c$
is dominated by the explicit isospin
symmetry breaking and spontaneous chiral symmetry breaking, and the pion superfluid with $\mu_I>\mu_I^c$ and the
corresponding BEC-BCS crossover is controlled by the spontaneous
isospin symmetry breaking and chiral symmetry restoration. From (\ref{tstu}),
the scattering amplitude is governed by the meson coupling constant,
${\cal T}_+\sim g_{\overline\pi_+ q \overline q}^4$. From Fig.\ref{figcou}, the meson mode $\overline\pi_+$ in the pion superfluid phase is always a bound state,
its coupling to quarks drops down with decreasing $\mu_I$,
and therefore the scattering amplitude $\left|{\cal T}_+\right|$
decreases when the system approaches to the phase transition and
reaches zero at the critical point $\mu^c_I$, due to
$g_{\overline\pi_+ q \overline q}=0$ at this point. The critical isospin chemical potential
is $\mu^c_I=m_\pi=134$ MeV at $T=0$ and $142$ MeV at $T=100$
MeV. After crossing the border of the phase
transition, the coupling constant changes its trend and starts to go up with decreasing isospin chemical potential
in the normal matter, and the scattering amplitude smoothly increases and finally approaches its vacuum value at $\mu_I\to 0$.

The above $\mu_I$-dependence of the meson-meson scattering amplitude
in the pion superfluid with $\mu_I>\mu_I^c$ can be understood well
from the point of view of BCS-BEC crossover. We recall that the BCS
and BEC states are defined in the sense of the degree of overlapping
among the pair wave functions. The large pairs in BCS state overlap
each other, and the small pairs in BEC state are individual objects.
Therefore, the cross section between two pairs should be large in
the BCS state and approach zero in the limit of BEC. From our
calculation shown in Fig.\ref{fig2m}, the $\pi-\pi$ scattering
amplitude is a characteristic quantity for the BCS-BEC crossover in
pion superfluid. The overlapped quark-antiquark pairs in the BCS
state at higher isospin density have large scattering amplitude,
while in the BEC state at lower isospin density with separable
pairs, the scattering amplitude becomes small. This provides a
sensitive observable for the BCS-BEC crossover at quark level,
analogous to the fermion scattering in cold atom systems.

\begin{figure}[hbt]
\centering
\includegraphics[width=7.5cm]{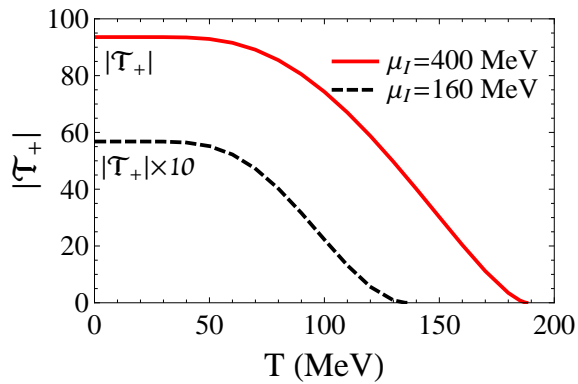}
\caption{(Color online) The scattering amplitude ${\cal T}_+$ as a
function of temperature $T$ at two values of isospin chemical
potential $\mu_I$ in the pion superfluid.} \label{fig3m}
\end{figure}
The minimum of the scattering amplitude at the critical point can generally be
understood in terms of the interaction between the two quarks. A strong interaction means a tightly
bound state with small meson size and small meson-meson cross section, and
a weak interaction means a loosely bound state with large meson size
and large meson-meson cross section. Therefore, the minimum of the meson scattering
amplitude at the critical point indicates the most strong quark interaction at the phase transition.
This result is consistent with theoretical calculations for the ratio $\eta/s$~\cite{kovtun,csernai} of
shear viscosity to entropy density and for the quark potential~\cite{mu2,jiang}, which show
a strongly interacting quark matter around the phase transition.

With increasing temperature, the pairs will gradually melt and the
coupling constant $g_{\overline\pi q\overline q}$ drops down in the
hot medium, leading to a smaller scattering amplitude at $T=100$ MeV
in the pion superfluid, in comparison with the case at $T=0$, as
shown in Fig.\ref{fig2m}. To see the continuous temperature effect on
the scattering amplitude in the BCS and BEC states, we plot in
Fig.\ref{fig3m} $\left|{\cal T}_+\right|$ as a function of $T$ at
$\mu_I=160$ and $\mu_I=400$ MeV, still keeping $\mu_B=0$. While the
temperature dependence is similar in both cases, the involved
physics is different. In the BCS state at $\mu_I=400$ MeV,
$\left|{\cal T}_+\right|$ is large and drops down with increasing
temperature and finally vanishes at the critical temperature
$T_c=188$ MeV. Above $T_c$ the system becomes a fermion gas with
weak coupling and without any pair. In the BEC state at $\mu_I=160$
MeV, the scattering amplitude becomes much smaller (multiplied by a
factor of $10$ in Fig.\ref{fig3m}). At a lower critical temperature
$T_c=136$ MeV, the condensate melts but the still strong coupling
between quarks makes the system be a gas of free pairs.

The meson scattering amplitude $\left|{\cal T}_+\right|$ shown in
Figs.\ref{fig2m} and \ref{fig3m} are obtained in a particular model,
the NJL model, which has proven to be rather reliable in the study
on chiral, color and isospin condensates at low temperature. Since
there is no confinement in the model, one may ask the question to
what degree the conclusions obtained here can be trusted. From the
general picture for BCS and BEC states, the feature that the meson
scattering amplitude approaches to zero in the process of BCS-BEC
crossover can be geometrically understood in terms of the degree of
overlapping between the two pairs. Therefore, the qualitative
conclusion of taking meson scattering as a probe of BCS-BEC
crossover at quark level may survive any model dependence. Our
result that the molecular scattering amplitude approaches to zero in
the BEC limit is consistent with our previous result in
Eq.(\ref{adda}) and the recent work for a general fermion
gas~\cite{he2}. Different from a system with finite baryon density
where the fermion sign problem~\cite{Lreview,Lreview1} makes it
difficult to simulate QCD on lattice, there is in principle no
problem to do lattice QCD calculations at finite isospin
density~\cite{Liso,Liso1,Liso2,Liso3}. From the recent lattice QCD
results~\cite{detmold} at nonzero isospin chemical potential in a
canonical approach, the scattering length in the pion superfluid
increases with increasing isospin density, which qualitatively
supports our conclusion here.

\section {Summary}
\label{s4}

In summary, we have presented in the article the studies of BCS-BEC crossover in relativistic Fermi systems, especially in QCD matter at finite density.

We studied the BCS-BEC crossover in a relativistic four-fermion interaction model. The relativistic effect
is significant: A crossover from nonrelativistic BEC to ultra relativistic BEC is possible, if the attraction can be strong enough.
In the relativistic theory, changing the density of the system can naturally induce a BCS-BEC crossover from high density to low density.
The mean field theory is generalized to including the contribution from uncondensed pairs. Applying the generalized mean field theory
to color superconducting quark matter at moderate density, the role of pairing fluctuations becomes important: The size of the pseudogap
at $\mu\sim 400$MeV can reach the order of $100$ MeV at the critical temperature.

We investigated two-color QCD at finite baryon density in the frame of NJL model. We can describe the weakly interacting diquark condensate at low density and the BEC-BCS crossover at high density. The baryon chemical potential for the predicted crossover is consistent with the lattice simulations of two-color QCD at finite $\mu_{\rm B}$. The study is directly applied to real QCD at finite isospin density. We proposed the meson-meson scattering in pion superfluid as a sensitive
probe of the BCS-BEC crossover.

{\bf Acknowledgement:} LH is supported by the Helmholtz International Center for FAIR within the framework of the LOEWE program launched by the State of Hesse, and SM and PZ are supported by the NSFC and MOST under grant Nos. 11335005, 2013CB922000 and 2014CB845400.

\appendix

\section {The One-Loop Susceptibilities}
\label{app}

In this appendix, we evaluate the explicit forms of the one-loop
susceptibilities $\Pi_{\text{ij}}(Q)$ (${\text i},{\text j}=1,2,3$)
and $\Pi_\pi(Q)$. At arbitrary temperature, their expressions are
rather huge. However, at $T=0$, they can be written in rather
compact forms. For convenience, we define
$\Delta=|\Delta|e^{i\theta}$ in this appendix.

First, the polarization functions $\Pi_{11}(Q)$ and $\Pi_{12}(Q)$ can be evaluated as
\begin{widetext}
\begin{eqnarray}
\Pi_{11}(Q)&=&N_cN_f\sum_{\bf k}\Bigg[\left(\frac{(u_{\bf
k}^-)^2(u_{\bf p}^-)^2}{i\nu_m-E_{\bf k}^- -E_{\bf
p}^-}-\frac{(v_{\bf k}^-)^2(v_{\bf p}^-)^2}{i\nu_m+E_{\bf
k}^-+E_{{\bf p}}^-}-\frac{(u_{\bf k}^+)^2(u_{\bf
p}^+)^2}{i\nu_m+E_{\bf k}^++E_{\bf p}^+}
+\frac{(v_{\bf k}^+)^2(v_{\bf p}^+)^2}{i\nu_m-E_{\bf k}^+-E_{\bf p}^+}\right){\cal T}_+\nonumber\\
&&\ \ \ \ \ \ \ \ \ \ \ \ \ \ +\left(\frac{(u_{\bf k}^-)^2(v_{\bf
p}^+)^2}{i\nu_m-E_{\bf k}^--E_{\bf p}^+}-\frac{(v_{\bf
k}^-)^2(u_{\bf p}^+)^2}{i\nu_m+E_{\bf k}^-+E_{\bf
p}^+}-\frac{(u_{\bf k}^+)^2(v_{\bf p}^-)^2}{i\nu_m+E_{\bf
k}^++E_{\bf p}^-}+\frac{(v_{\bf k}^+)^2(u_{\bf p}^-)^2}{i\nu_m-E_{\bf k}^+-E_{\bf p}^-}\right){\cal T}_-\Bigg],\nonumber\\
\Pi_{12}(Q)&=&N_cN_f\sum_{\bf k}\Bigg[\left(\frac{u_{\bf k}^-v_{\bf
k}^-u_{\bf p}^-v_{\bf p}^-}{i\nu_m+E_{\bf k}^-+E_{\bf
p}^-}-\frac{u_{\bf k}^-v_{\bf k}^-u_{\bf p}^-v_{\bf
p}^-}{i\nu_m-E_{\bf k}^--E_{\bf p}^-}+\frac{u_{\bf k}^+v_{\bf
k}^+u_{\bf p}^+v_{\bf p}^+}{i\nu_m+E_{\bf k}^++E_{\bf
p}^+}-\frac{u_{\bf k}^+v_{\bf k}^+u_{\bf p}^+v_{\bf
p}^+}{i\nu_m-E_{\bf k}^+-E_{\bf p}^+}\right){\cal T}_+
\nonumber\\
&&\ \ \ \ \ \ \ \ \ \ \ \ \ \ +\left(\frac{u_{\bf k}^-v_{\bf
k}^-u_{\bf p}^+v_{\bf p}^+}{i\nu_m+E_{\bf k}^-+E_{\bf
p}^+}-\frac{u_{\bf k}^-v_{\bf k}^-u_{\bf p}^+v_{\bf
p}^+}{i\nu_m-E_{\bf k}^--E_{\bf p}^+}+\frac{u_{\bf k}^+v_{\bf
k}^+u_{\bf p}^-v_{\bf p}^-}{i\nu_m+E_{\bf k}^++E_{\bf
p}^-}-\frac{u_{\bf k}^+v_{\bf k}^+u_{\bf p}^-v_{\bf
p}^-}{i\nu_m-E_{\bf k}^+-E_{\bf p}^-}\right){\cal
T}_-\Bigg]e^{2i\theta},
\end{eqnarray}
\end{widetext}
where ${\bf p}={\bf k}+{\bf q}$. Here ${\cal T}_\pm$ are factors
arising from the trace in spin space,
\begin{equation}
{\cal T}_\pm=\frac{1}{2}\pm\frac{{\bf k}\cdot{\bf p}+M^2}{2E_{\bf
k}E_{\bf p}},
\end{equation}
and $u_{\bf k}^{\pm}, v_{\bf k}^{\pm}$ are the BCS distribution
functions defined as
\begin{equation}
(u_{\bf k}^{\pm})^2=\frac{1}{2}\left(1+\frac{\xi_{\bf k}^\pm}{E_{\bf
k}^\pm}\right),\ \ \ \ \ (v_{\bf
k}^{\pm})^2=\frac{1}{2}\left(1-\frac{\xi_{\bf k}^\pm}{E_{\bf
k}^\pm}\right).
\end{equation}
At $Q=0$, we find
\begin{equation}
\Pi_{12}(0)=\Delta^2\frac{1}{4}N_cN_f\sum_{\bf
k}\left[\frac{1}{(E_{\bf k}^-)^3}+\frac{1}{(E_{\bf k}^+)^3}\right].
\end{equation}
Thus, near the quantum phase transition point, we have
$\Pi_{12}(0)=\Delta^2\beta_1+O(|\Delta|^4)$. On the other hand, a
simple algebra shows
\begin{equation}
\frac{1}{4G}+\Pi_{11}(0)-|\Pi_{12}(0)|=\frac{\partial\Omega_0}{\partial|\Delta|^2}.
\end{equation}
Therefore, the mean-field gap equation for $\Delta$ ensures the
Goldstone's theorem in the superfluid phase.

The term $\Pi_{13}$ standing for the mixing between the sigma meson and the diquarks reads
\begin{widetext}
\begin{eqnarray}
\Pi_{13}(Q)&=&N_cN_f\sum_{\bf k}\Bigg[\left(\frac{u_{\bf k}^+v_{\bf
k}^+(v_{\bf p}^+)^2+u_{\bf p}^+v_{\bf p}^+(v_{\bf
k}^+)^2}{i\nu_m-E_{\bf k}^+-E_{\bf p}^+} +\frac{u_{\bf k}^+v_{\bf
k}^+(u_{\bf p}^+)^2+u_{\bf p}^+v_{\bf p}^+(u_{\bf
k}^+)^2}{i\nu_m+E_{\bf k}^++E_{\bf p}^+}-\frac{u_{\bf k}^-v_{\bf
k}^-(u_{\bf p}^-)^2+u_{\bf p}^-v_{\bf p}^-(u_{\bf
k}^-)^2}{i\nu_m-E_{\bf k}^--E_{{\bf p}}^-} -\frac{u_{\bf k}^-v_{\bf
k}^-(v_{\bf p}^-)^2+u_{\bf p}^-v_{\bf p}^-(v_{\bf
k}^-)^2}{i\nu_m+E_{\bf k}^-
+E_{\bf p}^-}\right){\cal I}_+\nonumber\\
&&+\left(\frac{u_{\bf k}^+v_{\bf
k}^+(u_{\bf p}^-)^2+u_{\bf p}^-v_{\bf p}^-(v_{\bf
k}^+)^2}{i\nu_m-E_{\bf k}^+-E_{{\bf p}}^-} +\frac{u_{\bf k}^+v_{\bf
k}^+(v_{\bf p}^-)^2+u_{\bf p}^-v_{\bf p}^-(u_{\bf
k}^+)^2}{i\nu_m+E_{\bf k}^+ +E_{\bf p}^-}-\frac{u_{\bf k}^-v_{\bf
k}^-(v_{\bf p}^+)^2+u_{\bf p}^+v_{\bf p}^+(u_{\bf
k}^-)^2}{i\nu_m-E_{\bf k}^--E_{\bf p}^+} -\frac{u_{\bf k}^-v_{\bf
k}^-(u_{\bf p}^+)^2+u_{\bf p}^+v_{\bf p}^+(v_{\bf
k}^-)^2}{i\nu_m+E_{\bf k}^-+E_{\bf p}^+}\right){\cal
I}_-\Bigg]e^{i\theta},\nonumber\\
\end{eqnarray}
\end{widetext}
where the factors ${\cal I}_\pm$ are defined as
\begin{equation}
{\cal I}_\pm=\frac{M}{2}\left(\frac{1}{E_{\bf k}}\pm\frac{1}{E_{\bf
p}}\right).
\end{equation}
One can easily find $\Pi_{13}\sim M\Delta$, thus it vanishes
when $\Delta$ or $M$ approaches zero. At $Q=0$, we have
\begin{equation}
\Pi_{13}(0)=\Delta\frac{1}{2}N_cN_f\sum_{\bf k}\frac{M}{E_{\bf
k}}\left[\frac{\xi_{\bf k}^-}{(E_{\bf k}^-)^3}+\frac{\xi_{\bf
k}^+}{(E_{\bf k}^+)^3}\right].
\end{equation}
Thus the quantity $H_0$ defined in (\ref{bexp}) can be evaluated as
\begin{equation}
H_0=\frac{1}{2}N_cN_f\sum_{e=\pm}\sum_{\bf k}\frac{M_*}{E_{\bf
k}^*}\frac{1}{(E_{\bf k}^*-em_\pi/2)^2}=\frac{\partial^2\Omega_0(y,M)}{\partial M\partial y}\Bigg|_{y=0}.
\end{equation}

The polarization function $\Pi_{33}$ which stands for the sigma meson can be evaluated as
\begin{widetext}
\begin{eqnarray}
\Pi_{33}(Q)&=&
 N_cN_f\sum_{\bf k}
\Bigg[(v_{\bf k}^-u_{\bf p}^-+u_{\bf k}^-v_{\bf p}^-)^2\left(
\frac{1}{i\nu_m-E_{\bf k}^--E_{\bf p}^-}
-\frac{1}{i\nu_m+E_{\bf k}^-+E_{\bf p}^-}\right){\cal T}^\prime_- +(v_{\bf k}^+u_{\bf p}^++u_{\bf
k}^+v_{\bf p}^+)^2\left( \frac{1}{i\nu_m-E_{\bf k}^+-E_{\bf p}^+}
-\frac{1}{i\nu_m+E_{\bf k}^++E_{\bf p}^+}\right){\cal T}^\prime_-\nonumber\\
&&+(v_{\bf k}^+v_{\bf p}^-+u_{\bf
k}^+u_{\bf p}^-)^2\left( \frac{1}{i\nu_m-E_{\bf k}^+-E_{\bf p}^-}
-\frac{1}{i\nu_m+E_{\bf k}^++E_{\bf p}^-}\right){\cal T}^\prime_+ +(v_{\bf k}^-v_{\bf p}^++u_{\bf
k}^-u_{\bf p}^+)^2\left( \frac{1}{i\nu_m-E_{\bf k}^--E_{\bf p}^+}
-\frac{1}{i\nu_m+E_{\bf k}^-+E_{\bf p}^+}\right){\cal
T}^\prime_+\Bigg],\nonumber\\
\end{eqnarray}
\end{widetext}
where the factors ${\cal T}^\prime_\pm$ are defined as
\begin{equation}
{\cal T}^\prime_\pm=\frac{1}{2}\pm\frac{{\bf k}\cdot{\bf
p}-M^2}{2E_{\bf k}E_{\bf p}}.
\end{equation}
At $Q=0$ and for $\Delta=0$, we find
\begin{eqnarray}
{\bf M}_{33}(0)&=&\frac{1}{2G}-2N_cN_f\sum_{\bf k}\frac{1}{E_{\bf
k}^*}+2N_cN_f\sum_{\bf k}\frac{M_*^2}{E_{\bf k}^{*3}}\nonumber\\
&=&\frac{\partial^2\Omega_0(y,M)}{\partial M^2}\Bigg|_{y=0}.
\end{eqnarray}
Finally, the polarization function $\Pi_\pi(Q)$ for pions can be
obtained by replacing ${\cal T}^\prime_\pm\rightarrow{\cal T}_\pm$.
Thus, when $M\rightarrow 0$, the sigma meson and pions become
degenerate and chiral symmetry is restored.


\begin{thebibliography}{99}
\bibitem{note4}     The mesons we called here are the collective excitations which have the same quantum numbers as the pions and the sigma meson in the vacuum. They are also called ``pions" and ``sigma meson" in the chiral restored regime.
\end{thebibliography}
\end{document}